\documentclass[english,11pt,a4paper]{article}
\pdfoutput=1
\usepackage{jheppub}

\usepackage[T1]{fontenc}
\usepackage[utf8]{inputenc}
\usepackage{lmodern, amsmath, amssymb, xcolor, slashed, xfrac}
\usepackage{subfigure}
\usepackage{bm}
\usepackage{scalefnt}
\usepackage{comment}
\usepackage{latexsym}
\usepackage{relsize}
\usepackage[normalem]{ulem}
\usepackage{booktabs} 
\usepackage{colortbl} 
\usepackage{mathptmx} 
\DeclareMathAlphabet{\mathcal}{OMS}{cmsy}{m}{n} 
\graphicspath{{figs/}}

\def\figureautorefname~#1\null{Fig.\,#1\null}

\newcommand{\subappref}[1]{\hyperref[#1]{appendix~\ref{#1}}}

\def\equationautorefname~#1\null{Eq.\,(#1)\null}
\renewcommand{\eqref}[1]{\hyperref[#1]{(\ref{#1})}}

\definecolor{light_blue}{rgb}{0.15, 0.35, 0.95}
\definecolor{kit_green}{rgb}{0, 
0.58823 
, 0.50980 
}

\hypersetup{
  colorlinks = true,
  linkcolor = light_blue, 
  citecolor = light_blue, 
  urlcolor = kit_green 
}

\allowdisplaybreaks[4]
\def\beq#1\eeq{\begin{align}#1\end{align}}
\newcommand{\beqa}{\begin{eqnarray}}
\newcommand{\eeqa}{\end{eqnarray}}

\newcommand{\bpm}{\begin{pmatrix}}
\newcommand{\epm}{\end{pmatrix}}

\newcommand{\splitFunc}[1]{\mathrm{Split}}

\makeatletter
\def\EE{\@ifnextchar-{\@@EE}{\@EE}}
\def\@EE#1{\ifnum#1=1 \times10 \else \times10^{#1}\fi}
\def\@@EE#1#2{\times10^{-#2}}
\makeatother

\newcommand\unit[1]{\,\,\mathrm{#1}}

\newcommand\GeV{\unit{GeV}}
\newcommand\TeV{\unit{TeV}}

\newcommand\ifb{\unit{fb^{-1}}}

\newcommand{\eg}{{\em e.g.}}
\newcommand{\ie}{{\em i.e.}}

\def\Babar{{\mbox{\slshape B\kern-0.1em{\smaller A}\kern-0.1em B\kern-0.1em{\smaller A\kern-0.2em R}}}}
\newcommand{\tr}{_{\mathrm{T}}}
\newcommand{\nubar}{\overline{\nu}}

\RequirePackage{xspace}
\def\Bbar    {\kern 0.18em\overline{\kern -0.18em B}{}\xspace}
\def\Bb      {\ensuremath{\Bbar}\xspace}

\newcommand{\kf}{\negthinspace\relax}

\makeatletter\g@addto@macro\bfseries{\boldmath}\makeatother

\title{Non-resonant new physics search at the LHC 
for the $\boldsymbol{b \to c \tau \nu}$ anomalies}

\author[a,b,c]{Motoi Endo,}
\author[c,d,e,f]{Syuhei Iguro,}
\author[g,h]{Teppei Kitahara,}
\author[h,i]{Michihisa Takeuchi,}
\author[j]{and\\ Ryoutaro Watanabe}

\affiliation[a]{KEK Theory Center, IPNS, KEK, Tsukuba 305--0801, Japan}
\affiliation[b]{The Graduate University of Advanced Studies (Sokendai), Tsukuba 305--0801, Japan}
\affiliation[c]{Kavli Institute for the Physics and Mathematics of the Universe (WPI), The University of Tokyo Institutes for Advanced Study, The University of Tokyo, Kashiwa 277--8583, Japan}
\affiliation[d]{Department of Physics,
Nagoya University, Nagoya 464--8602, Japan}
\affiliation[e]{Institute for Theoretical Particle Physics (TTP), Karlsruhe Institute of Technology (KIT),
Engesserstra{\ss}e 7, 76131 Karlsruhe, Germany}
\affiliation[f]{Institute for Astroparticle Physics (IAP),
Karlsruhe Institute of Technology (KIT), 
Hermann-von-Helmholtz-Platz 1, 76344 Eggenstein-Leopoldshafen, Germany}
\affiliation[g]{Institute for Advanced Research, Nagoya University, Nagoya 464--8601, Japan}
\affiliation[h]{Kobayashi-Maskawa Institute for the Origin of Particles and the Universe, 
Nagoya University,  Nagoya 464--8602, Japan}
\affiliation[i]{Department of Physics, Osaka University, Toyonaka 560--0043, Japan}
\affiliation[j]{INFN, Sezione di Pisa, Largo Bruno Pontecorvo 3,  I-56127 Pisa, Italy}

\emailAdd{motoi.endo@kek.jp}
\emailAdd{igurosyuhei@gmail.com}
\emailAdd{teppeik@kmi.nagoya-u.ac.jp}
\emailAdd{m.takeuchi@het.phys.sci.osaka-u.ac.jp}
\emailAdd{wryou1985@gmail.com }

\abstract{Motivated by the $b \to c \tau \overline{\nu}$ anomalies, 
we study non-resonant searches for new physics at the large hadron collider (LHC) 
by considering final states with an energetic and hadronically decaying $\tau$ lepton, a $b$-jet and 
large missing transverse momentum ($pp \to \tau_h \overline{b} + E_{\rm T}^{\rm miss}$).
Such searches can be useful to probe new physics contributions to $b \to c \tau \overline{\nu}$.
They are analyzed not only within the dimension-six effective field theory (EFT) but also in explicit leptoquark (LQ) models with the LQ non-decoupled. 
The former is realized by taking a limit of large LQ mass in the latter. 
It is clarified that the LHC sensitivity is sensitive to the LQ mass for $\mathcal{O}(1)\TeV$ even in the search of $pp \to \tau_h \overline{b} + E_{\rm T}^{\rm miss}$. 
Although the LQ models provide a weaker sensitivity than the EFT limit, 
it is found that the non-resonant search of $pp \to \tau_h \overline{b} + E_{\rm T}^{\rm miss}$ can improve the sensitivity by 
$\approx 40\%$
versus a conventional mono-$\tau$ search ($pp \to \tau_h + E_{\rm T}^{\rm miss}$) in the whole LQ mass region.
Consequently, it is expected that most of the parameter regions suggested by the $b \to c \tau \overline{\nu}$ anomalies can be probed at the HL-LHC. 
Also, it is shown that $\text{R}_2$ LQ scenario is accessible entirely once the LHC Run 2 data are analyzed.
In addition, we discuss a charge selection of $\tau_h$ to further suppress the standard-model background, and investigate 
the angular correlations among $b,\, \tau$ and the missing transverse momentum to discriminate the LQ scenarios.
} 

\preprint{IPMU21--0074, KEK--TH--2366, OU--HET--1115, P3H--21--090, TTP21--046}

\begin{document}

\sloppy 

\maketitle

\renewcommand{\thefootnote}{\#\arabic{footnote}}
\setcounter{footnote}{0}

\section{\boldmath Introduction}
\label{Sec:introduction}

Semi-leptonic $B$-meson decay processes have been investigated
to test the Standard Model (SM) and to search for a hint for New Physics (NP). 
In the last decade, the BaBar~\cite{Lees:2012xj,Lees:2013uzd}, Belle~\cite{Huschle:2015rga,Hirose:2016wfn,Hirose:2017dxl,Abdesselam:2019dgh,Belle:2019rba} and LHCb collaborations~\cite{Aaij:2015yra,Aaij:2017uff,Aaij:2017deq} have reported exiting anomalies
in semi-leptonic decays of $B$ mesons, such as
$R_{D^{(\ast)}}=\text{BR}(B\to D^{(*)}\tau\nubar)/\text{BR}(B\to D^{(*)}\ell\nubar)$,
with $\ell=\mu$ for LHCb and an average of $e$ and $\mu$ for BaBar and Belle. 
Here, a ratio of the branching ratios is taken to reduce 
both experimental and theoretical (\ie, parametric and QCD) uncertainties significantly, so that 
$R_{D^{(\ast)}}$ is sensitive to NP that couples to quarks and leptons.
Although the latest result released by Belle becomes closer to the SM values~\cite{Abdesselam:2019dgh,Belle:2019rba}, 
the world average of $R_{D^{(\ast)}}$ measurements 
still deviates from the SM predictions at the $3\text{--} 4\,\sigma$ confidence level (CL)
(see Ref.~\cite{Iguro:2020cpg} for a recent summary of the SM predictions).

The $R_{D^{(\ast)}}$ discrepancy suggests violation of the lepton flavor universality (LFU) between $\tau$ and light leptons,
and has prompted many attempts of the NP introducing new scalar and vector mediators
(see, \eg, Ref.~\cite{London:2021lfn} for the very recent review). 
In terms of the low-energy effective Hamiltonian, 
their contributions are encoded as 
\begin{align}
 {\mathcal H}_{\rm{eff}}= 
 2 \sqrt 2 G_FV_{cb} \Bigl[ 
 & (1 + C_{V_1}) (\overline{c} \gamma^\mu P_Lb)(\overline{\tau} \gamma_\mu P_L \nu_{\tau})  +C_{V_2} (\overline{c} \gamma^\mu P_Rb)(\overline{\tau} \gamma_\mu P_L \nu_{\tau}) \notag \nonumber\\
 & +C_{S_1} (\overline{c} P_Rb)(\overline{\tau} P_L \nu_{\tau})  +C_{S_2}(\overline{c} P_Lb)(\overline{\tau} P_L \nu_{\tau}) \nonumber  \\
 & +C_{T} (\overline{c}  \sigma^{\mu\nu}P_Lb)(\overline{\tau} \sigma_{\mu\nu} P_L \nu_{\tau}) \Bigl] + \,\text{h.c.}\,,
  \label{Eq:effH}
\end{align} 
with $P_{L/R}=(1\mp\gamma_5)/2$.\kf\footnote{%
The Wilson coefficients are also shown as $C_{V_1}=C_{V}^L,\,C_{V_2}=C_{V}^R,\,C_{S_1} = C_{S}^R,$ and $C_{S_2} = C_{S}^L$~\cite{Blanke:2018yud}.}\footnote{
In this paper, right-handed neutrinos are not considered (or equivalently assumed to be heavier than the $B$ meson). 
See Refs.~\cite{Iguro:2018qzf,Asadi:2018wea,Greljo:2018ogz,Robinson:2018gza,Babu:2018vrl} for models with light right-handed neutrinos in the context of the $R_{D^{(\ast)}}$ anomaly.} 
Here, the Wilson coefficients (WCs), $C_X$, are normalized by the SM contribution, 
$\mathcal{H}_{\rm eff} = 2 \sqrt 2 G_F V_{cb} (\overline{c} \gamma^\mu P_Lb)(\overline{\tau} \gamma_\mu P_L \nu_{\tau})$, corresponding to $C_X=0$ for $X=V_{1,2},S_{1,2},T$.
Note that the SM contribution is suppressed by the Cabibbo-Kobayashi-Maskawa (CKM) matrix element $V_{cb}$~\cite{Cabibbo:1963yz,Kobayashi:1973fv}, where $V_{cb}=0.041$~\cite{Zyla:2020zbs} is set throughout this paper. 
One can see that a scale of NP implied by the $R_{D^{(\ast)}}$ anomaly is restricted as $\lesssim \mathcal{O}(10)\TeV$ by the perturbative unitarity limit on NP interactions~\cite{DiLuzio:2017chi}.

The large hadron collider (LHC) experiment has a great potential to test such NP contributions.
They can be probed, \eg, by resonant searches for new particles such as charged Higgs, $W^\prime$ (and related $Z^\prime$), and leptoquark (LQ), and 
by non-resonant searches for the contact interactions of Eq.~\eqref{Eq:effH}.
In addition to various flavor measurements,
\eg, $B_c \to \tau \nubar$, $\Lambda_b \to \Lambda_c \tau \nubar$, and polarization observables in $B \to D^{(\ast)}\tau\nubar$ in the near future,
which have been studied to check those contributions,
the collider searches provide independent information.
Moreover, they are free from uncertainties of the flavor observables especially inherent in $B\to D^{(\ast)}$ hadronic form factors.

In this paper, we examine non-resonant searches in light of the $R_{D^{(\ast)}}$ anomaly. 
Even if new particles are heavier than the LHC beam collision energy, 
their contributions could be detected indirectly by exchanging these particles in $t$-channel propagators. 
The ATLAS and CMS collaborations have performed non-resonant searches 
especially to probe $W^\prime$ boson (with assuming a decay $W^\prime \to \tau \nu$) in the sequential standard model.
They have done a $\tau\nu$ search, \ie, analyzed events with a hadronic $\tau$ jet and a large missing transverse momentum by using the Run 1 and 2 data~\cite{CMS:2015hmx,Aaboud:2018vgh,Sirunyan:2018lbg,ATLAS:2021bjk}. 
The results are consistent with the SM background (BG) expectations, and one can use them to set upper bounds on the NP interactions relevant to the $R_{D^{(\ast)}}$ anomaly, or the operators in Eq.~\eqref{Eq:effH}. 
References~\cite{Faroughy:2016osc,Iguro:2018fni,Mandal:2018kau,Greljo:2018tzh} have studied such an interplay, \ie, the relation between the high-$p\tr$ tail of the $\tau\nu$ events at the LHC and the $R_{D^{(\ast)}}$ anomaly in new physics models.

Recently, it has been pointed out that sensitivities to the NP may be improved versus the above non-resonant $\tau\nu$ search by requiring an additional $b$-jet in the final state~\cite{Altmannshofer:2017poe,Iguro:2017ysu,Abdullah:2018ets}. 
This can be understood from the fact that the genuine $\tau\nu+b$ final state is achieved by $gq \to b\tau\nu$ ($q=u,c$) within the SM.
Since this contribution is suppressed by $|V_{ub\,(cb)}|^2\sim \mathcal{O}(10^{-5\,(-3)})$, the main SM background comes from $\tau\nu+j$ events with mis-identifying a light-flavored jet as $b$ jet.   
This is in contrast to the $\tau\nu$ search, whose SM contributions, \eg, $\overline{u} d \to \tau \nubar$, are not suppressed by the CKM factors or mis-identifications. 
In addition, the additional $b$ quark allows us to study angular correlations among the final state particles,
which are potentially useful 
to distinguish the NP interactions.
Such a channel has been studied in Ref.~\cite{Marzocca:2020ueu} for general NP contact interactions, 
including those relevant to the $R_{D^{(\ast)}}$ anomaly.
They have argued that sensitivities to each WC searches can be improved by $\sim 30\%$ versus the $\tau\nu$ search. 
Moreover, it was argued that 
angular correlations between $b$ and $\tau$ or $\nu$ would be useful to distinguish possible NP scenarios working in the center of mass frame.

After the above analyses, there are significant developments within the context of the $\tau\nu$ search.
In the previous studies, the effective field theory (EFT) approach (\ie, the contact-interaction approximation) had been taken to describe the NP contributions. 
However, as pointed out in Ref.~\cite{Iguro:2020keo}, this prescription is not always appropriate to represent actual NP contributions when the LHC non-resonant search is studied. 
In fact, a transverse mass defined as 
\begin{align}
 m\tr = \sqrt{2p\tr^\tau E\tr^\textrm{miss} \left[1- \cos \Delta \phi (\vec{p}\tr^\tau, \vec{p}\tr^{\,\rm miss})\right]}  \,,
\end{align} 
is often introduced to analyze high-$p\tr$ events, where $\Delta \phi$ is a relative angle $(0\leq \Delta \phi\leq \pi)$ and the missing transverse momentum is expressed 
by $ \vec{p}\tr^{\,\rm miss}$ with magnitude $E_{\rm T}^{\rm miss}$.
Since a new particle appearing in the $t$-channel propagator is likely to carry a large momentum transfer to produce a high-$p\tr$ $\tau$ lepton and it produces an effective new particle mass (since $t <0$),
\beq
\mathcal{M}_{\rm LQ} \approx  \frac{g_{\rm LQ}^2}{\left|t-M_{\rm LQ}^2\right|} < \frac{g_{\rm LQ}^2}{M_{\rm LQ}^2}
\approx  \mathcal{M}_{\rm EFT} \,,
\eeq
the EFT description is no longer appropriate. 
We can see that sensitivities to the NP tend to become weaker than those in the EFT description, which is valid only for $M_\text{LQ} \gg m\tr$.
Although the study in Ref.~\cite{Iguro:2020keo} has been done for the non-resonant $\tau\nu$ search, a similar conclusion can hold for the $\tau\nu+b$ case. 
In this paper, it will be shown that 
the sensitivity to the WCs can be weakened
by up to $50\%$ even for $\tau\nu+b$.

Moreover, it is pointed out that the NP sensitivity can be improved by choosing negative-charge mono-$\tau$ events~\cite{Iguro:2020keo}.
This follows from the fact that the dominant SM background comes from $pp\to W^{\pm(\ast)} \to\tau^\pm\nu$, and
then the imbalance of $N(W^+)/N(W^-)>2$ is observed due to reflecting the proton charge~\cite{CMS:2016qqr,Hou:2019efy}.
This is in contrast to the NP case: the interaction in Eq.~\eqref{Eq:effH} predicts 
$N(\tau^+)/N(\tau^-)=1$
because the contribution is not generated from valence quarks.
In fact, in order to distinguish the charge of the $\tau$ jet, one has to observe a sagitta of the charged pion from the $\tau$ decay.
In the 
high-$p\tr$ region such as $p\tr^{\pi^\pm} = 1\TeV$, the sagitta in the CMS inner detector becomes $\mathcal{O}(100)\,\mu$m.
Since this is larger than the detector resolution, the charge of $\tau$ jet with $p\tr^{\tau} = \mathcal{O}(1)\TeV$ could be distinguished with good accuracy.
Therefore, it is important to study impacts of the charge selection.

In this paper, we perform a comprehensive analysis of the non-resonant $\tau\nu+b$ search as well as the $\tau\nu$ one with adopting the above developments.
We also discuss directions of further improvements of the NP sensitivity 
especially to distinguish the NP interaction operators, \eg, by utilizing the charge asymmetry of $\tau^\pm$ and the angular correlations among the final states.

This paper is organized as follows.
A model setup is explained in Sec.~\ref{Sec:Set_up}.
A strategy to generate the background and signal events is explained in Sec.~\ref{Sec:event_generation}. 
Numerical results and future prospects are explored in Sec.~\ref{Sec:numerical_results}.
Impacts of their sensitivities on the NP interpretation for the notorious $R_{D^{(*)}}$ anomaly are also given in this section.
Section~\ref{Sec:summary} is devoted to conclusions and discussion.

\section{New physics scenarios}
\label{Sec:Set_up}

In this paper, leptoquark (LQ) models are employed as an illustrative realization of the WCs of the effective Hamiltonian in Eq.~\eqref{Eq:effH}. 
They form the WCs at the NP scale $\Lambda \sim M_\text{LQ}$ as 
\begin{align}
 \label{Eq:WCEFT}
 2 \sqrt 2 G_F V_{cb} C_X (\Lambda) = N_X {h_1 h_2 \over M_\text{LQ}^2} \,,
\end{align}
with LQ mass $M_\text{LQ}$ and LQ couplings to the SM fermions $h_{1,2}$. 
The numerical factor $N_X$ depends on the Lorenz structure of the EFT operator ($X=V_{1,2},S_{1,2},T$). 

We are interested in NP scenarios that can explain the $R_{D^{(\ast)}}$ anomaly. 
Solutions to the anomaly are given in terms of $C_X$ in the  literature, {\it e.g.}, see Refs.~\cite{Iguro:2020keo,Blanke:2019qrx,Iguro:2018vqb}. 
A general consensus is, for instance, that scenarios with a single NP operator $X = V_{1,2}$ work well, which can be realized in particular LQ models.
Also, there are LQ models which contribute to multiple WCs. 

Given the LQ mass $M_\text{LQ}$, the high-$p\tr$ search puts an upper bound on the LQ couplings 
and the WCs in Eq.~\eqref{Eq:effH} at the $\Lambda_\text{LHC}$ scale,
which is encoded as $C_X (\Lambda_\text{LHC})$ in this paper.
The LHC scale reflects the high-$p\tr$ region sensitive to the NP signal. 
Hence, in the following analysis, we take a typical size as 
$\Lambda_\text{LHC} = 1\TeV$, which is the same as Ref.~\cite{Marzocca:2020ueu}.
In the flavor physics, the EFT limit $q^2 \ll \Lambda^2$ is a good approximation for $\Lambda \gtrsim \mathcal O(100)\GeV$. 
However, as mentioned in the introduction, this is not the case for the high-$p\tr$ searches at the LHC, where $p\tr$ can be $\mathcal{O}(1)\TeV$.
Thus, we will investigate explicit $M_\text{LQ}$ dependences
of the sensitivities 
of $C_X(\Lambda_\text{LHC})$.

In the following subsections, we show explicit LQ models to setup the NP scenarios of our interest and also give a brief explanation for collider signatures.

\subsection{$\text{U}_1$ LQ}
\label{Sec:U1LQ}

The $SU(2)_L$ singlet vector LQ ($\text{U}_1$) is one of the well-known candidates to explain several $B$ anomalies~\cite{Buttazzo:2017ixm,Angelescu:2018tyl,Angelescu:2021lln}. 
Its interaction is written as  
\begin{align}
 \label{Eq:U1LQ}
 \mathcal L_\text{$\text{U}_1$LQ} = 
 h_L^{ij} \Big(\overline u_i \gamma_\mu P_L \nu_j + \overline d_i \gamma_\mu P_L \ell_j \Big) \text{U}_{1}^\mu 
 + h_R^{ij} \Big( \overline d_i \gamma_\mu P_R \ell_j \Big) \text{U}_{1}^\mu
 + \text{h.c.}\,.
\end{align}
By integrating out the LQ, one can obtain two WCs as 
\begin{align}
 \label{Eq:U1LQWC}
 &2\sqrt{2}G_FV_{cb} C_{V_1} = + {\left(V_{\rm CKM} h_L\right)^{23} h_L^{*33} \over M_{\text{LQ}}^2}\,,&
 &2\sqrt{2}G_FV_{cb} C_{S_1} = -2 {\left(V_{\rm CKM} h_L\right)^{23} h_R^{*33} \over M_{\text{LQ}}^2}\,.&
\end{align}
It is noticed that these WCs depend on different couplings, \ie, are independent with each other. 
The couplings irrelevant to $b \to c \tau\nubar$ are assumed to be zero. 

The scenario with $C_{V_1} \neq 0$ and $C_{S_1} = 0$, so-called the single $C_{V_1}$ scenario, is realized by taking $h_R^{*33}=0$, which will be investigated later. 
Note that $C_{V_1} \neq 0$ can also be obtained by other LQ models such as the $SU(2)_L$ triplet vector ($\text{U}_3$), singlet scalar ($\text{S}_1$), and triplet scalar ($\text{S}_3$) LQs. 
However, these models confront a stringent constraint from $b \to s \nu\nubar$ unavoidably in single LQ scenarios at the tree level, see Appendix~\ref{Sec:App_flavor}. 
For instance, $|C_{V_1}| \lesssim 0.03$ is obtained for the $\text{S}_1$ LQ scenario, which is not consistent with the $R_{D^{(\ast)}}$ solution, $C_{V_1} = 0.09 \pm 0.02$. 
Hence, the $\text{U}_1$ LQ is the only possibility to realize this scenario (see Ref.~\cite{Crivellin:2017zlb} for alternative possibility by use of multiple LQs).
Note that 
the constraints from $\Delta M_s$ and $\Delta M_s/ \Delta M_d$ are UV-model dependent.
For $\text{U}_1$ LQ models, additional vector-like leptons are often incorporated in the UV models.
These constraints are weakened by incorporating light vector-like leptons contributions via a GIM-like mechanism~\cite{DiLuzio:2018zxy, Fuentes-Martin:2020hvc}.

Another scenario has been discussed in the context of a $U(2)$ flavor symmetry~\cite{Barbieri:1995uv,Barbieri:1997tu,Barbieri:2011ci,Barbieri:2011fc,Barbieri:2012uh,Blankenburg:2012nx,Barbieri:2015yvd,Fuentes-Martin:2019mun}. 
In this scenario, $h_L$ and $h_R$ are aligned, and the two WCs are related as 
\begin{align}
 C_{S_1} = -2 \beta_R C_{V_1} \,, 
 \label{Eq:betaR}
\end{align}
where $\beta_R = e^{i \phi_R}$ denotes a relative phase~\cite{Fuentes-Martin:2019mun}. 
Assuming $C_{V_1}$ to be real, the result to explain the $R_{D^{(\ast)}}$ anomaly is given as $\phi_R \sim 0.4\pi$ and $C_{V_1} \sim 0.09$. 
This scenario will also be investigated in this paper.
Note that the LHC study is less sensitive to the phase.

\subsection{$\text{R}_2$ LQ}
\label{Sec:R2LQ}

The $SU(2)_L$ doublet scalar LQ ($\text{R}_2$)
also provides distinctive solutions to the $R_{D^{(\ast)}}$ anomaly \cite{Becirevic:2018afm}.
A {\it practical} $\text{R}_{2}$ LQ model introduces two distinct 
LQ doublets, 
$\text{\bf R}_{2,1} 
= (\text{R}_{2,1}^{5/3},\,\text{R}^{2/3}_{2,1})$ and 
$\text{\bf R}_{2,2}
= (\text{R}^{2/3}_{2,2},\,\text{R}_{2,2}^{-1/3})$,
in the SM gauge invariant form, 
for which a large mixing between $\text{R}^{2/3}_{2,1}$ and $\text{R}^{2/3}_{2,2}$ is induced via an electroweak symmetry breaking term; $\text{R}^{2/3}_{2,1} \text{R}^{- 2/3}_{2,2} (H^{0\ast})^2$.
Then, the interaction of the mass eigenstate $\text{R}^{2/3}_2$
is picked out as 
\begin{align}
 \label{Eq:R2LQ}
 \mathcal L_\text{$\text{R}_2$LQ} = 
 \Big(
 h_{L}^{ij} \overline u_i P_L \nu_j 
 + h_{R}^{ij} \overline d_i P_R \ell_j 
 + \tilde h_{L}^{ij} \overline d_i P_L \ell_j 
 \Big) \text{R}_{2}^{2/3}
 + \text{h.c.}\,.
\end{align}
Then three WCs are generated, two of which are related, as  
\begin{align}
\label{Eq:R2LQWC}
 &2\sqrt{2}G_FV_{cb} C_{V_2} = + {h_L^{23} \tilde h_L^{*33} \over 2 M_{\text{LQ}}^2}\,,&
 &2\sqrt{2}G_FV_{cb} C_{S_2} = + {h_L^{23} h_R^{*33} \over 2M_{\text{LQ}}^2}\,,&
 &C_{S_2} = +4C_T\,.&
\end{align}
Both two scenarios, namely the one with the single $C_{V_2}$ and another for the specific combination $C_{S_2} = +4C_T$, can solve the $R_{D^{(\ast)}}$ anomaly.
Hence, collider studies will be performed for them in this paper.

Here, the coupling $\tilde h_{L}$ is generated from the mixing above the electroweak symmetry breaking scale. 
This implies that $C_{V_2}$ should have an additional suppression factor. 
See Ref.~\cite{Asadi:2019zja} for a UV completion of the $C_{V_2}$ scenario and its phenomenological bounds.
It will be shown that there are still viable parameter regions.
Nevertheless, our collider study provides a useful probe for the $C_{V_2}$ constraint as we will see in Sec.~\ref{Sec:single_operator}.

\subsection{$\text{S}_1$ LQ}
\label{Sec:S1LQ}

The $SU(2)_L$ singlet scalar LQ ($\text{S}_1$)
gives another solution to the $R_{D^{(\ast)}}$ anomaly.
The relevant Yukawa interactions with the SM fermions are described by
\begin{align}
\mathcal{L}_{{\text S}_1\text{LQ}} 
&= h_L^{ij} \, \overline{Q^C} i\tau_2 L_j\, {\rm S}_1 
+ h_R^{ij} \,\overline{u^C_{i}} P_R e_{j}\, {\rm S}_1 
+\mathrm{h.c.}\nonumber\\
&= \Big{[}\big{(}V_{\textrm{CKM}}^* h_L \big{)}^{ij}\, \overline{u^C_{L\,i}}\ell_{L\,j}-h_L^{ij}\,\overline{d^C_{L\,i}}\nu_{L\,j}+h_R^{ij}\, \overline{u^C_{R\,i}}\ell_{R\,j} \Big{]}\text{S}_1 + \mathrm{h.c.}\,.
\end{align}
The contribution to the relevant WCs are given by 
\begin{align}
\label{Eq:S1LQWC}
 &2 \sqrt{2} G_F V_{cb} C_{V_1}=
 \frac{h_L^{33} \big{(} V_{\text{CKM}} h_L^{*} \big{)}^{23}} {2M_{\text{LQ}}^2}\,,
 &2\sqrt{2} G_F V_{cb} C_{S_2}=-\frac{h_L^{33} h_R^{*23} } {2 M_{\text{LQ}}^2}\,,
 \hspace{15pt}C_{S_2} = - 4 \, C_{T}.
\end{align}
There are two sets of the WCs which are controlled by the different Yukawa couplings.
Although the single $C_{V_1}$ scenario looks promising, a stringent constraint from $b \to s \nu\nubar$ is unavoidable at the tree level.
We will discuss the relevant constraints in Sec.~\ref{Sec:numerical_results} and Appendix~\ref{Sec:App_flavor}. 

\section{\boldmath Event generation}
\label{Sec:event_generation}

Monte Carlo (MC) event generators are used to simulate both NP signal and SM background processes with a hard $\tau$ lepton and a large missing transverse momentum with/without an additional $b$-jet in the final states at $\sqrt{s}=13\TeV$. 
The NP models are implemented via {\sc\small FeynRules}~v2.3.34~\cite{Alloul:2013bka}.  
The model files are available in \href{https://arxiv.org/abs/2111.04748}{the arXiv web page}.
Event samples are generated by using {\sc\small MadGraph}5\_a{\sc\small MC}@{\sc\small NLO}~v2.8.3.2~\cite{Alwall:2014hca} 
interfaced with {\sc\small PYTHIA} v8.303~\cite{Sjostrand:2014zea} for hadronizations and decays of the partons.
The MLM merging is adopted in the five-flavor scheme~\cite{Alwall:2007fs}.
{\sc\small NNPDF}2.3 in {\sc\small LHAPDF} v6.3.0~\cite{Ball:2012cx} is used.
Detector effects are simulated by using {\sc\small Delphes} v3.4.2~\cite{deFavereau:2013fsa}. 
Here, we modified a prescription of the identification of the hadronic $\tau$ jet, as will be described below. 
The jets are reconstructed by using anti-$k_T$ algorithm~\cite{Cacciari:2008gp} with a radius parameter set to be $R=0.5$.
See Appendix~\ref{Sec:App_cutflow} for some details.

To investigate the non-resonant $\tau\nu$ and $\tau\nu+b$ searches, and especially to 
evaluate an impact of the latter,
the following two sets of kinematic cuts are compared: 
\begin{itemize}
 \setlength{\leftskip}{2.0em}
 \item[{\bf{cut a}}:] Kinematic cuts to select the $\tau\nu$ events by following Ref.~\cite{Marzocca:2020ueu}, originated from the CMS analysis~\cite{Sirunyan:2018lbg}:
\begin{itemize}
 \setlength{\leftskip}{2.0em}
 \item[-- 1.] require exactly one $\tau$-tagged jet, satisfying the transverse momentum of $\tau$, $p\tr^{\tau}\ge 200\GeV$, and the pseudo-rapidity of $\tau$, $|\eta_\tau| \le 2.1$,
 \item[-- 2.] veto the event if it includes any isolated electron or muon with $p\tr^{e,\mu}\ge 20\GeV$ within $|\eta_e| \le 2.5$ or $|\eta_\mu| \le 2.4$, where the lepton isolation criteria are the same as Ref.~\cite{Marzocca:2020ueu},
 \item[-- 3.] require large missing transverse momentum, $E\tr^{\text{miss}} \ge 200\GeV$, 
 to suppress the $W^{\pm}$ resonant contribution,
 \item[-- 4.] require that the missing momentum is balanced with the $\tau$-tagged jet with the back-to-back configuration as $\Delta\phi(\vec{p}\tr^\tau, \vec{p}\tr^{\,\rm miss}) \ge 2.4$ and $0.7 \le p\tr^{\tau}/E\tr^{\rm miss} \le 1.3$ to further suppress the SM backgrounds.
\end{itemize}
 \item[{\bf{cut b}}:] Additional kinematic cuts  to ``{\bf{cut a}}''
 for selecting the $\tau\nu+b$ events:
\begin{itemize}
 \setlength{\leftskip}{2.0em}
 \item[-- 1.] require exactly one $b$-tagged jet with $p\tr^b \geq 20\GeV$ and $|\eta^b| < 2.5$.
 \item[-- 2.] restrict the number of light-flavored jets, $N_j \leq 2\,$, to suppress the top-decay related backgrounds, where the jets satisfy $p\tr^j \geq 20\GeV$ and $|\eta^j| \leq 2.5$.
\end{itemize}
\end{itemize}

Energetic $\tau$ leptons can be emitted not only from the hard processes, 
but also from decays of energetic mesons,
\eg, $B \to \tau X$\,(at a branching ratio $\sim 3\%$) and $D_s \to \tau X$\,($\sim 5\%$).
Quantitatively, these secondary $\tau$ gives mild contributions to {\bf{cut~a}} and {\bf{cut~b}}. 
In reality, it is likely to be accompanied by nearby jets and vetoed by $\tau$ isolation conditions adopted in the ATLAS/CMS analyses. 
Since they do not use cut-based analyses, an implementation of their isolation procedure is complicated and beyond the scope of this paper. 
In our analysis, events with $\tau$ whose parent particle is mesons or baryons are vetoed, for simplicity.
Also, for a $\tau$-tagging efficiency, the ``VLoose'' working point is adopted for the hadronic decays; $\epsilon_{\tau \to \tau}=0.7$~\cite{CMS:2018jrd}.
As the mis-tagging efficiencies, 
we apply $p\tr^j$-dependent efficiency based on Ref.~\cite{CMS:2018jrd}.
For instance, $\epsilon_{j\to \tau}=3.7\times10^{-3}$ for $p\tr^j=100\GeV$ and $7.2\times10^{-4}$ for $p\tr^j=300\GeV$ or larger. 
The mis-tagging rare $\epsilon_{c,b\to\tau}$ is assumed to be 7.2$\times10^{-4}$ as a reference.
When one imposes the condition requiring an additional $b$-jet in the final state,
of crucial importance is which working point is chosen for the $b$-tagging efficiencies.
For instance lower mis-tagging efficiencies can suppress backgrounds coming from fake $b$-jets.
We adopt the following working point 
based on Table~4 of Ref.~\cite{ATLAS:2019bwq}, 
\begin{align}
 &\epsilon_{j\to b}=1/1300\,,&
 &\epsilon_{c \to b}=1/27\,,&
 &\epsilon_{b\to b}=0.6\,.&
\label{Eq:b-tag}
\end{align} 
Compared to the working point in Ref.~\cite{Marzocca:2020ueu},\kf\footnote{%
\label{footnote:b-tagging}
The reference~\cite{Marzocca:2020ueu} adopted
a different working point:
$\epsilon_{j\to b}=0.015$, $\epsilon_{c \to b}=0.3$, and $\epsilon_{b\to b}=0.7$.
} 
the mis-tagging rates, $\epsilon_{j \to b}$ and $\epsilon_{c \to b}$, are better by factors of $20$ and $8$, respectively,
while the $b$-tagging rate $\epsilon_{b\to b}$
is slightly worse.
Therefore,
it is expected that the number of background events originated from fake $b$-jets is reduced in our analysis for the {\bf cut b} category.

Note that the charge of the final-state $\tau$ lepton is not distinguished in {\bf{cut a}} or {\bf{cut b}}, though it may be possible at the LHC as mentioned in Sec.~\ref{Sec:introduction} and will be discussed later. 
In order to stress this point, the searches are described with a script $\pm$ as ``the $\tau^\pm \nu$ ($\tau^\pm \nu+b$) search'' hereafter.

\subsection{\boldmath Background simulation}

As for the SM background events generation,
we basically trace the method explored in Ref.~\cite{Marzocca:2020ueu}.
Nonetheless, since this is crucial to derive NP sensitivities, we dare to present all the essential steps in some details, though most of them may be familiar to experts.
The six categories of the background processes are considered:

\subsubsection*{$\boldsymbol{Wjj}$}

The event simulations in {\sc\small MadGraph}5\_a{\sc\small MC}@{\sc\small NLO} are performed up to {\tt QED=4}, which includes contributions from vector boson fusions.
The $W$ boson is assumed to decay as $W \to \tau \nubar$, and the events are matched allowing up to two jets. 
The $Wjj$ contribution dominates the SM background in the {\bf{cut a}} category,
and also one of the main sources of the backgrounds for {\bf{cut~b}} because light-flavored jets are mis-tagged as $b$ jets.
The working point of $b$-tagging efficiencies is given in Eq.~(\ref{Eq:b-tag}).
It is checked that the number of events of $W$ plus genuine $b$-jet is less than that of $Wjj$ by more than three orders of magnitude for ${\bf cut~b}$.
Therefore, improving the discrimination efficiency of the light-flavored jets from the genuine $b$ jets can result in suppressing  the SM background effectively.

\subsubsection*{$\boldsymbol{Zjj}$}

The $Z$ boson is assumed to decay as $Z \to \nu \nubar$, contributing to missing transverse momentum. 
The events are matched allowing up to two jets.
At least one fake $\tau$-jet is necessary to pass {\bf{cut~a}}. 
Namely, the final state should include associated QCD jets. 
This channel gives the subdominant contribution both for {\bf{cut~a}} and {\bf{cut~b}}. 
Note that the ATLAS and CMS analyses categorize $Zjj$ into ``QCD jet,'' and estimate them with a data-driven technique, 
\eg, extrapolating from $Zjj$ with $Z\to \mu^+\mu^-$ events and requiring $p\tr^\tau/E\tr^{\text{miss}} \le 0.7$.

\subsubsection*{$\boldsymbol{t \bar{t}}$}

The top quarks are assumed to decay as $t\overline{t}\to b W^+ \overline{b} W^-$ with both $W$ bosons decaying to $\tau$ or one of them decaying to $\tau$. 
The former contribution is larger by a factor of four than the latter after {\bf{cut a}}, while both are of similar size after {\bf{cut b}}.

\subsubsection*{Single $\boldsymbol{t}$}

The single top productions are divided by the following five sub-categories, $t+j$, $tW(1)$, $tW(2)$, $tZ(1)$, and $tZ(2)$.
The top quark decays into $bW$, and the number in the parentheses expresses how many gauge bosons decay leptonically, \ie,  $W \to \tau\nu$ or $Z \to \nu{\nubar}$. 
More explicitly 
$t+j \to b \tau \nu j$ is categorized as $t+j$.
$tW\to b \tau \nu jj$ and $tW\to b \tau\tau \nu \nu$ are classified into $tW(1)$ and $tW(2)$ respectively. 
$tZj \to b \tau \nu jjj $, $tZj\to b \tau \tau jjj$ and $tZj\to b \nu \nu jjj $ are denoted as $tZ(1)$, and 
$tZj \to b \tau \nu \tau \tau j $ and $tZj \to b \tau \nu \nu \nu j$ are classified into $tZ(2)$.

\subsubsection*{$\boldsymbol{Z,\gamma}$ Drell-Yan}

A pair of $\tau$ leptons are produced via Drell-Yan processes mediated by $Z$ or $\gamma$ 
in accompany with up to two jets.
Since the number of $\tau$ jets is required to be exactly one in {\bf{cut a}}, another $\tau$ lepton needs to be missed in the detectors. 
Although the efficiency of $\tau$ mis-tagged as other particles is not so small, 
it is unlikely to achieve a large missing momentum because jets are rarely overlooked or their momenta are hardly mis-reconstructed so largely in the detectors. 
Thus, the contribution will be found to be negligibly small.

\subsubsection*{$\boldsymbol{VV}$}

Pair-productions of vector bosons are classified by the species as $WW$, $ZZ(\gamma)$, and $WZ(\gamma)$. 
The events for $WW$ are simulated with both $W$'s decaying to $\tau$ and allowing up to two additional jets or one of $W$'s decaying into $\tau$ and allowing up to one additional jet.
The $ZZ(\gamma)$ events involve those with one of $Z$'s decaying as $Z\to\nu\overline\nu$ or into $\tau^+\tau^-$. 
As for $WZ(\gamma)$, the events are generated from a tauonic $W$ decay along with $\gamma\to\tau^+\tau^-$ or $Z\to\tau^+{\tau^-},\,\nu\overline\nu$. 
It will be shown that the resultant contribution is subdominant in {\bf{cut a}} and of $\mathcal{O}(1)\%$ in {\bf{cut b}}.\\

It is noted that pure QCD multi-jet backgrounds are not simulated in this paper.
In order to pass {\bf{cut a/b}}, one of energetic jets has to be mis-tagged as $\tau$.
Moreover, although another jet is required to be overlooked to pass the condition of large missing momentum, 
this rarely happens in the calorimeters.
Here we assume that the contributions are negligible, for simplicity, though one needs full detector simulations for further studies.
In fact, the CMS collaboration has checked that the QCD multi-jet background is 
smaller than that from $Zjj$ in their simulation, and shown that the simulated result agrees with the data in a control region~\cite{Sirunyan:2018lbg}.

\subsection{\boldmath Signal simulation}
\label{Sec:signal}

Here, we show our setup with respect to the NP scenarios of interest for investigating the LHC sensitivities in the  $\tau^\pm\nu(+b)$ search. 
Events of the NP signals are generated for each NP scenario by fixing the relevant LQ couplings and mass, and then matched by allowing up to two (five-flavored) jets. 
In turn, the couplings are encoded as in Eq.~\eqref{Eq:WCEFT} to present our output. 
As the high-$p_T$ tail is concerned, NP--SM interferences are tiny enough, \eg, see Ref.~\cite{Marzocca:2020ueu} showing that the interference effect is a few percent level.\kf\footnote{When one considers dimension-eight effective interactions, it is found that its NP--SM interference contribution is further smaller than the dimension-six NP--SM interference~\cite{Fuentes-Martin:2020lea}.} 
Note that a possible $s$-channel production is also suppressed by the requirement of the back-to-back condition between $\tau$ and $\nu$, see Appendix~\ref{Sec:App_cutflow}. 
A set of process cards for the {\sc\small MadGraph} event generation are available in \href{https://arxiv.org/abs/2111.04748}{the arXiv web page}.

As already mentioned, we proceed with the LQ models that generate the effective four-fermion interactions at the EFT limit. 
Our approach has a benefit to clarify difference between EFT and a practical model of interest, especially for the case of the $C_{V_1}$ type interaction as explained below. 

Motivated by the $R_{D^{(*)}}$ anomaly, NP contributions to $b\bar{c}\to\tau\bar{\nu}$ have been studied in the EFT limit.
However, one notices that there exist additional processes to be considered in realistic model setups.
In fact, the $V_1$ operator is constructed from the $\text{U}_1$ LQ model,  
and $C_{V_1}$ depends on the LQ couplings $h_L^{23}$ and $h_L^{33}$, as seen in Eq.~\eqref{Eq:U1LQWC}. 
Under the $SU(2)_L$ gauge invariance, the term, $\bar Q_L \gamma_\mu L_L \, \text{U}_1^\mu$, generates an interaction of $\overline{s}$--$\tau$--$\text{U}_1$ as well as that of $\overline{c}$--$\nu_\tau$--$\text{U}_1$ in presence of $h_L^{23}$. 
Therefore, 
additional production processes such as $s\overline{c} \to \tau\bar{\nu}$ should be taken into account even in the EFT limit. 
In this paper, this new process is considered via the following effective Lagrangian,
\begin{align}
 {\mathcal L}_{\rm{eff}}\supset 
 - 2 \sqrt 2 G_FV_{cb} \Bigl[ 
 & (1 + C_{V_1}) (\overline{c} \gamma^\mu P_Lb)(\overline{\tau} \gamma_\mu P_L \nu_{\tau})
 + R_{s/b} C_{V_1}(\overline{c} \gamma^\mu P_L s)(\overline{\tau} \gamma_\mu P_L \nu_{\tau})\Bigl] + \,\text{h.c.}\,.
  \label{Eq:effpp}
\end{align} 
The second term in the bracket corresponds to the new contribution, and $R_{s/b}$ is defined from Eq.~\eqref{Eq:U1LQ} as
\begin{align}
\label{Eq:Rsb}
R_{s/b} \equiv 
\frac{|\text{coupling~constant~of~}\overline{s}\text{--}\tau\text{--}\text{U}_1|}{|\text{coupling~constant~of~}\overline{b}\text{--}\tau\text{--}\text{U}_1|} 
= \left| \frac{h^{23}_{L}}{h^{33}_{L}}\right| \,. 
\end{align}
Hence, the $\text{U}_1$ LQ model possesses two parameters, $(C_{V_1}, R_{s/b})$, in the collider analysis, and the conventional EFT setup of $V_1$ is realized by taking $R_{s/b} \to 0$. 
Note that such an issue is not the case for the other operator scenarios. 
On the other hand, although the $\tau^\pm\nu+b$ search seems to be insensitive to it since the $b$ quark is required in the final state, it will be shown that the $\overline{s}$--$\tau$--$\text{U}_1$ interaction can affect the result through $g s \to c\tau\nubar$ with the final state $b$-jet mis-tagged from $c$-jet.

\subsubsection*{Single operator scenarios}

Here, we list NP scenarios which can be responsible for the $R_{D^{(*)}}$ anomaly and whose collider signals will be investigated in this paper. 
First, from the view point of the EFT limit in Eq.~\eqref{Eq:effH}, we consider LQ setups such that one of the WCs of $C_{V_1}$, $C_{V_2}$, $C_{S_1}$, $C_{S_2}$, and $C_{T}$ is non-vanishing. 
Let us call this setup as ``the single $C_X$ scenario.''
Note that $R_{s/b}=0$ is taken in the $C_{V_1}$ scenario. 
The signal events are generated for the following LQ masses,
\begin{align}
M_{\rm LQ}= \{ 1.5\,, \, 2.5\,, \, 4.0\,, \, 6.5\,, \, 10\,, \, 15\,, \, 20\}\TeV.
\label{eq:LQmass_range}
\end{align}
According to Ref.~\cite{Iguro:2020keo} the EFT approximation becomes valid for $M_{\rm LQ}\gtrsim 10\TeV$ in the $\tau^\pm\nu$ search. 
In this paper, $M_{\rm LQ}$ is taken up to $20\TeV$ to check the decoupling behavior in the $\tau^\pm\nu+b$ search. 
Then, we refer to the case of $M_{\rm LQ}=20\TeV$ as the EFT limit. 
It should be mentioned again that the LQ model which explains the $R_{D^{(\ast)}}$ anomaly is restricted as $M_{\rm LQ} < \mathcal{O}(10)\TeV$ due to the perturbative unitarity bound~\cite{DiLuzio:2017chi}. 

The above LQ masses satisfy the constraints from the LQ direct searches. 
The searches have been performed by studying LQ pair-production channels at the ATLAS~\cite{ATLAS:2021jyv} (CMS~\cite{CMS:2020wzx}) $\int\mathcal{L}\,dt=139\ifb$ and provided limits on the LQ mass as $M_{\rm LQ}\ge 1.2\,(1.0)\TeV$ for a scalar LQ, and $M_{\rm LQ}\ge 1.5\,(1.3)\TeV$ for a gauged-vector LQ at the $95\%$ CL.
\kf\footnote{The lower bound for a strongly-coupled sector originated vector LQ is given as $M_{\rm LQ}\ge 1.7\,(1.6)\TeV$.}
On the other hand, single-production channels can provide alternative bounds.
However, since they depend on LQ couplings irrelevant to the $R_{D^{(*)}}$ anomaly,
we do not take them into account. 

As we focus on the NP interactions responsible for the $R_{D^{(*)}}$ anomaly, the other LQ couplings, which are irrelevant for $b \to c \tau\nu_\tau$, are set to be zero, and thus, the LQ production process comes only from the initial partons of $cb,\, gc,\, gb,\, gg$.

\subsubsection*{Single LQ scenarios}

We also perform the analysis which is based on the LQ model rather than the EFT operator. 
In particular, multiple WCs can become non-vanishing simultaneously, or $R_{s/b}$ is not always zero. 
As aforementioned in Sec.~\ref{Sec:Set_up}, the following five scenarios can solve the $R_{D^{(*)}}$ anomaly by introducing a single LQ boson.
\begin{itemize}
\item 
The $\text{R}_2$ scalar LQ model induces the two independent WCs, $C_{V_2} (M_\text{LQ})$ and $C_{S_2}(M_\text{LQ}) = +4 C_T(M_\text{LQ})$, as given in Eq.~\eqref{Eq:R2LQWC}. 
Thus, we study these two scenarios, called as single-$\text{R}_2 (C_{V_2})$ and single-$\text{R}_2 (C_{S_2})$ scenarios, respectively. 
Note that the former is identical to the $C_{V_2}$ scenario (unlike $C_{V_1}$ in the $\text{U}_1$ LQ scenario). 
\item
The $\text{S}_1$ scalar LQ model induces the two independent WCs, $C_{V_1} (M_\text{LQ})$ and $C_{S_2}(M_\text{LQ}) = -4 C_T(M_\text{LQ})$, as seen in Eq.~\eqref{Eq:S1LQWC}. 
In contrast to the $\text{R}_2$ LQ case, however, the single $C_{S_2} = -4 C_T$ case cannot address the $R_{D^{(*)}}$ anomaly within $1\,\sigma$, though the tension can be relaxed. 
Thus, we investigate the scenario in a two-dimensional parameter space, $(C_{V_1}, C_{S_2})$, with assuming real WCs, simply called as $\text{S}_1$ LQ scenario. 
\item
The U$_1$ vector LQ model possesses the two independent WCs, $C_{V_1} (M_\text{LQ})$ and $C_{S_1} (M_\text{LQ})$. 
In this paper, we investigate two scenarios in terms of the WCs of Eq.~\eqref{Eq:U1LQWC}; 
the single $C_{V_1}$ scenario assuming $C_{S_1}=0$, 
and the scenario satisfying $C_{S_1}=-2e^{i \phi_R} C_{V_1}$ under the $U(2)$ flavor symmetry, as introduced in Sec.~\ref{Sec:U1LQ}. 
Hereafter, they are referred as single-$\text{U}_1$ and $U(2)$-$\text{U}_1$ scenarios, respectively. 
One can easily find that the relative phase $\phi_R$ is almost irrelevant for the following collider analysis and taken to be zero.
On the other hand, both two scenarios involve the aforementioned $R_{s/b}$. 
In our analysis, the region of $1/16 \leq R_{s/b} \leq  16$ 
 is searched to see its effect in detail, 
in addition to the case of $R_{s/b}=0$ that corresponds to the $C_{V_1}$ scenario. 
\end{itemize}
In the analysis, the LQ mass region in Eq.~\eqref{eq:LQmass_range} is studied. 
Besides, since we are interested only in the LQ couplings relevant to the $R_{D^{(*)}}$ anomaly, the LQ production processes come from the initial partons of $cb,\, gc,\, gb,\, gg$ for $\text{R}_2$ and $\text{S}_1$ LQs, while the additional production from $cs$ is taken into account for $\text{U}_1$ LQ.

\section{\boldmath Numerical results}
\label{Sec:numerical_results}

In this section, we present numerical results of the LHC simulations for the $\tau\nu (+b)$ processes. 
Also, it is argued how the requirement of an additional $b$-jet improves NP sensitivities and gives an impact on the NP solutions to the $R_{D^{(*)}}$ anomaly.

\subsection{\boldmath Event numbers after selection cuts}
\label{Sec:selection_cut}

\begin{table}[t!]
\centering
\newcommand{\bhline}[1]{\noalign{\hrule height #1}}
\renewcommand{\arraystretch}{1.3}
   \scalebox{0.94}{
  \begin{tabular}{c| ccc ccc | c} 
  \bhline{1 pt}
  \rowcolor{white}
  BG ($\bf{cut~a}$) &$Wjj$ & $Zjj~(Z\to\nu\overline\nu)$ & $t\overline{t}$ & $Z,\gamma$ DY & $VV$ & single $t$ & total  \\  \hline  
 $0.7$ $< m\tr < 1\TeV$  & 70.5 & 20.1 & 0.34 & 3.03 & 1.30 & 0.02 & 95.3\\ 
 $1\TeV$ $ < m\tr$ & 16.9 & 5.1 & 0.06 & 0.56 & 0.32 & 0.02 & 23.0\\ \hline \hline
$1\TeV$ $ < m\tr$~\cite{Sirunyan:2018lbg} & $22\pm 6.2$ & $0.9\pm0.5$ & $< 0.1$ & $<  0.1$ & $0.7\pm0.1$ & $< 0.1$ & $23.4\pm7.2$ \\ 
$1\TeV < m\tr$~\cite{Marzocca:2020ueu} & $18$ & $5.2$ & $0.44$ & $0.0025$ & $1.7$ & $0.1$ & 25.4\\
\bhline{1 pt}
   \end{tabular}
   }
    \caption{\label{table:BG_cuta} 
    Expected number of SM background events after {\bf{cut a}} (the $\tau^\pm\nu$ search) for 
    $\int\mathcal{L}\,dt=35.9\ifb$ and $\sqrt{s}=13\TeV$ in each background category. 
    The last two rows show the results obtained by
    Refs.~\cite{Sirunyan:2018lbg} and~\cite{Marzocca:2020ueu}.
    Also shown are the total systematic uncertainties for the former.
    Detailed cut flows are shown in Table~\ref{table:AP_BG_cutaflow}.
}
\end{table}
\begin{table}[t!]
\centering
\newcommand{\bhline}[1]{\noalign{\hrule height #1}}
\renewcommand{\arraystretch}{1.3}
   \scalebox{0.89}{
  \begin{tabular}{c| ccc ccc |c} 
  \bhline{1 pt}
  \rowcolor{white}
 BG ($\bf{cut~b}$) &$Wjj$ & $Zjj~(Z\to\nu\overline\nu)$ & $t\overline{t}$ & $Z,\gamma$ DY & $VV$ & single $t$ & total   \\  \hline  
   $0.7$ $< m\tr < 1\TeV$  & 0.58 & 0.37 & 0.056 & 0.28 & 0.018 & 0.029 & 1.33\\ 
  $1\TeV$ $ < m\tr$ & 0.16 & 0.06 & 0.01 & 0.007 & 0.005 & 0.005 & 0.25 \\ \hline\hline
  $1\TeV$ $ < m\tr$~\cite{Marzocca:2020ueu} & 0.18(5) & 0.21(12) & 0.29(3) & $4.2(4)\times10^{-5}$ & 0.35(5) & 0.067(7) & 1.10(14)\\ 
\bhline{1 pt}
   \end{tabular}
   }
    \caption{\label{table:BG_cutb} 
    Expected number of SM background events after $\bf{cut~b}$ (the $\tau^\pm\nu+b$ search) for
    $\int\mathcal{L}\,dt=35.9\ifb$ and $\sqrt{s}=13\TeV$ in each background category. 
    The last row shows the result by Ref.~\cite{Marzocca:2020ueu}, where a number in the parenthesis represents uncertainties. 
    Note that their $b$-tagging efficiencies are different from ours (see the footnote \ref{footnote:b-tagging}).
    Detailed cut flows are shown in Table~\ref{table:AP_BG_cutbflow}. 
}
\end{table}

The expected numbers of SM background events after the cuts, {\bf{cut a}} and {\bf{cut b}}, are shown in Tables~\ref{table:BG_cuta} and \ref{table:BG_cutb}, respectively. 
They correspond to the result at the integrated luminosity of $35.9\ifb$ and $\sqrt{s}=13\TeV$,
which is equivalent to the CMS result~\cite{Sirunyan:2018lbg}.
Note that we imposed a pre-cut given in Eq.~\eqref{Eq:pre_cut} of Appendix \ref{Sec:App_cutflow} at the generator level in the analysis to reduce the simulation cost. 
The cut can affect the event distributions for $m\tr \lesssim 500\GeV$, while the result is insensitive to it for $m\tr > 700\GeV$. 
Detailed cut flows of the SM background are shown in Tables~\ref{table:AP_BG_cutaflow} and \ref{table:AP_BG_cutbflow} in Appendix~\ref{Sec:App_cutflow}.

From Table \ref{table:BG_cuta}, it is found that the main background of {\bf{cut a}} (specified for the $\tau^\pm\nu$ search, \ie, without requiring $b$-jets in the final state) comes from the $Wjj$ channel. 
Our result is consistent with those obtained by Refs.~\cite{Sirunyan:2018lbg} and \cite{Marzocca:2020ueu}. 
The next-to-leading contribution is provided by the $Zjj$ channel and consistent with Ref.~\cite{Marzocca:2020ueu}, while it is larger by a factor of $5$ than the CMS result.
Note that the CMS $Zjj$ result is based on a data driven analysis.
It should be mentioned that, although the background events are categorized by the channels, their criteria are not unique and not shown explicitly in the literature.
Nevertheless, the total number of the SM background is consistent with those in Refs.~\cite{Sirunyan:2018lbg} and~\cite{Marzocca:2020ueu} for $m\tr > 1\TeV$, which may validate our analysis. 

Let us comment on a preliminary result of the $\tau^\pm\nu$ search by the ATLAS collaboration with $\int\mathcal{L}\,dt=139\ifb$~\cite{ATLAS:2021bjk}.
It has not observed any significant excess, and hence, constrained the $W'$ mass as $\gtrsim 5\TeV$. 
To be precise, the observed event number is smaller than the expected SM background, and thus, one can infer (much) stronger upper bounds on the EFT operators in Eq.~\eqref{Eq:R2LQWC} than 
the results based on the CMS analysis. 
Nevertheless, our result cannot be compared with it straightforwardly because the ATLAS has not provided enough information for this purpose and the tagging efficiency of hadronic $\tau$ is different from ours. 

In Table \ref{table:BG_cutb}, an additional $b$-jet is required, corresponding to {\bf{cut b}} (specified for the $\tau^\pm\nu+b$ search).
It is shown that the total number of the background is suppressed by about two orders of magnitude versus the result for {\bf{cut a}}. 
In detail, the event number after the cut decreases in every channel, particularly in $Wjj$ and $Zjj$.
Here, the range of reduction depends on whether the event involves genuine $b$-jets or not.
Also, it is noticed from Table~\ref{table:AP_BG_cutbflow} that the condition on the number of $b$-jets is effective to suppress the background when it does not come from the top quarks, while the back-to-back condition reduces those via the top quarks. 
Our result is also compared with that given by Ref.~\cite{Marzocca:2020ueu}, where the $b$-tagging efficiencies, especially those for fake $b$-jets, are different from ours (see the footnote \ref{footnote:b-tagging}).
The total number of the background becomes smaller by $\sim\,40\%$ than their result. 
The difference is prominent in $Zjj$, $t\bar{t}$ and $VV$, because a large number of events include fake $b$-jets.

\begin{table}[t!]
\centering
\newcommand{\bhline}[1]{\noalign{\hrule height #1}}
\renewcommand{\arraystretch}{1.3}
  \scalebox{0.83}{
  \begin{tabular}{c|cccc|cccc|c } 
  \bhline{1 pt}
  signal ($\bf{cut~a}$) &$C_{V_1, 1.5\TeV}$ & $C_{V_1, \text{EFT}}$ &$C_{V_1, 1.5\TeV}^{R_{s/b}=1}$ & $C_{V_1, \text{EFT}}^{R_{s/b}=1}$ &$\text{U}_{1, 1.5\TeV}^{R_{s/b} =0}$ & $\text{U}_{1, \text{EFT}}^{R_{s/b} =0}$ & $\text{U}_{1, 1.5\TeV}^{R_{s/b} =1}$ & $\text{U}_{1, \text{EFT}}^{R_{s/b} =1}$& BG \\  \hline 
 $0.7$ $< m\tr < 1\TeV$   & 90.0 & 139.4 & 225.9 & 351.4 & 361 & 582 & 502 & 809 & 95.3\\ 
 $1\TeV$ $ < m\tr$& 54.4 & 123.6 & 146.9 &345.8 & 204 & 571 & 279 & 799 & 23.0\\ 
\bhline{1 pt}
   \end{tabular}
   }
    \caption{\label{table:CV1_cuta} 
    Expected numbers of signal events after $\bf{cut~a}$ (the $\tau^\pm\nu$ search) for
    $\int\mathcal{L}\,dt=35.9\ifb$ and $\sqrt{s}=13\TeV$ in the single  $C_{V_1}$ scenario (from second to fifth columns) and the $U(2)$-$\text{U}_1$ scenario ($C_{S_1} = -2\beta_R\,C_{V_1}$) (from sixth to ninth columns). 
    In all cases, $C_{V_1} =1$ is fixed, while the LQ mass is $M_{\rm LQ} = 1.5\TeV$ or $20\TeV$ (the EFT limit). 
    The $s$-quark contribution, parameterized by $R_{s/b}$, is also studied. 
    The last column is the expected number of SM background (see Table~\ref{table:BG_cuta}).
    Detailed cut flows are shown in Tables~\ref{table:AP_CV1_cutaflow} and \ref{table:AP_U2_cutaflow}.
}
\end{table}
\begin{table}[t!]
\centering
\newcommand{\bhline}[1]{\noalign{\hrule height #1}}
\renewcommand{\arraystretch}{1.3}
   \scalebox{0.83}{
  \begin{tabular}{c|cccc|cccc|c} 
  \bhline{1 pt}
  signal ($\bf{cut~b}$) &$C_{V_1, 1.5\TeV}$ & $C_{V_1, \text{EFT}}$ &$C_{V_1, 1.5\TeV}^{R_{s/b}=1}$ & $C_{V_1, \text{EFT}}^{R_{s/b}=1}$ &$\text{U}_{1, 1.5\TeV}^{R_{s/b} =0}$ & $\text{U}_{1, \text{EFT}}^{R_{s/b} =0}$ & $\text{U}_{1, 1.5\TeV}^{R_{s/b} =1}$ & $\text{U}_{1, \text{EFT}}^{R_{s/b} =1}$& BG  \\  \hline 
   $0.7$ $< m\tr < 1\TeV$  &11.6 & 16.6 & 13.9 & 21.7 & 53.9 & 86.4 & 55.8 & 92.0 & 1.33\\
 $1\TeV$ $ < m\tr$ & 6.51 &14.6 & 9.39 & 21.6  & 26.0 & 71.6 & 30.7 & 101 & 0.25\\ 
\bhline{1 pt}
   \end{tabular}
   }
    \caption{\label{table:CV1_cutb} 
    Expected numbers of signal events after $\bf{cut~b}$ (the $\tau^\pm\nu+b$ search) for
    $\int\mathcal{L}\,dt=35.9\ifb$ and $\sqrt{s}=13\TeV$. 
    See the caption of Table~\ref{table:CV1_cuta} for further information.
    The last column is the expected number of SM background (see Table~\ref{table:BG_cutb}). 
     Detailed cut flows are shown in Tables~\ref{table:AP_CV1_cutbflow} and \ref{table:AP_U2_cutbflow}.
}
\end{table}

The expected numbers of signal events after the selections, {\bf{cut~a}} and {\bf{cut~b}}, are shown in Tables~\ref{table:CV1_cuta} and \ref{table:CV1_cutb}, respectively. 
As a reference, the single $C_{V_1}$ scenario and  
the $U(2)$-$\text{U}_1$ scenario $(C_{S_1} = -2\beta_R\,C_{V_1})$
are evaluated at $\int\mathcal{L}\,dt=35.9\ifb$ and $\sqrt{s}=13\TeV$. 
Here, $C_{V_1} =1$ is fixed, 
while the LQ mass is set to be $M_{\rm LQ} = 1.5\TeV$ and $20\TeV$, where the latter corresponds to the EFT limit. 
By varying $R_{s/b}$, effects of the $s$-quark contribution are also studied. 
Detailed cut flows for those scenarios are given in Tables~\ref{table:AP_CV1_cutaflow}, \ref{table:AP_U2_cutaflow}, \ref{table:AP_CV1_cutbflow}, and \ref{table:AP_U2_cutbflow} of Appendix~\ref{Sec:App_cutflow}.

From the tables, it is confirmed that the event number after the cut depends on the LQ mass for $M_{\rm LQ}=\mathcal{O}(1)\TeV$. 
For instance, according to the results for $m\tr > 1\TeV$ the event number with $C_{V_1, 1.5\TeV}=1$ is less than a half of that in the EFT limit, $C_{V_1, \text{EFT}}=1$. 
Such a feature is valid for both {\bf{cut~a}} and {\bf{cut~b}}.

Let us comment that the signal event numbers in our results are smaller than those in Ref.~\cite{Marzocca:2020ueu}, \eg, 25.6 events are expected for $C_{V_1, \text{EFT}}~(m\tr > 1\TeV)$ in their analysis. 
This is mainly because the $b$ mis-tagging rate is different; $\epsilon_{c\to b}=0.3$ in Ref.~\cite{Marzocca:2020ueu}, while it is $1/27\simeq 0.04$ in our case. 
With their set up, we checked that almost a half of the signal events come from this fake $b$-jet in the simulation. 

Further systematic uncertainties 
can stem from a charm-quark PDF.
We have checked that the PDF uncertainty including the scale variations and estimated by comparing different PDF sets is of order $20\%$ in the number of signal events.
This corresponds to $\sim 10\%$ uncertainty for the sensitivities to the NP model in terms of the WCs.
Moreover, although the total number of background events in our analysis is consistent with the experimental result \cite{Sirunyan:2018lbg},
the number of events in each category does not match perfectly (see Table~\ref{table:BG_cuta}).
This might be due to a lack of sufficient information on the criteria of the categories.
Nonetheless, if one adopted, for instance, the result of the $Zjj$ background in Ref.~\cite{Sirunyan:2018lbg}, which is based on the data-driven estimation, the total number of background events would be reduced by $20\%$.
This would amplify the signal sensitivity.
Besides, an uncertainty in the hadronic $\tau$-tagging efficiency
could affect our results quantitatively.
Therefore, dedicated studies especially with experimental information are required to improve the analysis. 

\subsection{\boldmath Test of background-only hypothesis}
\label{Sec:signal_sensitivity}

In order to study sensitivities to the NP contributions, the background-only hypothesis is tested;
under the hypothesis, the result is identified to be consistent with the SM if the total number of events, \ie, the sum of the signal and background event numbers (denoted as $N_{Sig}$ and $N_{BG}$, respectively), is smaller than an upper bound $(U_{\rm tot})$. 
In this paper, the bound is determined as follows.
Let us first turn off the systematic uncertainty to focus on the statistical one. 
Under the background-only hypothesis, $U_{\rm tot}$ satisfies the relation,
\begin{align}
 \sum_{n=0}^{N_{BG}} P(n;U_{\rm tot}^{\rm stat}) = p\,,
 \label{eq:def_upperbound}
\end{align}
where $P(n;\mu)$ is the probability function of the Poisson distribution for observing $n$ events with the mean value $\mu$, and $p=0.05$ is taken, corresponding to 95\% confidence level (CL).
Here, $U_{\rm tot}$ is denoted with the superscript ``stat,'' since the systematic uncertainty is ignored. 
Then, the systematic uncertainty is taken into account.
Although it is unknown, we assign $60\%$ relative to the mean value at 95\% CL for $\int\mathcal{L}\,dt=35.9\ifb$ as inferred from the CMS result~\cite{Sirunyan:2018lbg}.\kf\footnote{
As shown in Table~\ref{table:BG_cuta}, the SM background is dominated by $Wjj$. 
The CMS analysis obtains the total systematic uncertainty of $28\%$ on this channel at $68\%$ CL~\cite{Sirunyan:2018lbg}.
}
Furthermore,
it is supposed to be scaled with $1/\sqrt{L}$ for the integrated luminosity $L$.
Hence, the systematic uncertainty is assigned as $\sigma_{\rm syst}^{95\%}=N_{BG} \times 60\% \times  \sqrt{35.9\,[\text{fb}]/L\,[\text{fb}]}$.
In this paper, we combine the systematic uncertainty with the statistical one linearly, and then, $U_{\rm tot}$ is obtained as $U_{\rm tot} = U_{\rm tot}^{\rm stat} + \sigma_{\rm syst}^{95\%}$.
Finally, the upper bound on the NP signal event number $N_{Sig}$ is derived as $S^{95\%} = U_{\rm tot}-N_{BG}$; the result is regarded as the SM consistent if $N_{Sig} < S^{95\%}$ is satisfied.

In the analysis, the expected number of events is not always integers, as shown in the tables of the previous subsection. 
Then, $N_{BG}$ in Eq.~\eqref{eq:def_upperbound} is replaced with $\lfloor N_{BG}\rfloor$ corresponding to the mode for the Poisson distribution.
Here, $\lfloor x \rfloor$ is the floor function, \ie, returns the maximum integer not exceeding $x$.
The background event number for $\int\mathcal{L}\,dt=35.9\ifb$ is given in Tables~\ref{table:BG_cuta} and \ref{table:BG_cutb}.
For higher luminosity, the event numbers are obtained by scaling the results in the tables with corresponding to the integrated luminosity.
Note that, although the HL-LHC (LHC Run 4 and 5) will be operated at $\sqrt{s}=14\TeV$, we ignore differences between the results at $\sqrt{s}=13$ and $14\TeV$, for simplicity.

Before proceeding to the results, let us mention about the $m\tr$ dependence of the NP sensitivities. 
From the tables it can be found that the category of $m\tr>1\TeV$ provides higher sensitivities to the NP contributions than the category of $0.7 < m\tr < 1\TeV$.
Similarly, among the three different $m\tr$ bins in the CMS analysis, $m\tr<500\GeV$, $500\GeV < m\tr < 1\TeV$ and $m\tr>1\TeV$, 
the last one provides the most stringent constraints.
Hence, we will present the results obtained from $m\tr>1\TeV$ in the following.

\subsection{\boldmath Single operator scenarios}
\label{Sec:single_operator}

\begin{table}[p]
\centering
\newcommand{\bhline}[1]{\noalign{\hrule height #1}}
\renewcommand{\arraystretch}{1.5}
   \scalebox{1.1}{
  \begin{tabular}{c ccccc } 
  \bhline{1 pt}
  \rowcolor{white}
  $\tau^\pm\nu$ search  &$C_{V_1}$ & $C_{V_2}$ & $C_{S_1}$ & $C_{S_2}$ & $C_{T}$  \\  \hline  
      \multicolumn{6}{c}{current upper bound on EFT~\cite{Iguro:2020keo}: LHC $36\ifb$}  \\
      \hline
      $\mu=m_b$ &  0.32 &  0.33 & 0.55 & 0.55 & 0.17 \\ 
      \hline
   \multicolumn{6}{c}{sensitivity: LHC $139\ifb$}  \\
   \hline
 $\mu=\Lambda_{\rm LHC}$ & 0.30 (0.46) &  0.32 (0.68) &  0.32 (0.54) & 0.32 (0.59) & 0.18 (0.46) \\ 
$\mu=m_b$ & 0.30 (0.46) &  0.32 (0.68)& 0.55 (0.93) &0.55 (1.02)  &0.15 (0.39) \\ 
\hline
\multicolumn{6}{c}{sensitivity: HL-LHC $1000\ifb$}  \\
\hline
 $\mu=\Lambda_{\rm LHC}$ & 0.18 (0.28) & 0.20 (0.41) &  0.19 (0.33) &  0.19 (0.35) & 0.11 (0.28) \\ 
$\mu=m_b$ & 0.18 (0.28) & 0.20 (0.41) &  0.33 (0.56) & 0.33 (0.61) & 0.09 (0.24) \\ 
\hline
\multicolumn{6}{c}{sensitivity: HL-LHC $3000\ifb$}  \\
   \hline
 $\mu=\Lambda_{\rm LHC}$ & 0.14 (0.21) & 0.15 (0.31) &  0.15 (0.25)& 0.15 (0.27) & 0.08 (0.21) \\ 
$\mu=m_b$ & 0.14 (0.21) & 0.15 (0.31)  & 0.25 (0.43) &  0.25 (0.47) & 0.07 (0.18) \\ 
\bhline{1 pt}
   \end{tabular}
   }
    \caption{\label{table:cutasummary} 
  Expected sensitivities to the absolute value of the WCs in the single-operator scenarios.
  They are evaluated at the scale of $\mu=\Lambda_{\rm LHC}$ and $m_b$ based on {\bf cut a} (the $\tau^\pm\nu$ search).
  The number without (inside) the parenthesis correspond to the EFT limit ($M_{\rm LQ}=1.5\TeV$).
  Also shown are the current upper bounds~\cite{Iguro:2020keo} based on the dataset of CMS $35.9\ifb$~\cite{Sirunyan:2018lbg}.}
  \vspace{0.6cm}
   \scalebox{1.1}{
  \begin{tabular}{c ccccc } 
  \bhline{1 pt}
  \rowcolor{white}
   $\tau^\pm\nu+b$ search &$C_{V_1}$ & $C_{V_2}$ & $C_{S_1}$ & $C_{S_2}$ & $C_{T}$  \\  \hline  
   \multicolumn{6}{c}{sensitivity: LHC $139\ifb$}  \\
   \hline
 $\mu=\Lambda_{\rm LHC}$ & 0.20 (0.31) &  0.20 (0.41) &  0.20 (0.33) &0.18(0.32)  & 0.11 (0.22)\\ 
$\mu=m_b$ &  0.20 (0.31) &  0.20 (0.41) &  0.33 (0.57) & 0.31 (0.56) & 0.09 (0.19)\\ 
\hline
\multicolumn{6}{c}{sensitivity: HL-LHC $1000\ifb$}  \\
\hline
 $\mu=\Lambda_{\rm LHC}$ & 0.12 (0.18) & 0.12 (0.24) & 0.11 (0.20)  &  0.11 (0.19) & 0.06 (0.13) \\ 
$\mu=m_b$ & 0.12 (0.18) & 0.12 (0.24)  &   0.20 (0.34)  & 0.18 (0.33) &  0.05 (0.11) \\ 
\hline
\multicolumn{6}{c}{sensitivity: HL-LHC $3000\ifb$}  \\
   \hline
 $\mu=\Lambda_{\rm LHC}$ & 0.09 (0.14) &0.09 (0.18) &  0.09 (0.15) &0.08 (0.14)  & 0.05 (0.10)\\ 
$\mu=m_b$ &0.09 (0.14) &0.09 (0.18)  &  0.15 (0.26) & 0.14 (0.25) & 0.04 (0.08) \\ 
\bhline{1 pt}
   \end{tabular}
   }
    \caption{\label{table:cutbsummary}
  Same as Table~\ref{table:cutasummary} but for {\bf cut b} (the $\tau^\pm\nu+b$ search).
    }
\end{table}

In Tables~\ref{table:cutasummary} and \ref{table:cutbsummary}, the expected sensitivities to $|C_X (\Lambda_\text{LHC})|$ are shown for each single operator scenario in the $\tau^\pm\nu$ and $\tau^\pm\nu+b$ searches, respectively.
They are determined from $S^{95\%}$ defined in the previous subsection. 
The number without (inside) the parenthesis is obtained in the EFT limit (at $M_{\rm LQ}=1.5\TeV$).
The integrated luminosities are $\int\mathcal{L}\,dt=139\ifb$ (for the current sensitivity) and $1000/3000\ifb$ (for future). 
The results are provided at two scales; one is a scale of flavor experiments, $m_b = 4.2\GeV$, and another is that of the collider search, $\Lambda_\text{LHC}$, which is fixed to be $1\TeV$ in this paper. 
The WCs at the scale of $m_b = 4.2\GeV$ are derived by taking RG running corrections into account.
For the $\tau^\pm\nu$ search, the current upper bounds on $|C_X (m_b)|$ are also listed in Table~\ref{table:cutasummary}.
They are obtained in Ref.~\cite{Iguro:2020keo} based on the CMS result with $\int\mathcal{L}\,dt=35.9\ifb$~\cite{Sirunyan:2018lbg}. 
Similar limits are provided in Refs.~\cite{Greljo:2018tzh,Marzocca:2020ueu}. 
It is noted that there are no experimental studies in the $\tau^\pm\nu+b$ search. 

The LQ mass dependence of the sensitivities are shown in Fig.~\ref{Fig:WC_bound_MLQ} for the integrated luminosities of $\int\mathcal{L}\,dt=139$ (solid line) and $3000\ifb$ (dashed). 
The scale is set to be $\Lambda_\text{LHC}$.
The blue (red) lines correspond to the $\tau^\pm\nu$ ($\tau^\pm\nu+b$) search.
In the figure, the upper plot for each scenario shows a sensitivity to the WC based on {\bf{cut~a}} or {\bf{cut~b}}.

As mentioned above, the charge of the final-state $\tau$ lepton is not identified in {\bf{cut~a}} or {\bf{cut~b}}.
If the event selections are performed with distinguishing the $\tau$-lepton charge, the sensitivities may be affected. 
The lower plot in each scenario displays $\delta C_{X}^{95\%}/C_{X}^{95\%}|_{\tau^\pm}$ for $\int\mathcal{L}\,dt=3000\ifb$ with $\delta C_{X}^{95\%} = C_{X}^{95\%}|_{\tau^\pm} - C_{X}^{95\%}|_{\tau^-}$, where $C_{X}^{95\%}|_{\tau^{-(\pm)}}$ is the sensitivity to the WC with (without) selecting negative charged $\tau$ leptons, $\tau^-$. 
A positive value means that the sensitivities are improved by selecting $\tau^-$ versus the result collecting both $\tau^+$ and $\tau^-$.

\begin{figure*}[p]
\begin{center}
\includegraphics[width=38em]{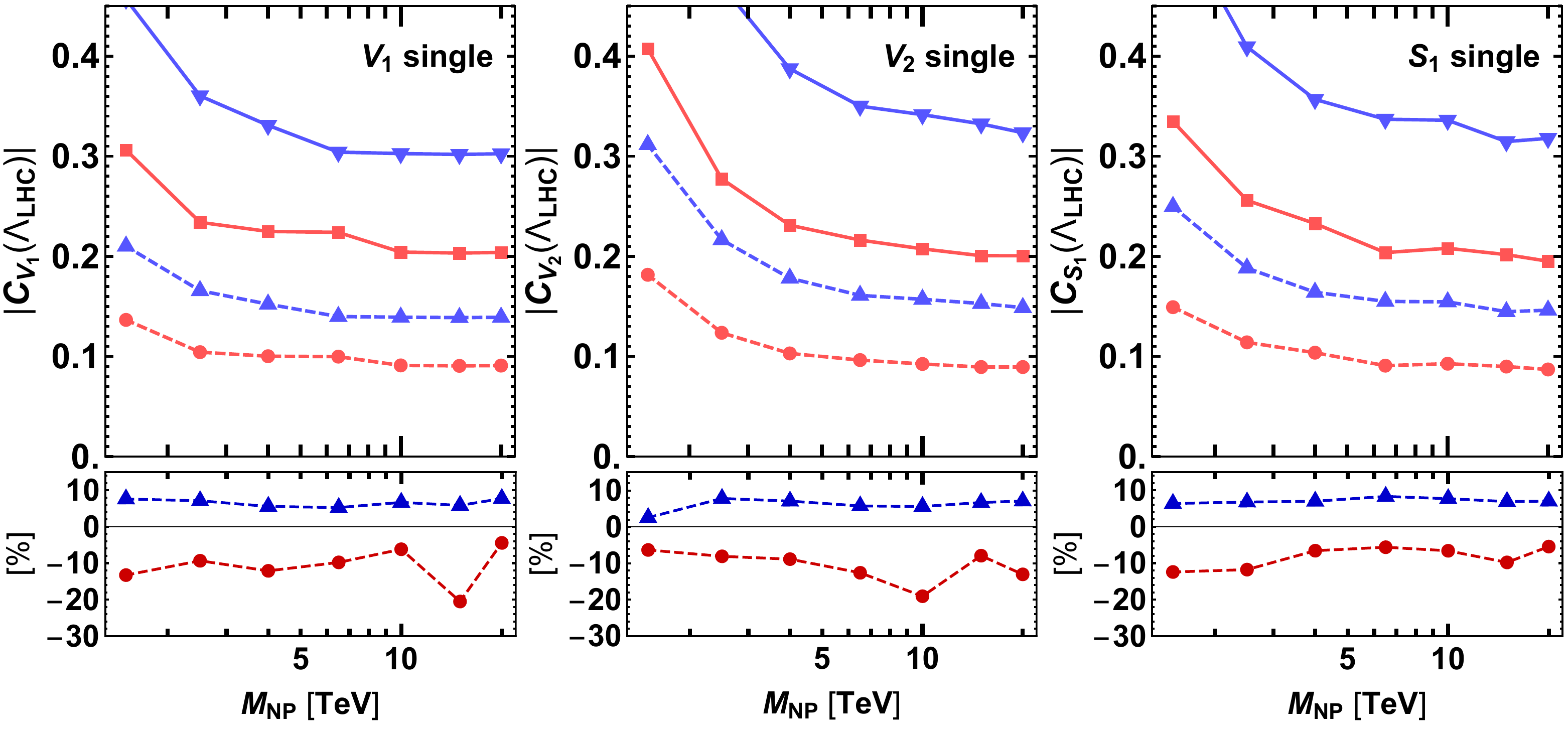} \\[0.5em]
\includegraphics[width=38em]{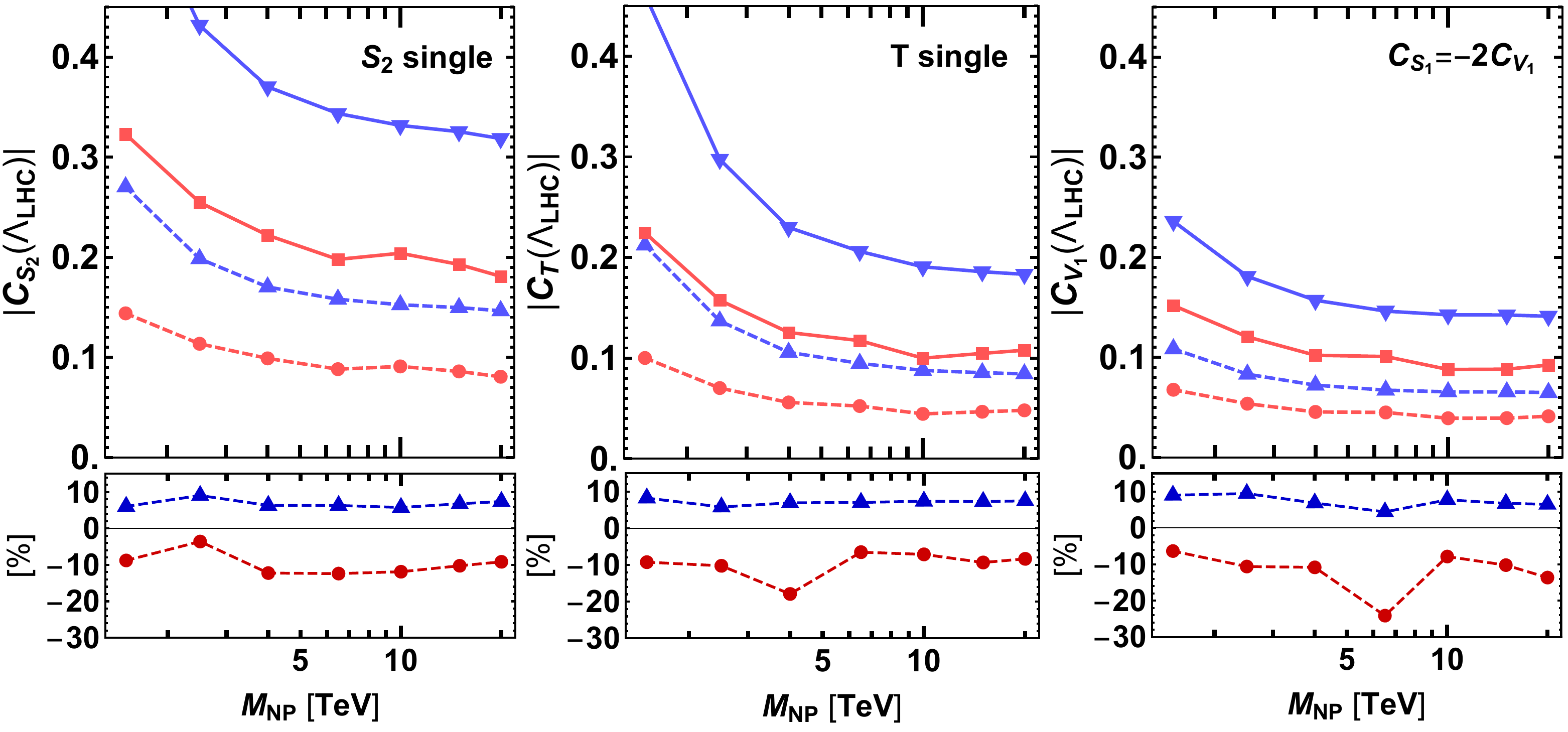} \\[0.5em]
\includegraphics[width=38em]{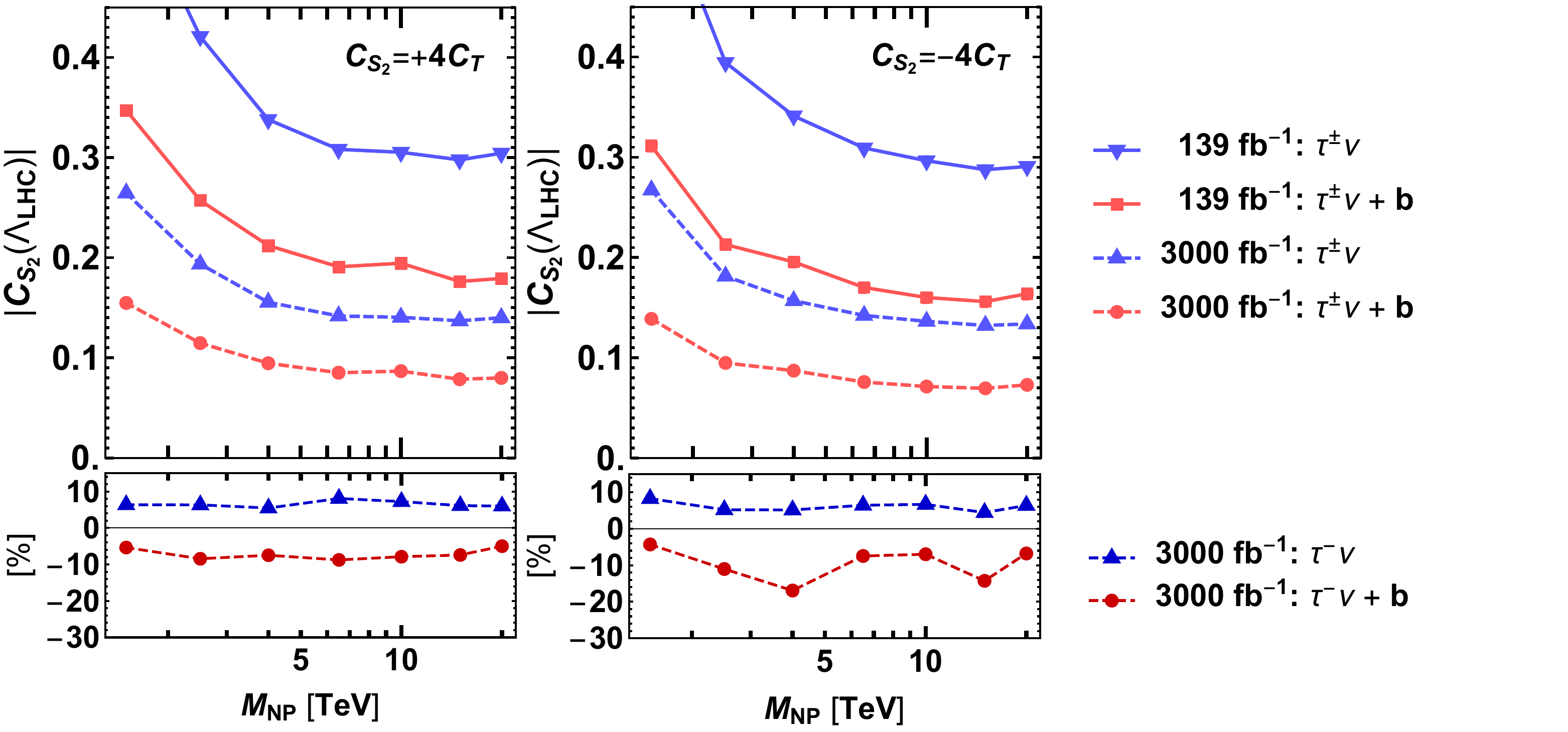} 
\caption{
\label{Fig:WC_bound_MLQ}
Expected sensitivities to the absolute value of $C_X (\Lambda_\text{LHC})$ (upper in each scenario) with $\int\mathcal{L}\,dt=139\ifb$ (solid) and $3000\ifb$ (dashed) in the $\tau^\pm\nu$ (blue) and $\tau^\pm\nu+b$ (red) searches.
In the lower plot of each scenario, $\delta C_{X}^{95\%}/C_{X}^{95\%}|_{\tau^\pm}$ is displayed for $\int\mathcal{L}\,dt=3000\ifb$ with $\delta C_{X}^{95\%} = C_{X}^{95\%}|_{\tau^\pm} - C_{X}^{95\%}|_{\tau^-}$, where $C_{X}^{95\%}|_{\tau^{-(\pm)}}$ is the sensitivity to the WC with (without) selecting $\tau^-$. }
\end{center}
\end{figure*}

Let us summarize our observations from the figure and tables:
\begin{itemize}
 \item 
 In the $\tau^\pm\nu$ search, compared with the current bounds, some of the sensitivities are not improved even with the larger dataset of $\int\mathcal{L}\,dt=139\ifb$. This is mainly because the observed data at CMS in Ref.~\cite{Sirunyan:2018lbg} are smaller than the expected SM background,
 probably due to unexpected (statistical) fluctuations.
 \item
 In the $\tau^\pm\nu$ search, the sensitivities to the WCs can be improved by a factor of two at $\int\mathcal{L}\,dt=3000\ifb$ compared with the current bounds ($36\ifb$) or sensitivities ($139\ifb$).
 \item 
 By requiring an additional $b$-jet in the final states, the NP sensitivities can be improved by 
 $\approx 40\%$
 versus those in the $\tau^\pm\nu$ search.
 Note that this is beyond the statistical uncertainty of our MC; 
 in the analysis, we generate 100K events for each NP model point, and the number of signal events after the cut could become $\lesssim 100$ for $\tau^-\nu+b$, 
 leading to $\mathcal{O}(10)\%$ MC-uncertainty at most.
 \item 
 The sensitivities depend on the LQ mass obviously; they become better as the mass increases.
 The dependence for $\tau^\pm\nu+b$ is similar to that for $\tau^\pm\nu$. 
 It is found that the sensitivity from the $\tau^\pm\nu+b$ search is better than that from $\tau^\pm\nu$ in the whole mass region. 
 (Note that the conclusion is valid for $R_{s/b} \lesssim 1$. See Fig.~\ref{Fig:U1contours}.)
 \item 
 In the case of the $\tau^-\nu$ search, by selecting the negative-charged $\tau$ leptons, the sensitivities can be improved by $\approx 10\%$ compared with the result obtained without selecting $\tau^-$. 
 However, the selection is not effective to improve the sensitivity for $\tau^-\nu+b$,
 especially because the number of events after the cut is not large enough even at $\int\mathcal{L}\,dt=3000\ifb$; 
 the number of signal events is halved by the charge selection.
 Then, with the number of background events $N_{BG} = \mathcal{O}(10)$, 
 the reduction of $N_{BG}$ due to the charge selection is not sufficient for improving $U_{\rm tot}$. 
 In other words, we found that larger $N_{BG}$ or better reduction is necessary to improve the sensitivity. 
 \item 
 In small LQ mass regions, the sensitivity for $C_{V_1}$ is better than that for $C_{V_2}$ at $\int\mathcal{L}\,dt=3000\ifb$, because of differences of $\tau$ angular distributions; the signal acceptance for the former is better than that for the latter (see Ref.~\cite{Iguro:2020keo}).
\end{itemize}

\subsection{\boldmath Single LQ scenarios}
\label{Sec:single_LQ}

We discuss an impact of the LHC searches on the single LQ scenarios that can solve the $R_{D^{(*)}}$ anomaly.
There are three single LQ fields, $\text{U}_1$, $\text{R}_2$, and $\text{S}_1$, introduced in 
Secs.~\ref{Sec:U1LQ}, \ref{Sec:R2LQ}, and \ref{Sec:S1LQ}, respectively. 
For calculating the flavor observables such as $R_{D^{(*)}}$, 
we used the formulae of Ref.~\cite{Iguro:2018vqb} with updating the form factors~\cite{Iguro:2020cpg}.

Let us first summarize the expected sensitivities based on the $\tau^\pm\nu$ and $\tau^\pm\nu+b$ searches in Tables~\ref{table:LQcutasummary} and \ref{table:LQcutbsummary}, respectively.
Here, the sensitivity to $C_{S_2}$ is shown for $\text{R}_2$ and $\text{S}_1$, while that to $C_{S_1}$ is given for the scenario of $\text{U}_1$ LQ with $U(2)$ flavor symmetry.
The interplay with the $R_{D^{(*)}}$ anomaly is discussed in the following subsections. 

In discussing the LHC search for the NP contributions and its interplay with the flavor observables, 
there are three renormalization scales; $\mu = m_b$, $\Lambda_\text{LHC}$, and $M_\text{LQ}$. 
The WCs in different energy scales should be evaluated by taking the RG corrections into account~\cite{Alonso:2013hga,Gonzalez-Alonso:2017iyc,Blanke:2018yud}. 
Although all WCs are to be input at $\mu = M_\text{LQ}$, 
we show the results with discarding the RG corrections between $\Lambda_\text{LHC}=1\TeV$ and $M_\text{LQ}=\mathcal{O}(1)\TeV$, because they are found to be negligible (a few percent level for WCs).
Nonetheless, the corrections are taken into account for $\mu = m_b$.

\begin{table}[ptb!]
\centering
\newcommand{\bhline}[1]{\noalign{\hrule height #1}}
\renewcommand{\arraystretch}{1.5}
   \scalebox{1.1}{
  \begin{tabular}{c ccccc } 
  \bhline{1 pt}
  \rowcolor{white}
  $\tau^\pm\nu$ search & $\text{R}_2(C_{S_2})$&$\text{S}_1 (C_{S_2})$ & $\text{U}_1^{R_{s/b}=0} (C_{S_1})$  & $\text{U}_1^{R_{s/b}=1} (C_{S_1})$& $C_{V_1}^{R_{s/b}=1}$  \\  \hline  
   \multicolumn{6}{c}{sensitivity: LHC $139\ifb$}  \\
   \hline
 $\mu=\Lambda_{\rm LHC}$  & 0.30 (0.58) & 0.29 (0.58)& 0.28 (0.47) & 0.24 (0.40) & 0.18 (0.28) \\ 
$\mu=m_b$  & 0.51 (0.96) &0.52 (1.04)& 0.48 (0.81) &
0.41 (0.69) & 0.18 (0.28) \\ 
\hline
\multicolumn{6}{c}{sensitivity: HL-LHC $1000\ifb$}  \\
\hline
 $\mu=\Lambda_{\rm LHC}$  &0.18 (0.35) & 0.18 (0.35)& 0.17 (0.28) & 0.14 (0.24) &  0.11 (0.17) \\ 
$\mu=m_b$ &0.31 (0.58)& 0.32 (0.63)  &0.29 (0.49) &
0.25 (0.42) &  0.11 (0.17)  \\ 
\hline
\multicolumn{6}{c}{sensitivity: HL-LHC $3000\ifb$}  \\
   \hline
 $\mu=\Lambda_{\rm LHC}$  &0.14 (0.26) & 0.13 (0.27)& 0.13 (0.22) & 0.11 (0.19) & 0.08 (0.13) \\ 
$\mu=m_b$  & 0.23 (0.44) & 0.24 (0.48)& 0.22 (0.37)& 
0.19 (0.32) & 0.08 (0.13) \\ 
\bhline{1 pt}
   \end{tabular}
   }
    \caption{\label{table:LQcutasummary} 
    Expected sensitivities of the absolute value of the WCs for {\bf cut a} in the single LQ scenarios.
    The sensitivity to $C_{S_2}$ is shown for $\text{R}_2$ and $\text{S}_1$, while that to $C_{S_1}$ is given for the scenario of $\text{U}_1$ LQ with $U(2)$ flavor symmetry.
    See Table~\ref{table:cutasummary} for details of the descriptions. }
    \vspace{0.6cm}
   \scalebox{1.1}{
  \begin{tabular}{c ccccc } 
  \bhline{1 pt}
  \rowcolor{white}
  $\tau^\pm\nu+b$ search & $\text{R}_2 (C_{S_2})$  &$\text{S}_1 (C_{S_2})$ & $\text{U}_1^{R_{s/b}=0} (C_{S_1})$  & $\text{U}_1^{R_{s/b}=1} (C_{S_1})$& $C_{V_1}^{R_{s/b}=1}$  \\  \hline  
   \multicolumn{6}{c}{sensitivity: LHC $139\ifb$}  \\
   \hline
 $\mu=\Lambda_{\rm LHC}$  &0.18 (0.35) &0.16 (0.31) & 0.18 (0.30) &0.16 (0.28)  & 0.17 (0.25) \\ 
$\mu=m_b$  & 0.30 (0.58) & 0.29 (0.56)& 0.32 (0.52) &
0.27 (0.48) & 0.17 (0.25)  \\ 
\hline
\multicolumn{6}{c}{sensitivity: HL-LHC $1000\ifb$}  \\
\hline
 $\mu=\Lambda_{\rm LHC}$ & 0.11 (0.20)  & 0.10 (0.18)& 0.11 (0.18) & 0.09 (0.17) &  0.10 (0.15)\\ 
$\mu=m_b$  & 0.18 (0.34) & 0.17 (0.33)& 0.19 (0.31) & 
0.16 (0.28) &  0.10 (0.15) \\ 
\hline
\multicolumn{6}{c}{sensitivity: HL-LHC $3000\ifb$}  \\
   \hline
 $\mu=\Lambda_{\rm LHC}$ & 0.08 (0.15)& 0.07 (0.14)  & 0.08 (0.14) & 0.07 (0.13) & 0.07 (0.11) \\ 
$\mu=m_b$  & 0.13 (0.26)& 0.13 (0.25) & 0.14 (0.23) & 
0.12 (0.21) &  0.07 (0.11) \\ 
\bhline{1 pt}
   \end{tabular}
   }
    \caption{\label{table:LQcutbsummary} 
    Same as Table~\ref{table:LQcutasummary} but for 
   {\bf cut b} (the $\tau^\pm \nu + b$ search).
        }
\end{table}

\subsubsection{R$_2$ LQ scenarios}

In the $\text{R}_2$ LQ model, two sets of WCs, $C_{V_2} (M_\text{LQ})$ and $C_{S_2}(M_\text{LQ}) = +4 C_T(M_\text{LQ})$, are induced independently, as explained in Eq.~\eqref{Eq:R2LQWC}. 
Thus, we study the following two scenarios; single-$\text{R}_2 (C_{V_2})$ and single-$\text{R}_2 (C_{S_2})$ scenarios separately.
For each scenario, in order to solve the $R_{D^{(*)}}$ anomaly, we obtain that the WCs are favored to be 
\begin{align}
 &\text{single-$\text{R}_2 (C_{V_2})$}: C_{V_2} (\Lambda_\text{LHC}) \approx \pm i\, 0.42 \,,&  
 &\text{single-$\text{R}_2 (C_{S_2})$}: C_{S_2} (\Lambda_\text{LHC}) \approx -0.07 \pm i\, 0.35 \,,&
\end{align}
where the measured values of $R_{D}$ and $R_{D^{*}}$ are fitted. 
Note that $\pm$ does not mean an uncertainty but represents two solutions. 
Since the LHC study is almost insensitive to the phase of WCs, it is set to be zero in the collider analysis. 

Such large WCs are expected to be probed at the LHC.\kf\footnote{Interference with the SM part is preferred to be small by 
a fit for the $R_{D^{(\ast)}} $ anomaly
in the $\text{R}_2$ LQ model. 
Therefore, 
resultant WCs have large imaginary components, and their absolute values tend to be large enough to be able to probed at the LHC.}
In the single-$\text{R}_2 (C_{V_2})$ scenario, by comparing Tables~\ref{table:LQcutasummary} and \ref{table:LQcutbsummary} with the background results in Fig.~\ref{Fig:WC_bound_MLQ},
it is found that the LHC sensitivity of the $\tau^\pm\nu$ search 
is marginal at $\int\mathcal{L}\,dt=139\ifb$ to test the $R_{D^{(*)}}$ explanation depending on the LQ mass, 
whereas that of the $\tau^\pm\nu+b$ search is enough to probe
the parameter region
in all ranges of the LQ mass.
We would like to stress that the scenario can be probed with use of the current data samples at the LHC ($139\ifb$) for $\tau^\pm\nu+b$. 
On the other hand, in the single-$\text{R}_2 (C_{S_2})$ scenario, it is also shown that the $R_{D^{(*)}}$-favored value of $|C_{S_2} (\Lambda_\text{LHC})| \approx 0.36$ can be fully probed by the $\tau^\pm\nu+b$ search at $139\ifb$, but not by $\tau^\pm\nu$.  
Therefore, it is concluded that requiring an additional $b$-jet is significant to test the LQ scenarios in light of the $R_{D^{(*)}}$ anomaly.
%
\begin{figure*}[tb!]
\begin{center}
\subfigure[$M_{\text{R}_2\,{\rm LQ}}=1.5\TeV$]{
\includegraphics[width=19em]{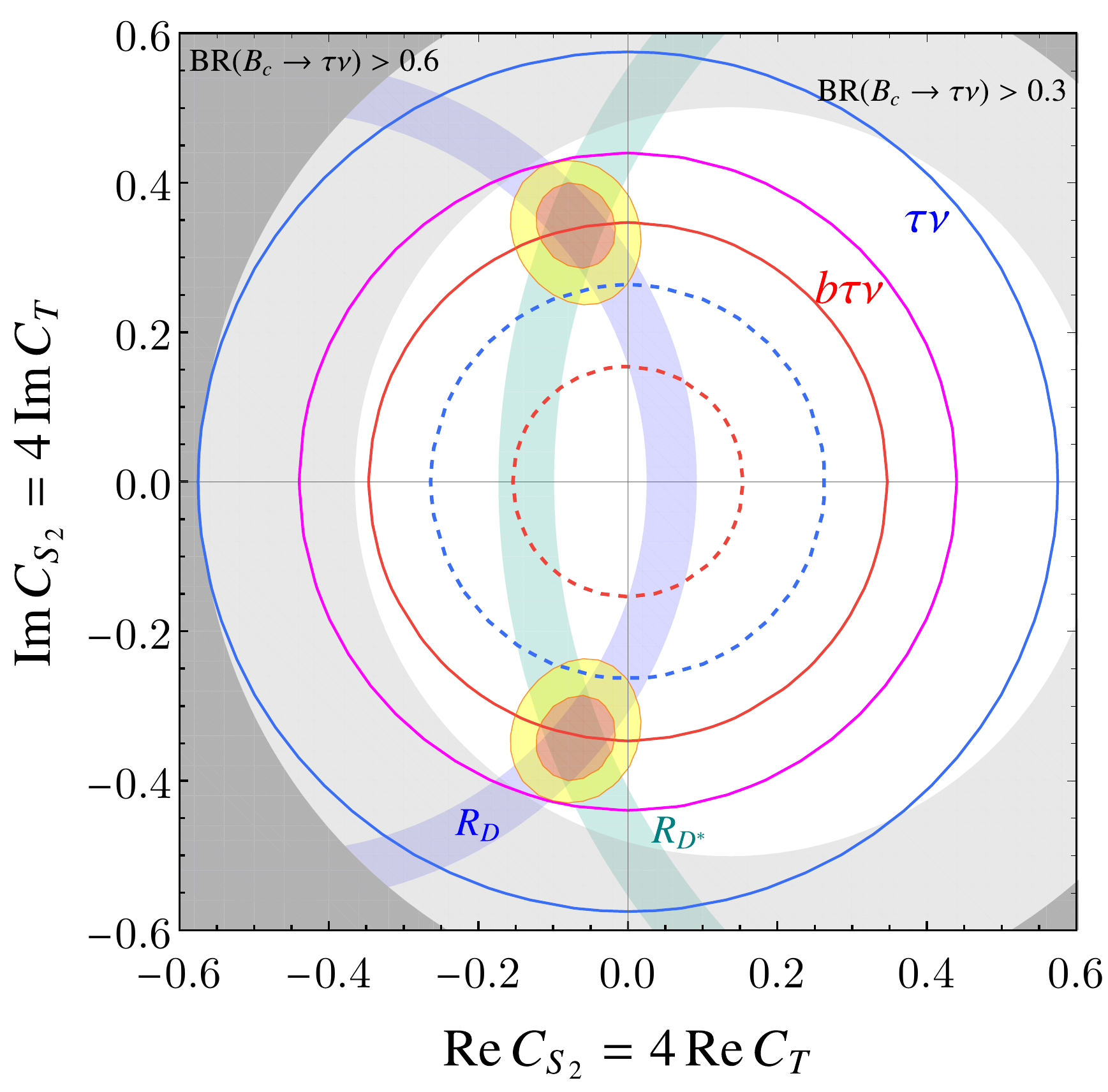}}~~
\subfigure[$M_{\text{R}_2\,{\rm LQ}}=2.5\TeV$]{
\includegraphics[width=19em]{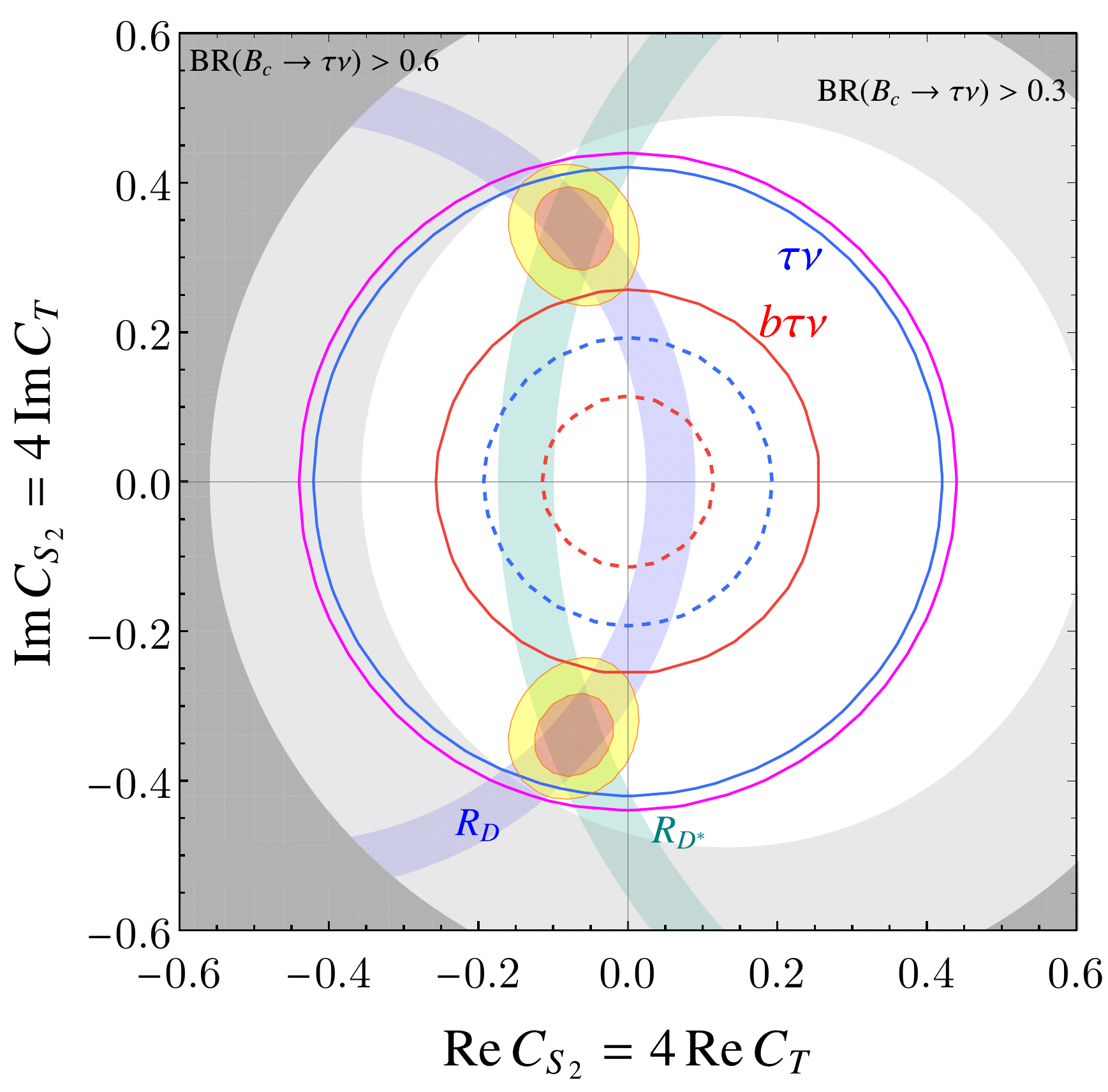}}
\caption{
\label{Fig:WC_bound_R2}
The $\text{R}_2$ LQ scenario with $M_{\rm LQ}=1.5$ and $2.5\TeV$.
The regions outside the blue and red lines are probed by the $\tau^\pm \nu$ and $\tau^\pm \nu+b$ searches, respectively,
where the solid (dashed) lines correspond to $\int\mathcal{L}\,dt=139\ifb$ ($3000\ifb$).
The magenta line shows the current bound from the experimental data with $\int\mathcal{L}\,dt=36\ifb$~\cite{Iguro:2020keo}.  
The (lighter) gray shaded regions are constrained by BR$(B_c \to \tau \nu)>0.6$ ($>0.3$).
The $R_D$ and $R_{D^\ast}$ anomalies are explained at $1\,\sigma$ in the blue and green shaded regions, respectively, 
while the combined fit at $1/2\,\sigma$ is shown in orange/yellow. 
} 
\end{center}
\end{figure*}

The combined summary plot for the LHC sensitivity and the allowed region from the flavor observables is shown in Fig.~\ref{Fig:WC_bound_R2} for the case of the single-$\text{R}_2 (C_{S_2})$ scenario with $M_\text{LQ} = 1.5\TeV$ and $2.5\TeV$.
The sensitivity at $\int\mathcal{L}\,dt=139\ifb$ from the $\tau^\pm\nu$ and $\tau^\pm\nu+b$ channels are denoted by solid blue and red lines, respectively. 
Their HL-LHC prospects at $\int\mathcal{L}\,dt=3000\ifb$ are shown in dashed lines with the same color.
The magenta lines are the current constraint from the CMS $36\ifb$ data, taken from Ref.~\cite{Iguro:2020keo} assuming $M_{\text{LQ}}=2\TeV$. 
The blue and green bands show the region favored by the measured $R_{D}$ and $R_{D^*}$, respectively. 
Then, the combined $1\,\sigma$ ($2\,\sigma$) regions are shown in red (yellow). 
We also put the $B_c$ constraint as $\text{BR}(B_c\to\tau\nu)< 60\%$ ($30\%$) shown in (light) gray as references.
Here, an updated study for the $B_c\to\tau\nu$ constraint is available in Refs.~\cite{Aebischer:2021ilm,Aebischer:2021eio}.
We can clearly check from this figure that the $\tau^\pm\nu+b$ search fully (partially) covers the single-$\text{R}_2 (C_{S_2})$ scenario 
with $M_\text{LQ}>2.5\TeV$ ($1.5\TeV<M_\text{LQ}<2.5\TeV$) responsible for the $R_{D^{(*)}}$ anomaly.

\subsubsection{$\text{S}_1$ LQ scenario} 
In the $\text{S}_1$ LQ model, two sets of WCs, $C_{V_1} (M_\text{LQ})$ and $C_{S_2}(M_\text{LQ}) = -4 C_T(M_\text{LQ})$, are induced independently as given in Eq.~\eqref{Eq:S1LQWC}. 
In the latter case, although the $R_{D^{(*)}}$ discrepancy can be reduced, the experimental result cannot be addressed within $1\,\sigma$. 
Thus, the study is performed in the two-dimensional parameter space, $(C_{V_1}, C_{S_2})$.

In Fig.~\ref{Fig:WC_bound_S1}, the LHC sensitivity and the region favored by the $R_{D^{(*)}}$ anomaly are shown for the $\text{S}_1$ LQ scenario on the plane of $(C_{V_1}, C_{S_2})$. Here, the imaginary components are fixed to be zero. See Fig.~\ref{Fig:WC_bound_R2} for the color convention. 
As briefly mentioned in Sec.~\ref{Sec:S1LQ}, unlike the cases for $\text{R}_2$ and $\text{U}_1$ LQs, the $\text{S}_1$ LQ scenario inevitably produces a tree-level contribution to $b \to s \nu \nubar$ in addressing $R_{D^{(*)}}$. 
Thus, the parameter space is constrained from precision measurements of $B \to K^{(\ast)} \nu \nubar$, which is shown in the figure with the cyan-shaded region. 
Its evaluation formula is given in Appendix~\ref{Sec:App_flavor}.
In addition, a more robust flavor bound comes from the $B_s$--$\Bb_s$ mixing ($\Delta M_s$). 
Based on the studies of Refs.~\cite{Crivellin:2019dwb,Crivellin:2021lix}, the $\Delta M_s$ constraint is provided in the figure with the red-shaded region. 
This bound is comparable to or severer than $B \to K^{(\ast)}\nu\nubar$ depending on the LQ mass. 
See again Appendix~\ref{Sec:App_flavor} for its detail. 
Although $\Delta M_d /\Delta M_s$ can give more stringent bound in general since QCD uncertainties are partially canceled, this constraint is avoidable if additional LQ contributions to $\Delta M_d$ are introduced properly.

From the figure, we see that the $\text{S}_1$ LQ mass larger than $4\TeV$ is disfavored by the $\Delta M_s$ and $B \to K^{(\ast)} \nu \nubar$ constraints. 
Since this implies that the smaller LQ mass $M_\text{LQ} < 4\TeV$ is viable for the $R_{D^{(\ast)}}$ anomaly, the LQ mass dependence on the NP sensitivity of the present LHC searches is important. 
Then, one can see that the $\tau^\pm\nu+b$ search at $\int\mathcal{L}\,dt=3000\ifb$ can reach the sensitivity to probe this scenario, while $\tau^\pm\nu$ cannot.

\begin{figure*}[th]
\begin{center}
\subfigure[$M_{\text{S}_1\,{\rm LQ}}=1.5\TeV$]{
\includegraphics[width=19em]{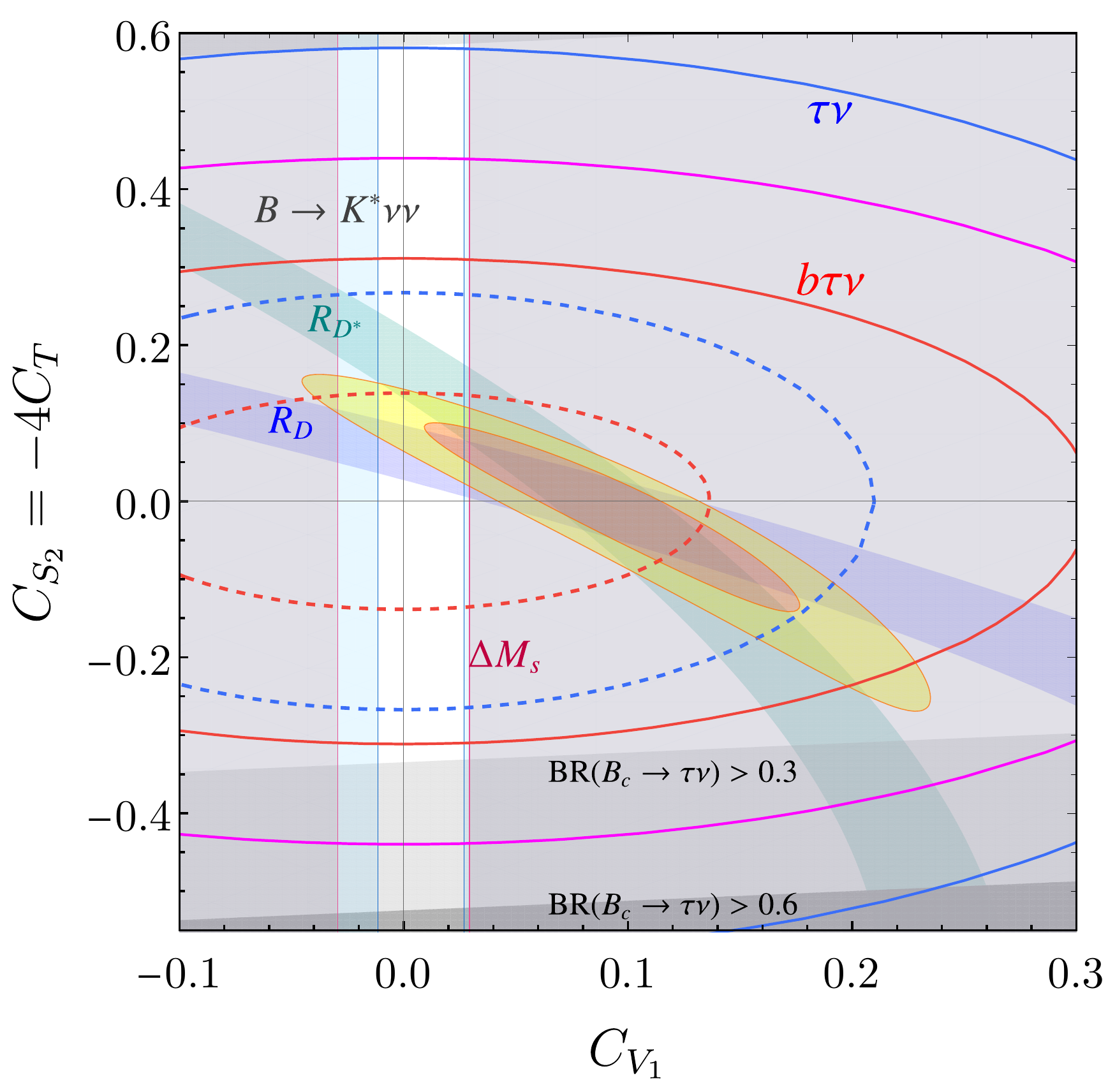}}~~
\subfigure[$M_{\text{S}_1\,{\rm LQ}}=4.0\TeV$]{
\includegraphics[width=19em]{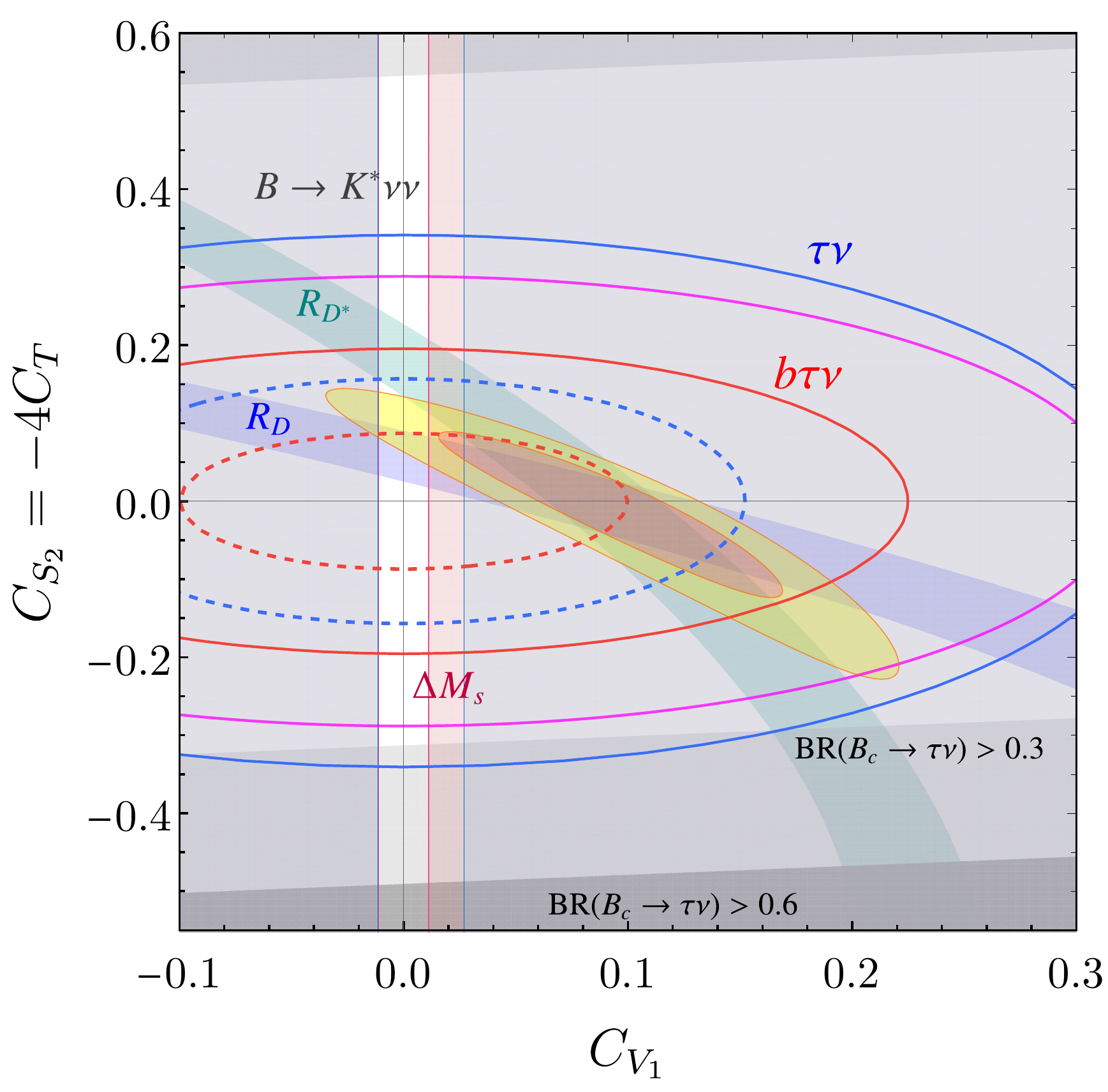}}
\caption{
\label{Fig:WC_bound_S1}
The $\text{S}_1$ LQ scenario with $M_{\rm LQ}=1.5\TeV$ and $4.0\TeV$ on the $(C_{V_1}, C_{S_2})$ plane. 
The color convention is the same as in Fig.~\ref{Fig:WC_bound_R2}. 
The magenta lines are the current bound from the experimental data with $\int\mathcal{L}\,dt=36\ifb$
by assuming $M_{\text{LQ}}=2\TeV$ (left panel) and the EFT limit (right panel)~\cite{Iguro:2020keo}. 
In addition, the cyan-shaded region is excluded by the $B \to K^\ast \nu \nubar $ measurement at the $90\%$ CL, and the red-shaded region is excluded by $\Delta M_s$.
} 
\end{center}
\end{figure*}

\subsubsection{U$_1$ LQ scenarios}

The U$_1$ vector LQ model introduced in Sec.~\ref{Sec:U1LQ} is one of the most promising candidates to solve the $B$ anomalies. 
In fact, unlike the above two scalar LQ scenarios,
flavor constraints can be suppressed or avoided. 
Therefore, the LHC search is significant to probe the model. 
In this paper, we investigate two scenarios in terms of the WCs of Eq.~\eqref{Eq:U1LQWC}; 
the single $C_{V_1}$ scenario (setting $C_{S_1}=0$), 
and the scenario satisfying $C_{S_1}=-2e^{i \phi_R} C_{V_1}$ with the $U(2)$ flavor symmetry, referred as the single-$\text{U}_1$ and $U(2)$-$\text{U}_1$ scenarios, respectively. 
By performing a parameter fit for these two scenarios to the $R_{D^{(*)}}$ measurement, we obtain the following WCs, 
\begin{align}
 &\text{single-$\text{U}_1$}: C_{V_1} (\Lambda_\text{LHC}) \approx 0.09 \,,&  
 &\text{$U(2)$-$\text{U}_1$}: C_{V_1} (\Lambda_\text{LHC}) \approx 0.09 \,,~ \phi_R \approx \pm0.42\pi\,.&
\end{align}
Note again that $\pm$ does not mean an uncertainty. 
Also, the phase $\phi_R$ for $U(2)$-$\text{U}_1$ is almost irrelevant for the present LHC study.

\begin{figure*}[t!]
\begin{center}
\includegraphics[width=17em]{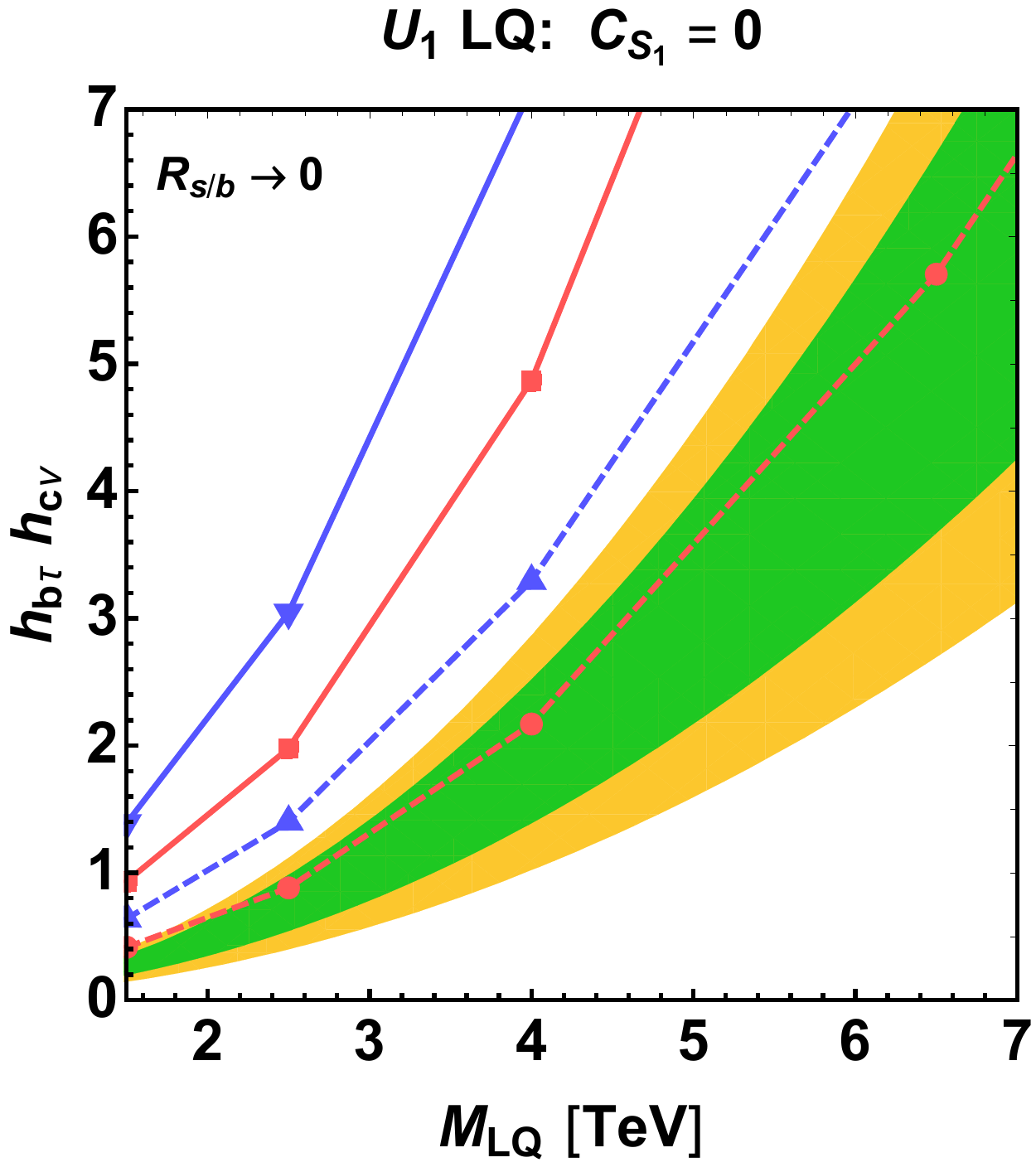}~
\includegraphics[width=17em]{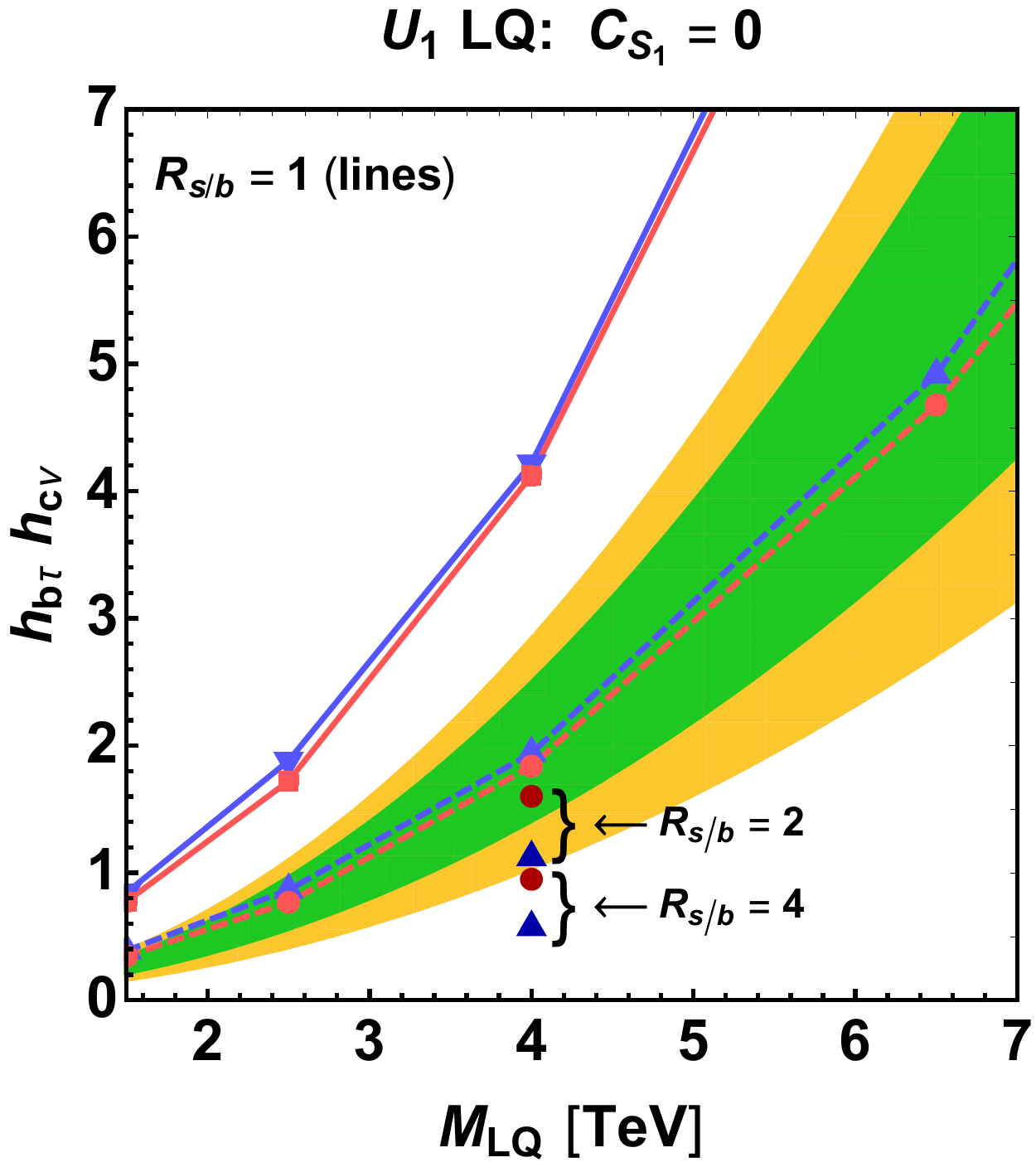} \\[1.5em]
\includegraphics[width=17em]{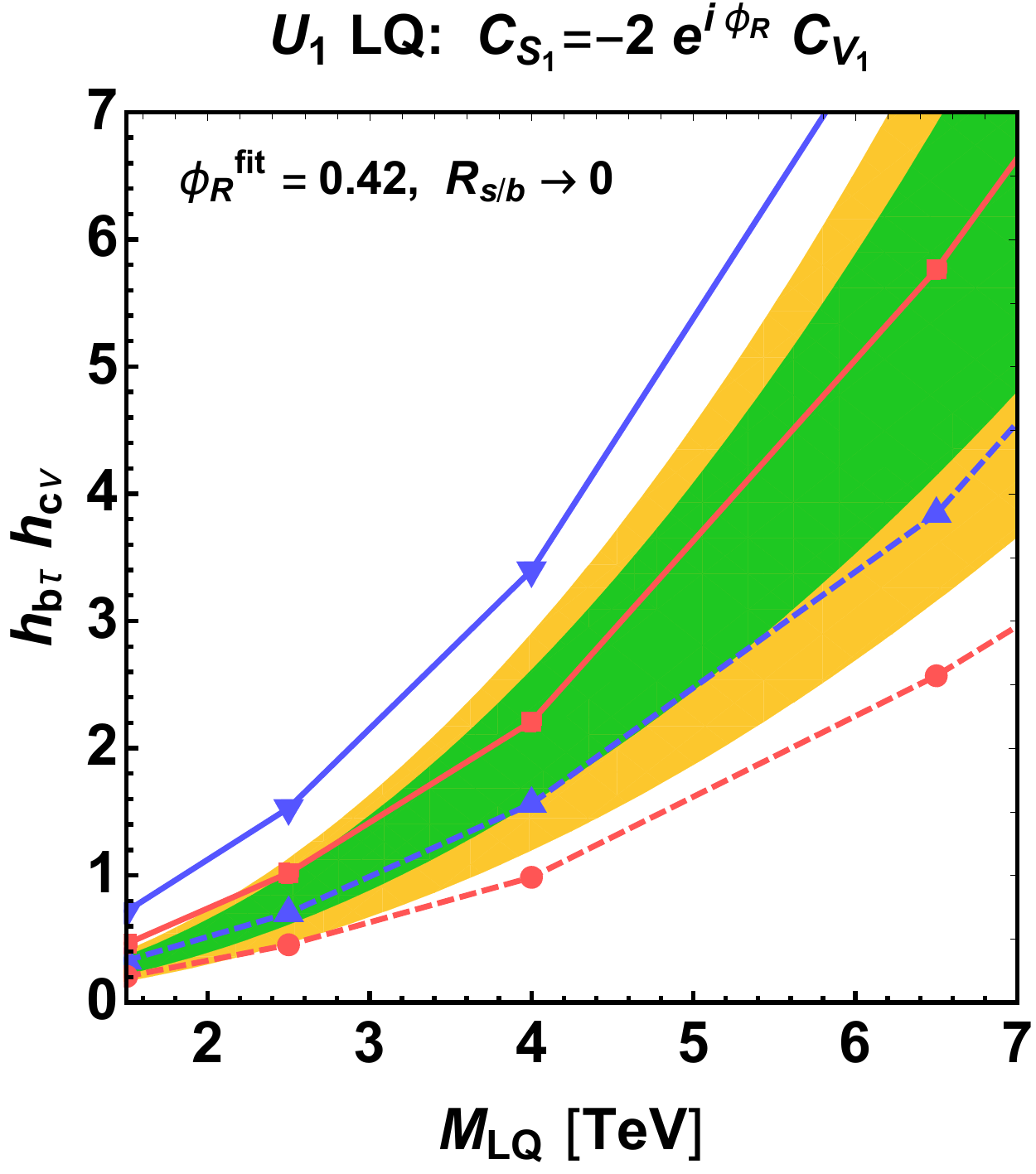}~
\includegraphics[width=17em]{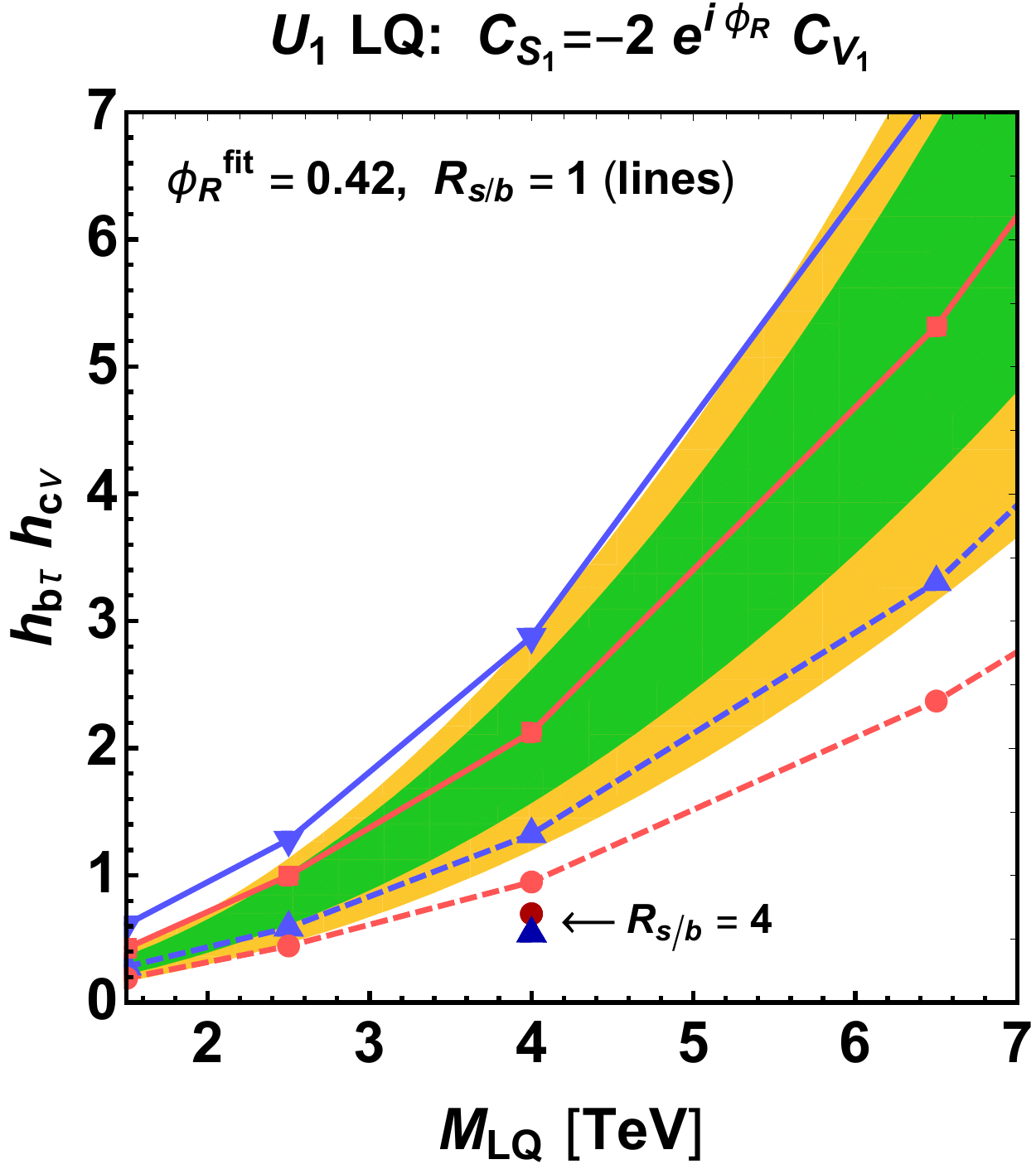} 
\caption{
\label{Fig:WC_bound_CV1}
Expected sensitivities to the $\text{U}_1$ LQ scenario.
The vertical axis is a product of the $\text{U}_1$ couplings, $h_L^{33}h_L^{23} \equiv h_{b\tau} h_{c\nu}$, and the horizontal one is the LQ mass, $M_\text{LQ}$.
Here, $R_{s/b} \to 0$ and $R_{s/b} = 1$ in the left and right panels, respectively.
The $R_{D^{(*)}}$ anomaly is solved at the $1\,\sigma$ (green) and $2\,\sigma$ (yellow) levels.
See Fig.~\ref{Fig:WC_bound_MLQ} for the conventions of the plot markers and colors. 
The sensitivities for $R_{s/b} = 2,4$ and $4$ are also shown in the single-$\text{U}_1$ and $U(2)$-$\text{U}_1$ scenarios, respectively, at $M_\text{LQ}=4\TeV$. 
} 
\end{center}
\end{figure*}

In Fig.~\ref{Fig:WC_bound_CV1}, the NP sensitivities are shown in the $\tau^\pm\nu$ ($\tau^\pm\nu + b$) search by the blue (red) lines. The solid (dashed) lines correspond to $\int\mathcal{L}\,dt=139\ifb$ ($3000\ifb$).
The vertical axis is a product of the $\text{U}_1$ couplings, $h_L^{33}h_L^{23} \equiv h_{b\tau} h_{c\nu}$, and the horizontal one is the LQ mass, $M_\text{LQ}$. 
The region favored by $R_{D^{(*)}}$ at the $1\,\sigma$ ($2\,\sigma$) level is also given in the green (yellow) color. 
Regarding the $U(2)$-$\text{U}_1$ scenario, the relative phase is fixed as $\phi_R =0.42\pi$. 

In the figure, the results are shown for $R_{s/b} \to 0$ and $R_{s/b} = 1$ in the left and right panels, respectively.
The former corresponds to the single $C_{V_1}$ scenario, and hence, the NP sensitivity is exactly the same as that given in the previous section. 
On the other hand, since $h_L^{23} = h_{s\tau} = h_{c\nu}$ in the $\text{U}_1$ LQ model as aforementioned in Sec.~\ref{Sec:signal}, 
the latter indicates how $h_{s\tau} \neq 0$ contribution to the signal production affects the NP sensitivity. 
For $R_{s/b} = 1$, it is found that the $\tau^\pm\nu$ search can be competitive to that of $\tau^\pm\nu + b$. 
We also show the sensitivities for larger $R_{s/b}$ as $= 2,4$ and $4$ in the single-$\text{U}_1$ and $U(2)$-$\text{U}_1$ scenarios, respectively, at $M_\text{LQ}=4\TeV$ for further comparison.
It is concluded from the figures that both scenarios can be tested at HL-LHC with $\int\mathcal{L}\,dt=3000\ifb$. 
Regarding the $U(2)$-$\text{U}_1$ scenario, the present LHC data sample is large enough to probe the scenario if the $\tau^\pm\nu + b$ analysis is performed.  
Also, it should be mentioned that the result depends on $R_{s/b}$ significantly. 
It is shown that the sensitivities are enhanced by larger $R_{s/b}$ even in the EFT limit. 
Its contribution will be investigated in detail later.

\begin{figure*}[t!]
\begin{center}
\includegraphics[width=18em]{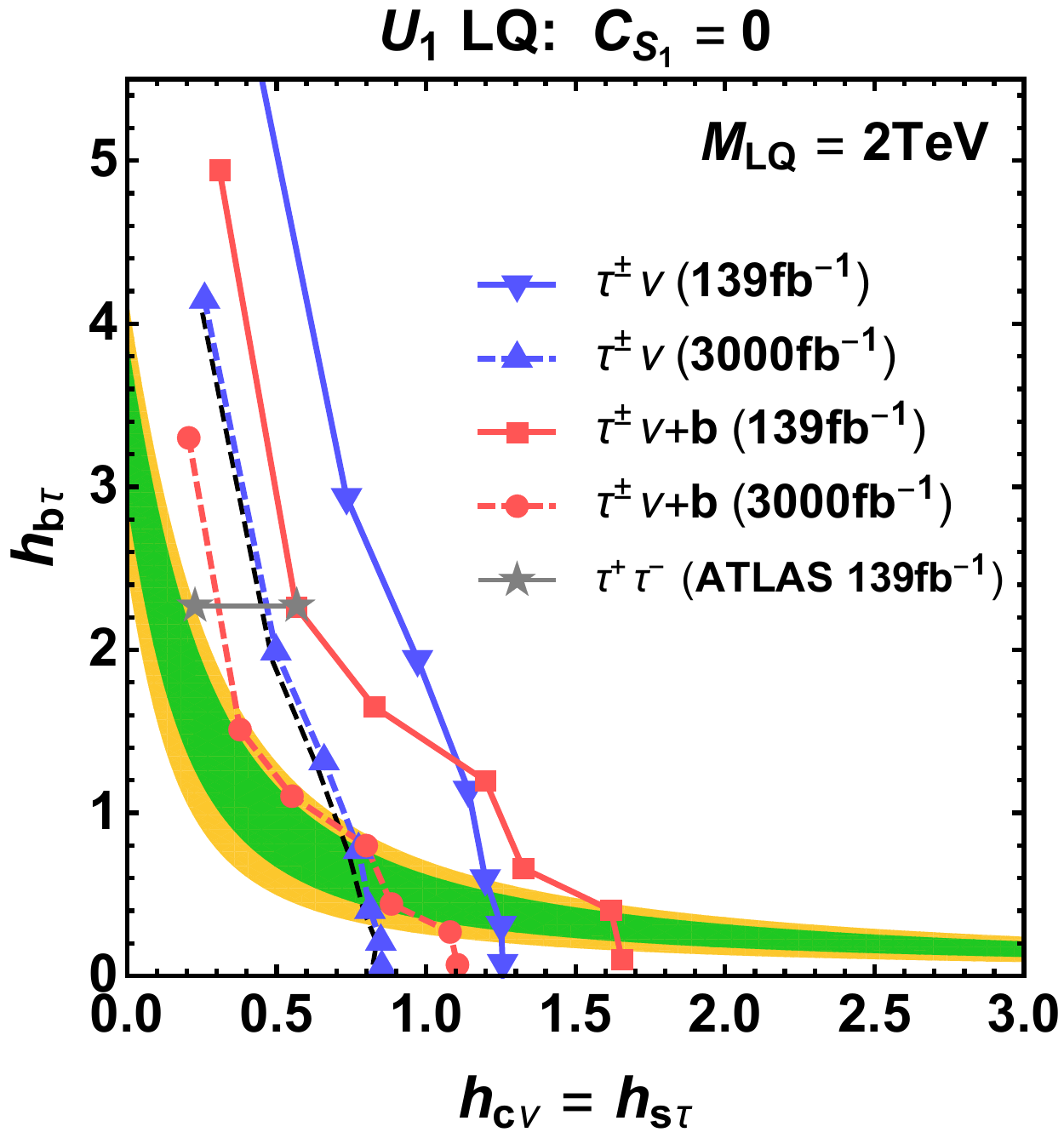}
\includegraphics[width=18em]{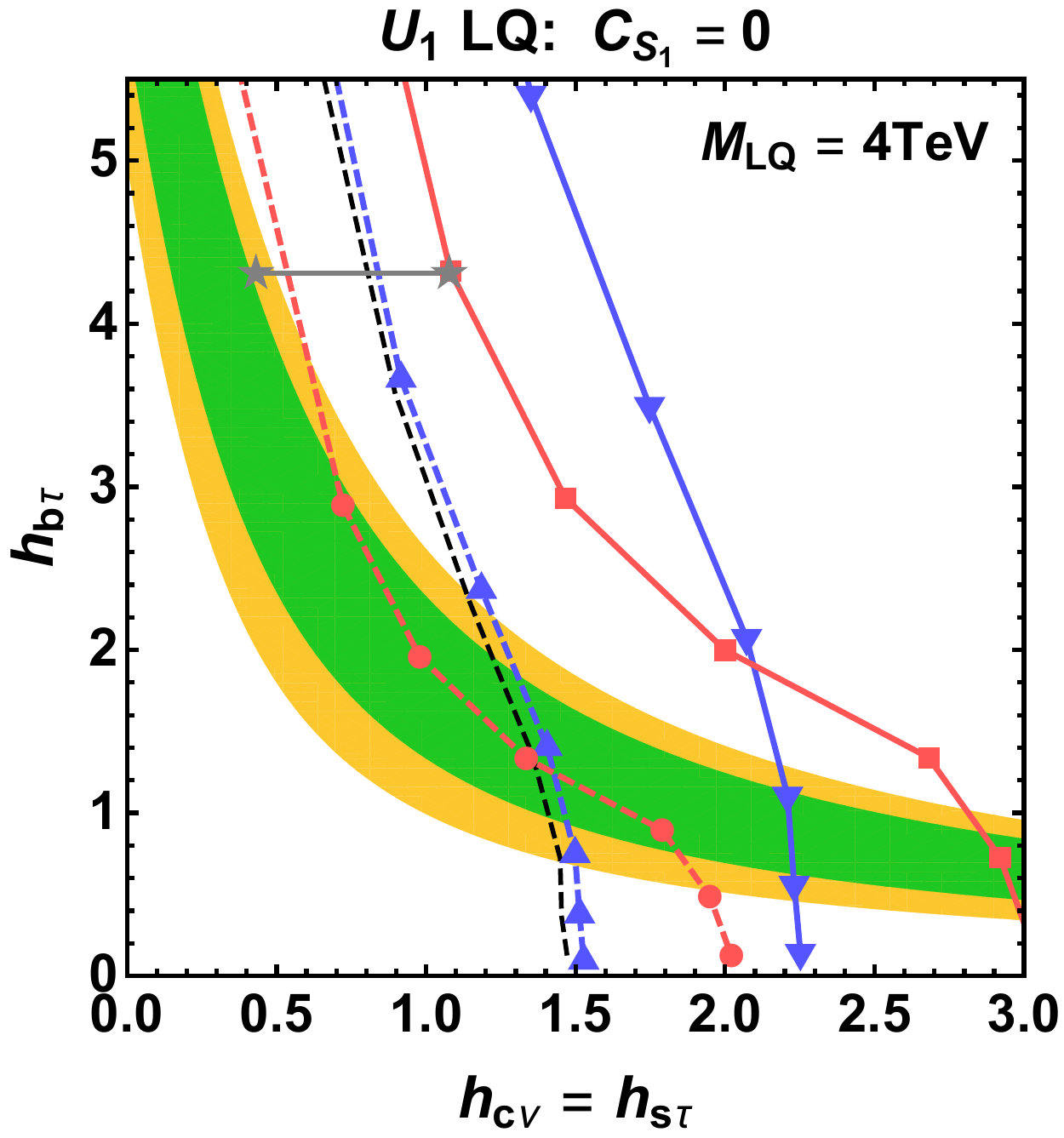}\\[1.5em]
\includegraphics[width=18em]{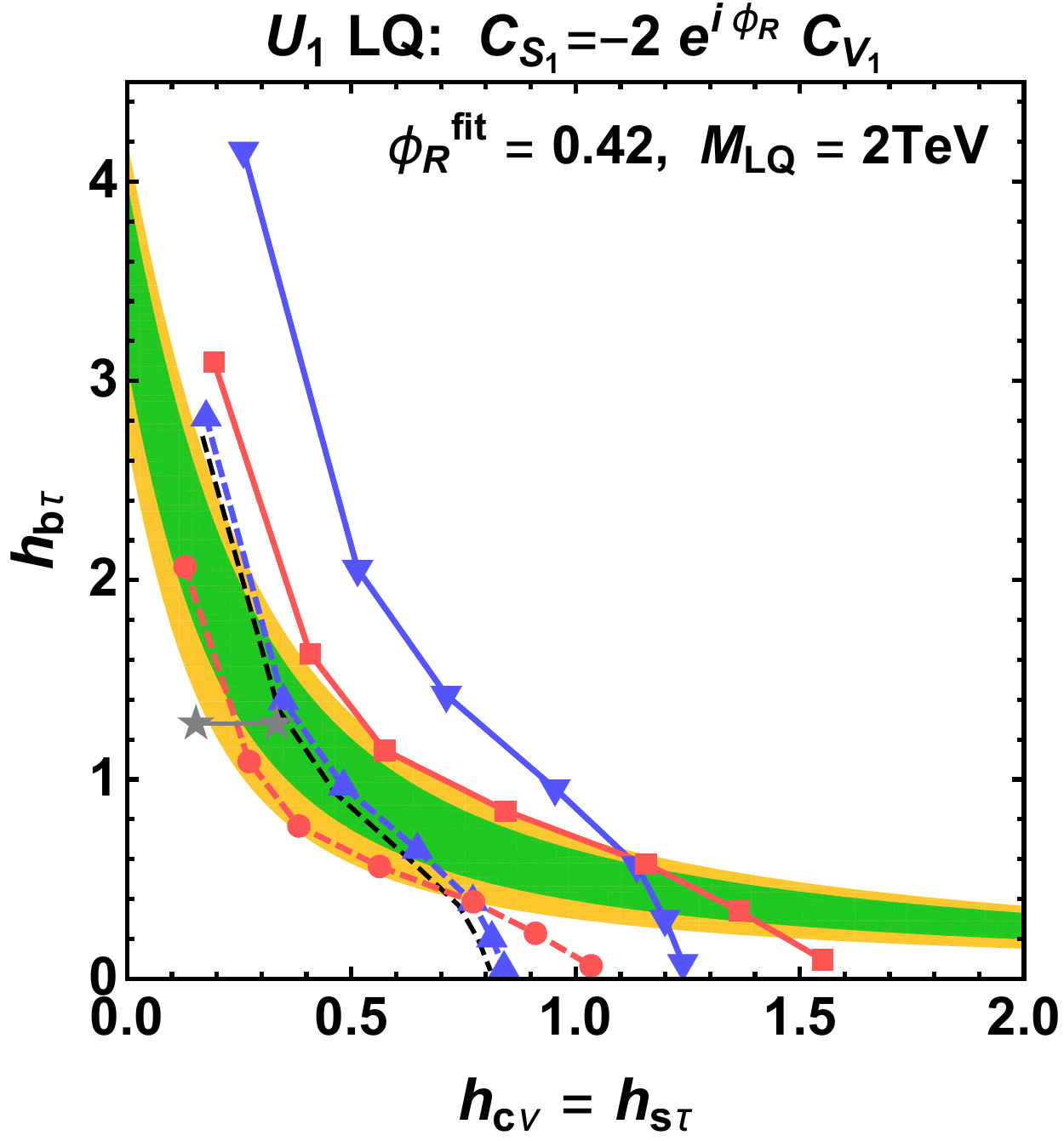}
\includegraphics[width=18em]{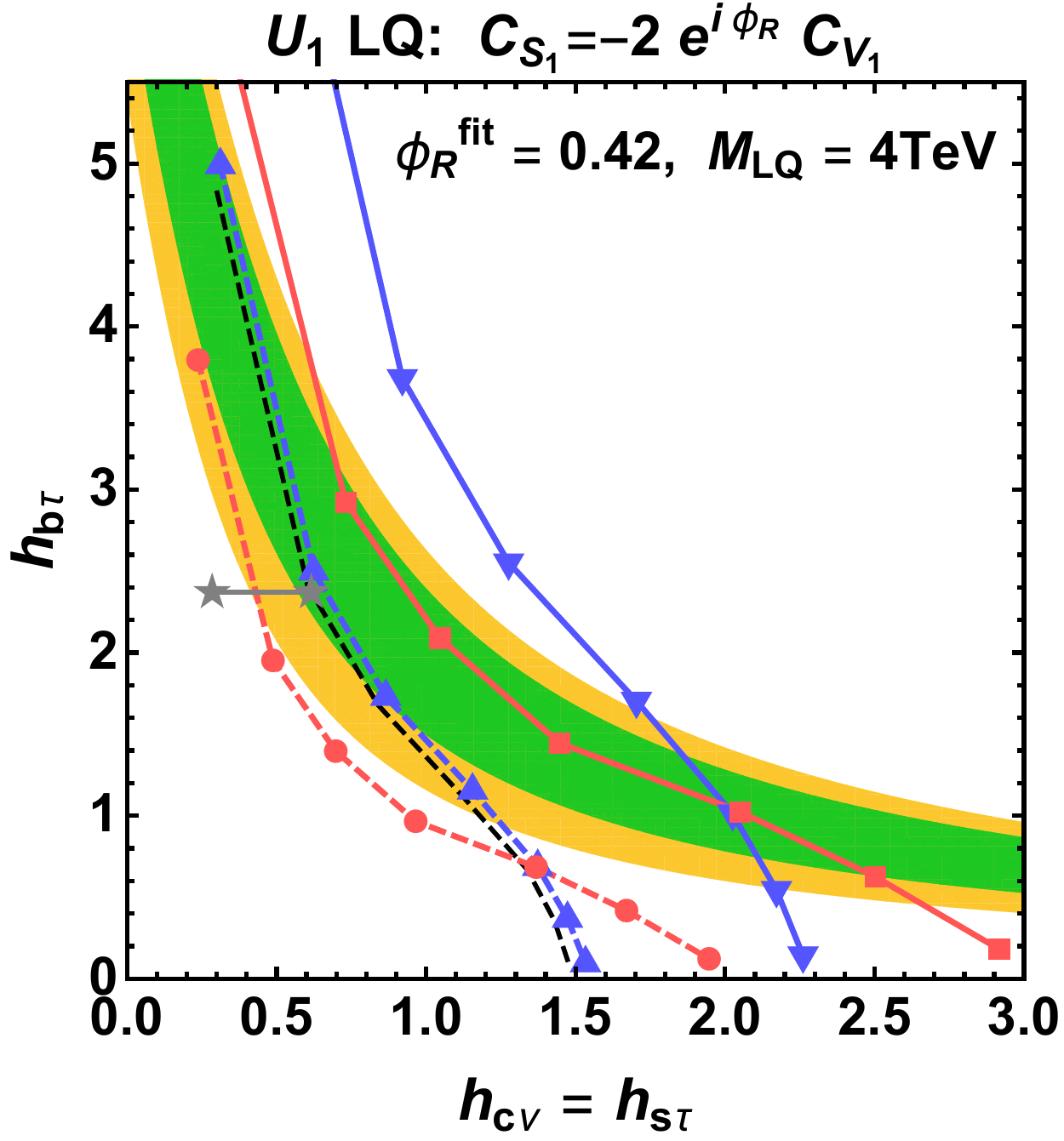}
\caption{
\label{Fig:U1contours}
Expected sensitivities to the $\text{U}_1$ LQ scenario as functions of the LQ couplings with the LQ mass of $2\,\text{TeV}$ (left) and $4\,\text{TeV}$ (right). 
The black dashed lines show the results with selecting $\tau^-\nu$.
The gray horizontal lines correspond to the current LHC bound recast from the ATLAS $pp \to \tau^+\tau^-$ search. 
See Fig.~\ref{Fig:WC_bound_CV1} for other color conventions. 
} 
\end{center}
\end{figure*}

Figure~\ref{Fig:U1contours} shows a dependence of the LHC sensitivity on the LQ couplings, $h_L^{23} (= h_{s\tau} = h_{c\nu})$ and $h_L^{33} (= h_{b\tau})$ for $M_\text{LQ} = 2$ and $4\TeV$. 
One can see that, for both scenarios, the result in the $\tau^\pm\nu + b$ search is sensitive to large $h_{b\tau}$ and small $h_{s\tau}$, namely small $R_{s/b}$, whereas that of $\tau^\pm\nu$ is sensitive to larger $R_{s/b}$. 
For the single-$\text{U}_1$ scenario, the region favored by $R_{D^{(*)}}$ can be tested at $\int\mathcal{L}\,dt=3000\ifb$ by the $\tau^\pm\nu (+ b)$ search only 
for $h_{s\tau} \gtrsim 0.8 (1.1)$ with $M_\text{LQ}=2\TeV$, and for $h_{s\tau} \gtrsim 1.5 (2.0)$ with $M_\text{LQ}=4\TeV$. 
As for $U(2)$-$\text{U}_1$, on the other hand, the $R_{D^{(*)}}$-favored region is fully probed by $\tau^\pm\nu + b$ at $\int\mathcal{L}\,dt=3000\ifb$ for both $M_\text{LQ}=2,\,4\TeV$.

In Ref.~\cite{Cornella:2021sby}, the $pp \to \tau^+\tau^-$ search by the ATLAS~\cite{ATLAS:2020zms} has been used to constrain the present two $\text{U}_1$ LQ scenarios.\kf\footnote{%
References~\cite{Aydemir:2019ynb,Bhaskar:2021pml,Angelescu:2021lln} also provide bounds on the LQ scenarios from the $pp \to \tau^+\tau^-$ search.} 
Their definition of the LQ couplings are related to ours as $h_{b\tau} \equiv g_U \beta_L^{b\tau}$ and $h_{s\tau} \equiv g_U \beta_L^{s\tau}$ by taking $\beta_L^{b\tau}=1$. 
Then, the upper limit on $(g_U, M_\text{LQ})$ has been recast from the ATLAS result at $\int\mathcal{L}\,dt=139\ifb$, where $\beta_L^{s\tau} \supset [0.10, 0.25]$ ($[0.12,0.26]$) is fitted from relevant flavor observables for single-$\text{U}_1$ ($U(2)$-$\text{U}_1$). 
Although it is unclear how to implement the sub-leading contributions from $bs/ss \to \tau^+\tau^-$ in their study, we naively translate their result into the $(h_{s\tau},h_{b\tau})$ plane as shown in the figure with gray lines. 
It is found that the $\tau^\pm\nu + b$ search is complementary to that of $\tau^+\tau^-$, though further discussions are needed on the analysis. 

\begin{figure*}[tb!]
\begin{center}
\subfigure[$M_{\text{U}_1\,\text{LQ}}=1.5\TeV$]{
\includegraphics[width=19em]{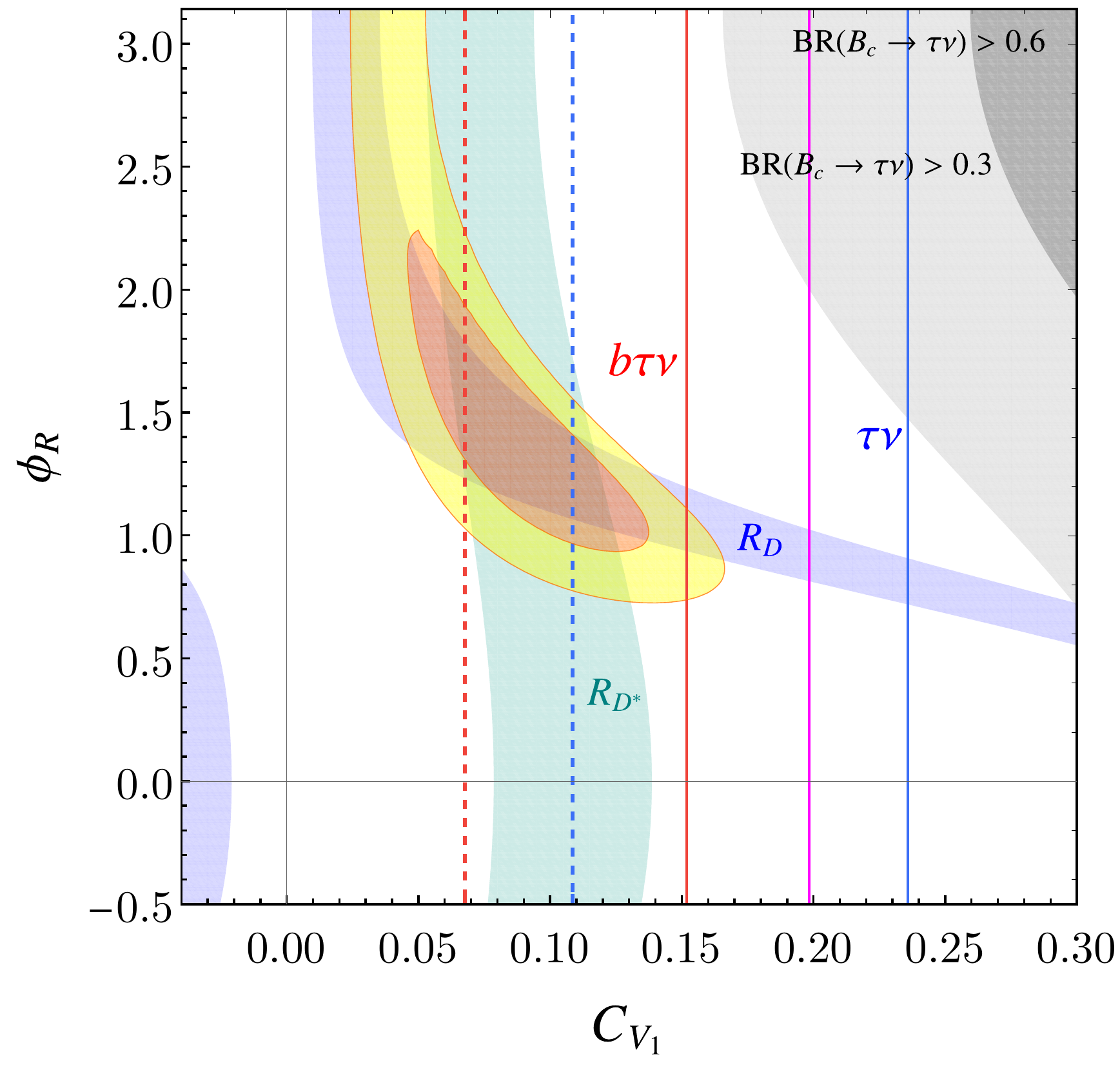}}~~
\subfigure[$M_{\text{U}_1\,\text{LQ}}=4.0\TeV$]{
\includegraphics[width=19em]{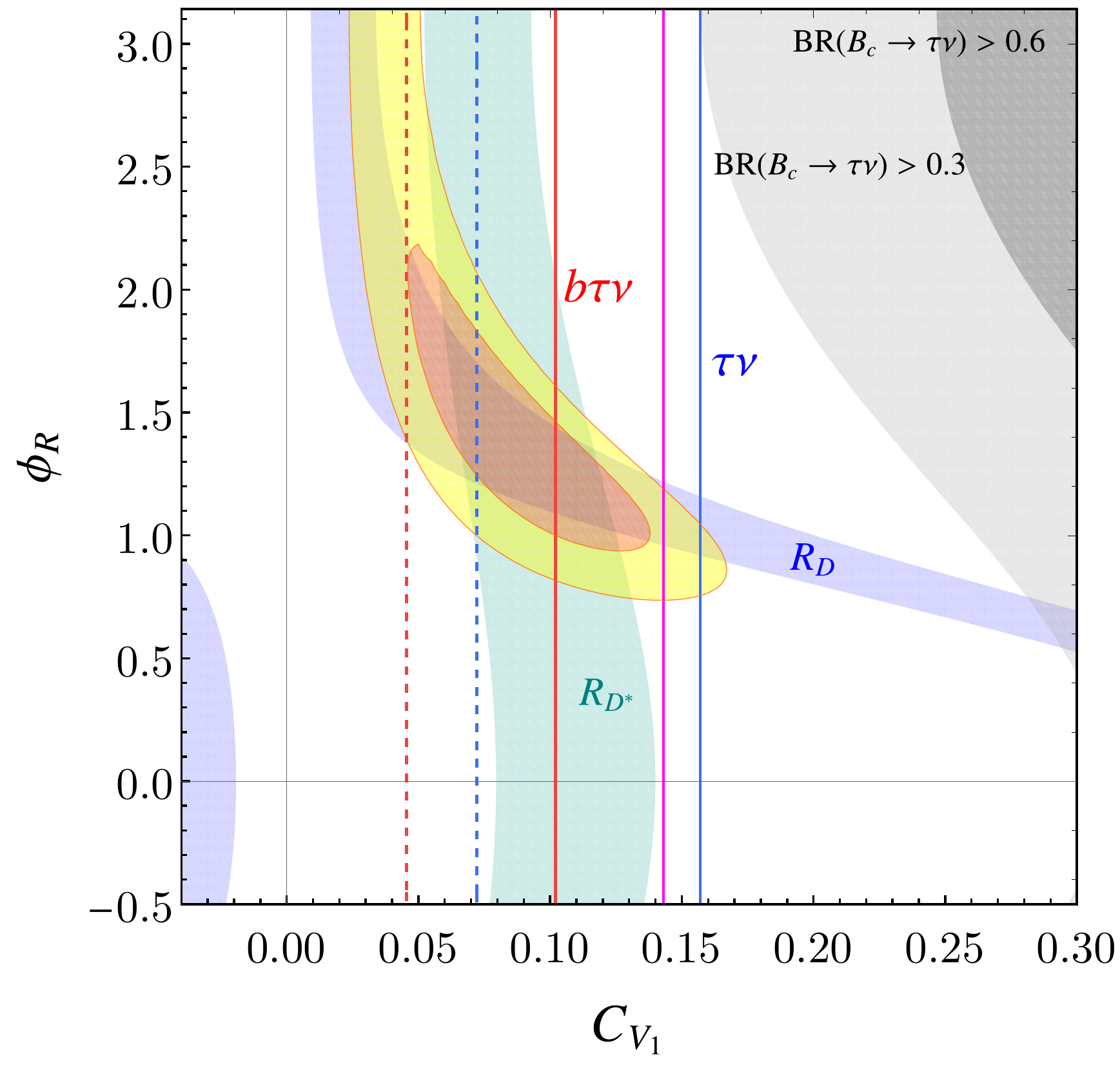}}
\caption{
\label{Fig:U1LQU2summary}
The $U(2)$-$\text{U}_1$ scenario with $M_{\text{LQ}}=1.5$ and $4.0\TeV$ 
on the $(C_{V_1},\,\phi_R)$ plane.
The color convention is the same as in Fig.~\ref{Fig:WC_bound_R2}. 
The magenta lines correspond to the current bound from the experimental data with $\int\mathcal{L}\,dt=36\ifb$
by assuming $M_{\text{LQ}}=2\TeV$ (left panel) and the EFT limit (right panel)~\cite{Iguro:2020keo}.  
} 
\end{center}
\end{figure*}

In Fig.~\ref{Fig:U1LQU2summary}, the $R_{D^{(*)}}$-favored region is compared with the LHC sensitivities and flavor constraints for $M_{\text{LQ}}=1.5\TeV$ (left) and $4\TeV$ (right) in the $U(2)$-$\text{U}_1$ scenario on the $(C_{V_1},\,\phi_R)$ plane. 
The region in the right-hand side of the vertical (solid/dashed) lines is probed or constrained by the LHC searches. 
The orange (yellow) region is favored by the measured $R_{D^{(*)}}$ at the $1\,\sigma$ ($2\,\sigma$) level. 
Note that the best fit is given at $\phi_R \simeq \pm 0.42\pi$, 
implying 
$C_{S_1}/C_{V_1}
\simeq -0.50 \mp 1.94 i
$.
Similar to the $\text{R}_2$ LQ model,
imaginary component is favored to be large.\kf\footnote{Phase degrees of freedom are not taken into account in the parameter fit in literature~\cite{Cornella:2019hct,Cornella:2021sby}. 
}
From this figure, we found that
the $R_{D^{(\ast)}}$-favored region can be fully (mostly) probed by
 $\tau^\pm \nu +b$ at $\int\mathcal{L}\,dt=3000\ifb$ for  $M_\text{LQ} > 4\TeV$ ($< 4\TeV$).

Similar to the $\text{S}_1$ LQ scenario, 
there is a strong bound from $\Delta M_s$, as briefly mentioned in Sec.~\ref{Sec:U1LQ}.
In realistic model setups of the $\text{U}_1$ LQ scenario,
vector-like leptons are introduced 
to realize a model flavor structure appropriately~\cite{Calibbi:2017qbu,DiLuzio:2018zxy,Cornella:2019hct,Fuentes-Martin:2020hvc,Iguro:2021kdw}.
Their mass scale is comparable to the LQ one
up to a factor depending on gauge and Yukawa couplings.
Then, the GIM-like mechanism does work 
and the box contributions to $\Delta M_s$ are suppressed. 
Since the vector-like lepton mass determines an energy scale of the breakdown of the GIM-like cancellation,
it cannot become too large, \ie, must be around the TeV scale at most.\kf\footnote{In such a case, three-body decay branching ratios (mediated by $\text{U}_1$ LQ) of the vector-like leptons
become dominant, and conventional searches~\cite{Kumar:2015tna,CMS:2019hsm} are not applicable directly.
The dedicated search for such a vector-like lepton at the LHC 
would be, therefore, important to probe a footprint of the 
LQ scenarios behind the $B$ anomalies.}
To summarize, model predictions of $\Delta M_s$ are quite model-dependent in the $\text{U}_1$ LQ scenarios, 
and dedicated studies are necessary.
In Figs.~\ref{Fig:WC_bound_CV1} and \ref{Fig:U1contours}, 
we do not draw the bounds from $\Delta M_s$, for simplicity.

\subsection{\boldmath Angular correlations}
\label{Sec:angular}

We investigate the angular distributions in the $\tau^\pm \nu +b$ searches, which would be helpful to distinguish new physics scenarios and to further suppress the background.
Requiring an additional $b$-jet not only amplifies sensitivity of new physics search,
but also provides us information of the angular observables.
Since the LQ models are characterized by the Lorentz structure of new physics interactions and 
the angular distributions of the final state are sensitive to them according to the analytic formulae of the scattering cross sections in Ref.~\cite{Marzocca:2020ueu}, 
they are useful to discriminate the models.

\begin{figure*}[t!]
\begin{center}
\includegraphics[width=17em]{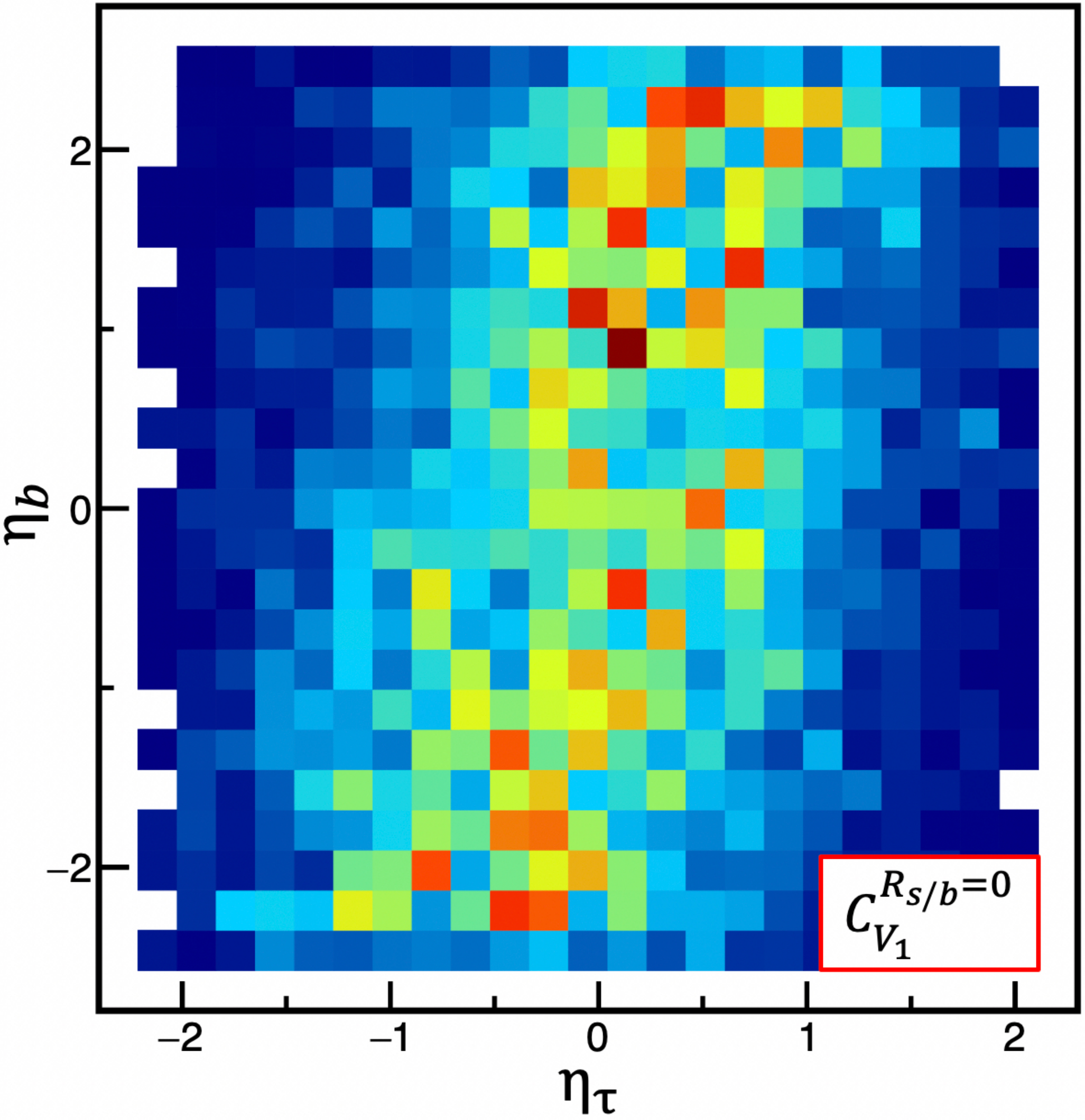}\qquad 
\includegraphics[width=17em]{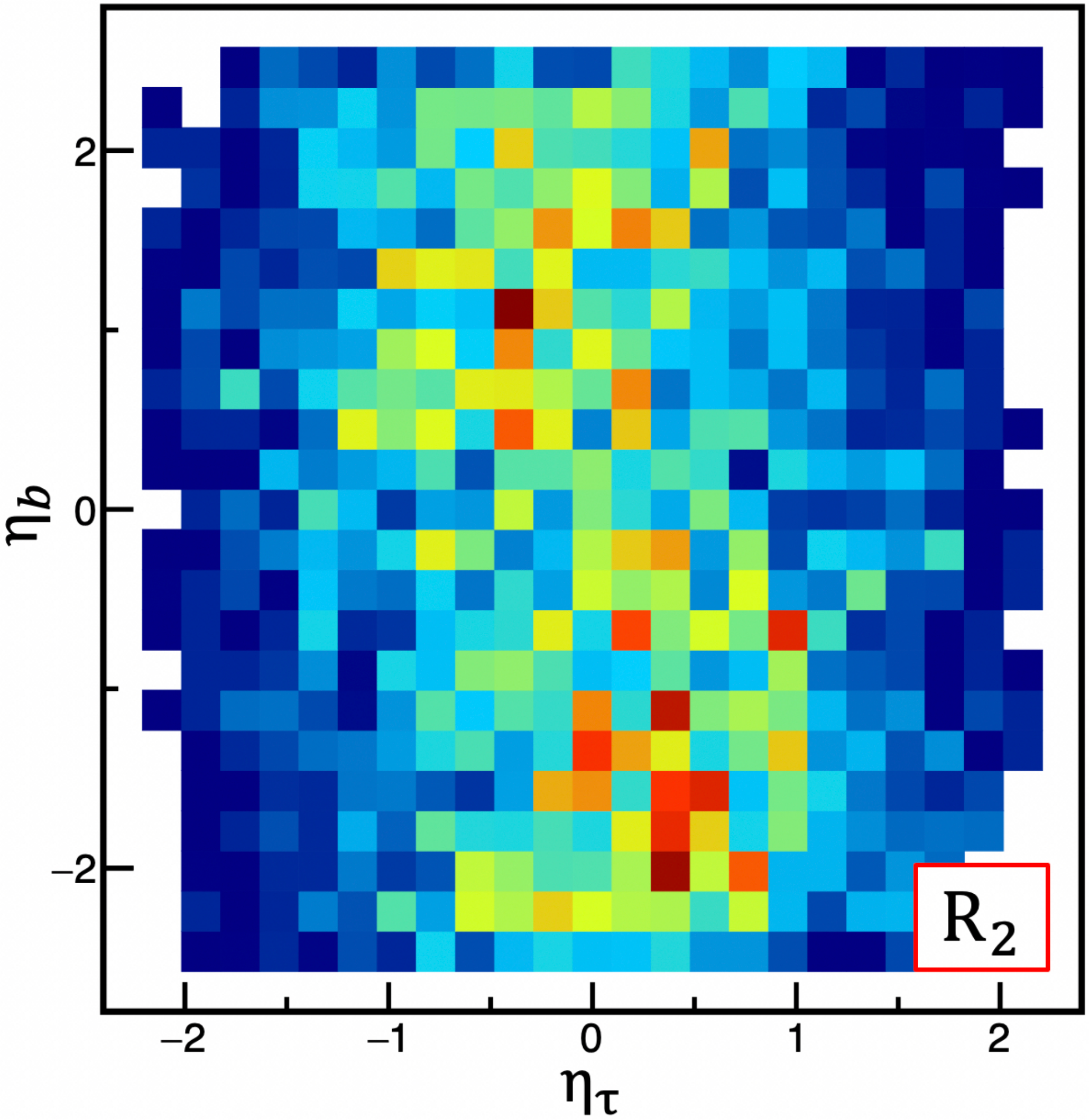} 
\caption{
\label{Fig:2d_eta}
Distribution of signal number density in  $\eta_\tau$--$\eta_b$ plane for
the single-$\text{U}_1$ ($C_{V_1}^{R_{s/b}=0}$) (left) and single-$\text{R}_2$ ($C_{S_2}$) (right).
In both scenarios, the LQ mass is set to be $1.5\TeV$.
The warmer/reddish (colder/bluish) colors represent larger (smaller) number of signal events.
} 
\end{center}
\end{figure*}

Let us first demonstrate a correlation between the pseudorapidities of the bottom quark ($\eta_b$) and of the $\tau$ lepton ($\eta_\tau$).
Figure~\ref{Fig:2d_eta} shows a pseudorapidity correlation 
in the single-$\text{U}_1 (C_{V_1})$ scenario with $R_{s/b}=0$ (left) 
and the single-$\text{R}_2 (C_{S_2})$ scenario (right) for $M_{\text{LQ}}=1.5\TeV$. 
Here, the LQ signal events passing {\bf cut b} with $m\tr\ge 700\GeV$ are exhibited.
There are larger (smaller) number of signal events left after the cut in the reddish (blueish) points. 
As observed in the single-$\text{U}_1 (C_{V_1})$ scenario (left panel),
their positive correlation indicates that $b$ and $\tau$ jets tend to be emitted in the same direction in the detector.
On the other hand, the single-$\text{R}_2 (C_{S_2})$ scenario (right panel) predicts a mild opposite correlation.
Since the signal distribution on the ($\eta_\tau, \eta_b$) plane depends on the NP scenarios, they could be distinguished by measuring the pseudorapidity correlation. 
It is noted that 
the same tendency is observed for $M_{\text{LQ}}=20\TeV$.
Moreover, 
it is found that 
distributions in a case of  $R_{s/b}=1$ are similar to those for $R_{s/b}=0$.
This is 
because a contribution from  the $\overline{s}$--$\tau$--${\text{U}}_1$ interaction, 
which comes from the $b$-jet mis-tagged from $c$-jet,
is negligible in the $\tau^\pm\nu+b$ events  for $R_{s/b}\lesssim 1$ (see Fig.~\ref{Fig:WC_bound_CV1}).

With having these observations, we propose the following quantities to probe the pseudorapidity correlation:
\beq
\eta^\prime_\tau=
\text{sgn}(\eta_b) \times \eta_\tau\,,~~~~~~
\eta^\prime_b=
\text{sgn}(\eta_\tau) \times \eta_b\,.
\label{Eq:etaprime}
\eeq
For instance,
the former is a modification of $\eta_{\tau}$ according to the $b$-jet direction.
If a distribution of the $b$-jet is isotropic, a peak of  $\eta^\prime_\tau$ distribution must be placed at zero.
However, events in the quadrants I and III of Fig.~\ref{Fig:2d_eta} provide
a positive $\eta^\prime_\tau$,
while the others yield a negative $\eta^\prime_\tau$.
As a result, when there is the positive (negative) pseudorapidity correlation, a peak of  $\eta^\prime_\tau$ distribution shifts 
in a positive (negative) direction.
Figure~\ref{Fig:etaprime} shows the signal event distribution against $\eta^\prime_\tau$ (left) and $\eta^\prime_b$ (right) in 
the scenarios of single-$\text{R}_2 (C_{S_2})$ (red), $\text{S}_1$ with $C_{V_1}=0$ (black), single-$\text{U}_1$ ($C_{V_1}$) (blue), and $U(2)$-$\text{U}_{1}$ (light green).
The event normalization for each histogram is taken to be arbitrary.
As expected from Fig.~\ref{Fig:2d_eta},
it is found that 
the single-$\text{R}_2  (C_{S_2})$ and single-$\text{U}_1$ ($C_{V_1}$) scenarios 
have a peak in a negative and positive $\eta^\prime_\tau$ (and also $\eta^\prime_b$) region, respectively.
In fact, the condition $m\tr\ge700\GeV$ leads to large amount of events around $\eta_{\tau} =0$, while  $\eta_b$ tends to be isotropic. As a result, it is found that
modification of the shape is clearer in the $\eta^\prime_{b}$ plane than $\eta^\prime_{\tau}$.
It is also observed that for the $\text{S}_1$ and $U(2)$-$\text{U}_{1}$ scenarios predict larger numbers of signal events in the $\eta^\prime_{\tau(b)}>0$ region compared to $\eta^\prime_{\tau(b)}<0$.

\begin{figure*}[t!]
\begin{center}
\includegraphics[width=17em]{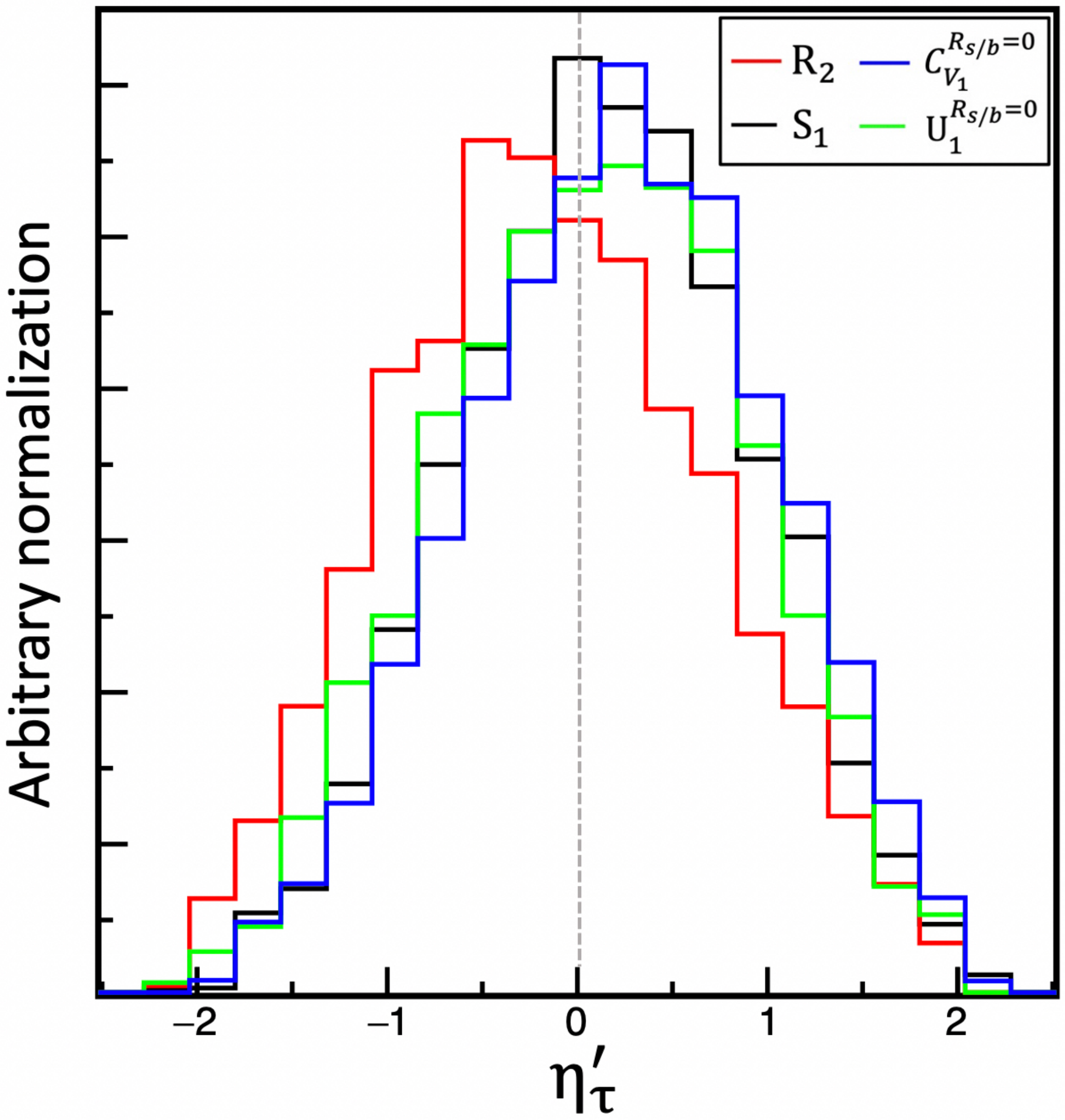} \qquad
\includegraphics[width=17em]{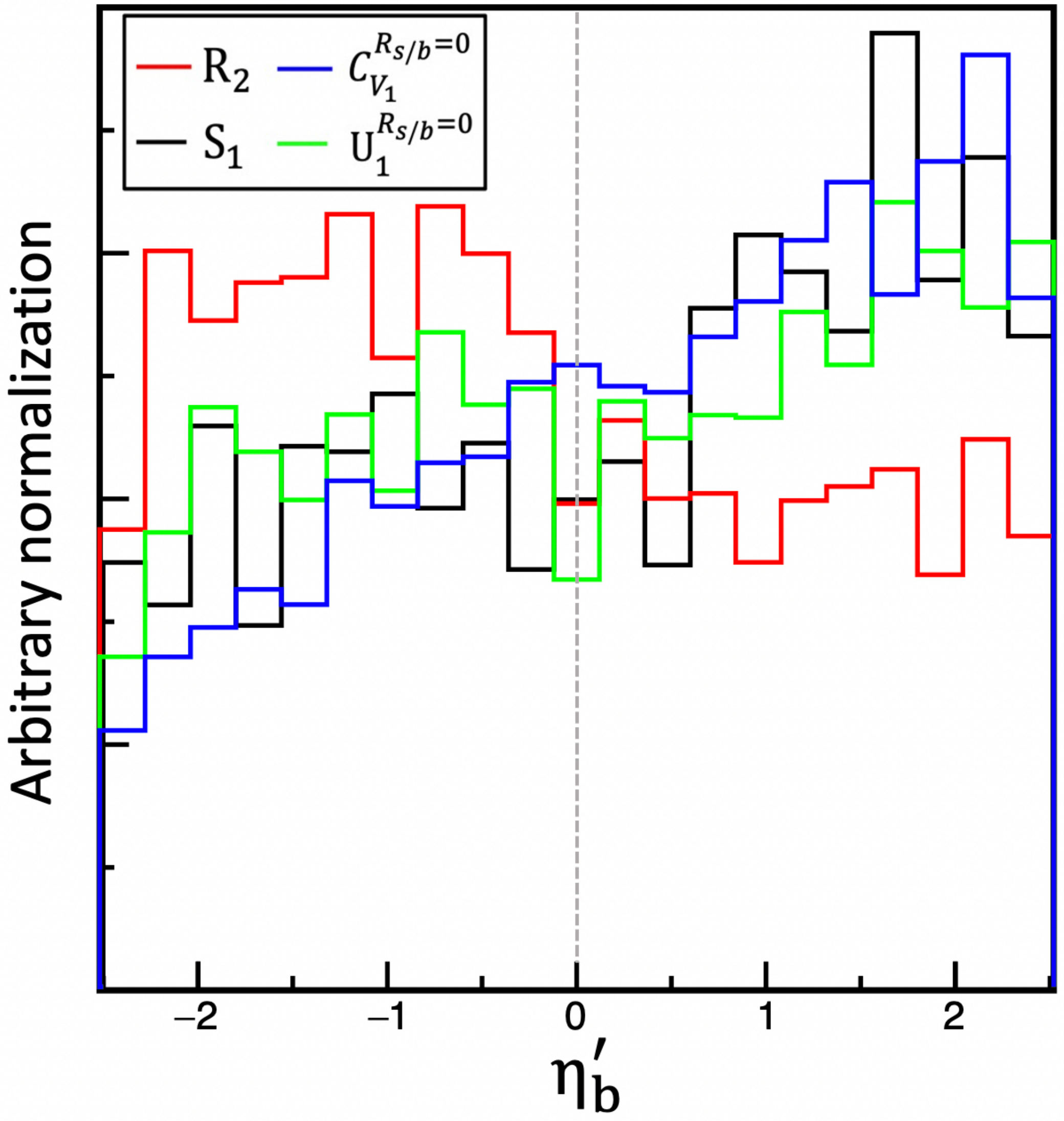}
\caption{
\label{Fig:etaprime}
Distribution of signal event numbers against the modified pseudorapidity $\eta^\prime_\tau$ (left) and $\eta^\prime_b$ (right) defined by Eq.~\eqref{Eq:etaprime}
in the scenarios of single-$\text{R}_2 (C_{S_2})$ (red), $\text{S}_1$ (black), single-$\text{U}_1$ ($C_{V_1}^{R_{s/b}=0}$) (blue), and $U(2)$-$\text{U}_{1}^{R_{s/b} =0}$ (light green).
Here, $M_{\text{LQ}}=1.5\TeV$ is taken, and
the normalization of each histogram is arbitrary.
} 
\end{center}
\end{figure*}

The azimuthal angle could also provide a tool to discriminate the UV models.
We study the relative azimuthal angles among $\tau$, $\nu$ (missing transverse momentum) and $b$ to distinguish the NP scenarios. 
We show $\Delta \phi(\vec{p}\tr^{\,\tau},\,\vec{p}\tr^{\,b})$ (left) and $\Delta \phi(\vec{p}\tr^{\,\rm miss},\,\vec{p}\tr^{\,b})$ (right) in Fig.~\ref{Fig:phi}. 
Note that $\Delta \phi(\vec{p}\tr^{\,\tau},\,\vec{p}\tr^{\,\rm miss})$ distribution is already used in the cut as the back-to-back configuration: $\Delta \phi(\vec{p}\tr^{\,\tau},\,\vec{p}\tr^{\,\rm miss}) \geq 2.4$.
The color convention is the same as the  Fig.~\ref{Fig:etaprime}. 
It is observed that the single-$\text{R}_2 (C_{S_2})$ scenario has more events in $0\le\Delta \phi(\vec{p}\tr^{\,\tau},\,\vec{p}\tr^{\,b})\le\pi/2$ than the rest of that region, 
while the $\text{S}_1$ scenario has more in $\pi/2\le\Delta \phi(\vec{p}\tr^{\,\tau},\,\vec{p}\tr^{\,b})\le \pi$. 
As for $\Delta \phi(\vec{p}\tr^{\,\rm miss},\,\vec{p}\tr^{\,b})$,
it is found that the single-$\text{R}_2 (C_{S_2})$ scenario has more events in $\pi/2\le\Delta \phi(\vec{p}\tr^{\,\rm miss},\,\vec{p}\tr^{\,b})\le \pi$. 

In conclusion, once signal events are measured, they would be helpful to discriminate the LQ models.

\begin{figure*}[t!]
\begin{center}
\includegraphics[width=17em]{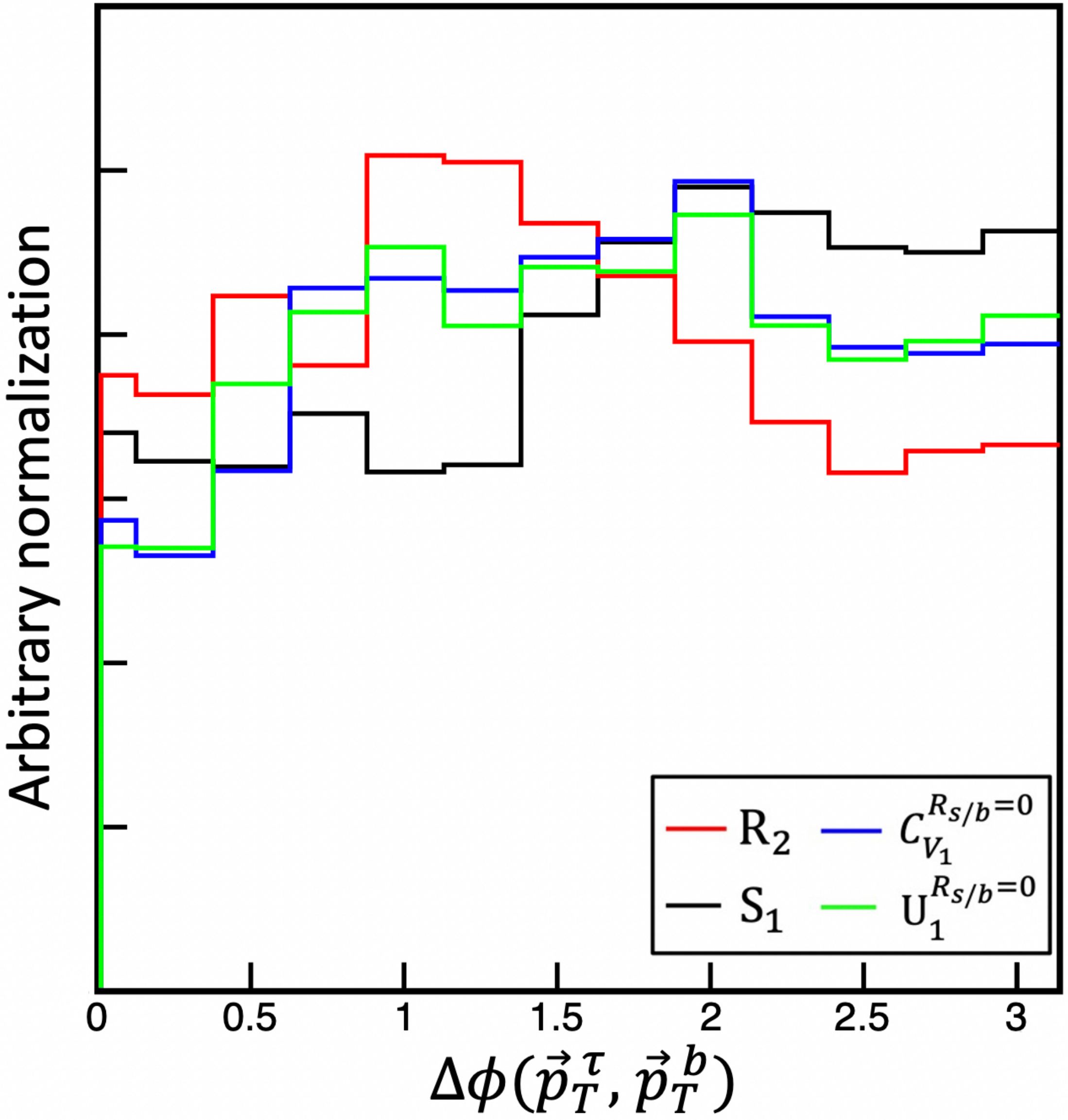} \qquad 
\includegraphics[width=17em]{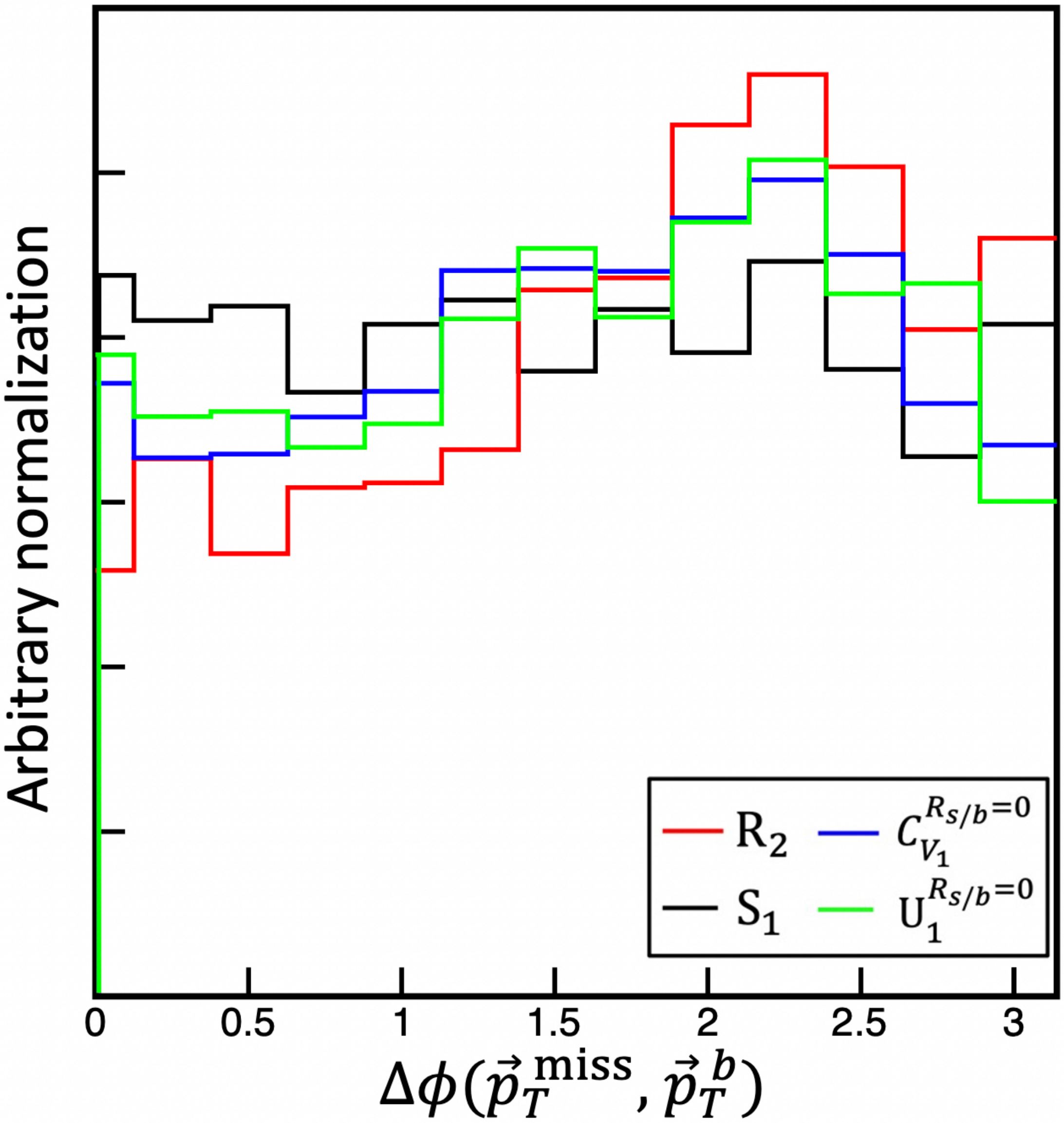} 
\caption{
\label{Fig:phi}
Distribution of signal event numbers against $\Delta \phi(\vec{p}\tr^{\,\tau},\,\vec{p}\tr^{\,b})$ (left) and $\Delta \phi(\vec{p}\tr^{\,\rm miss},\,\vec{p}\tr^{\,b})$ (right) in each LQ scenario. 
The color convention is the same as in Fig.~\ref{Fig:etaprime}. 
Here, $M_{\text{LQ}}=1.5\TeV$ is taken, and
the normalization of each histogram is arbitrary. 
} 
\end{center}
\end{figure*}

\section{\boldmath Conclusions and discussion}
\label{Sec:summary}

The $R_{D^{(*)}}$ anomaly is one of the hottest topics in the flavor physics from early in the last decade.
Since the relevant $b\to c\tau\nubar$ process is induced by exchanging the $W$ boson at the tree level within the SM and
the observed deviation is $+\mathcal{O}(10)\%$ at the amplitude level,
the NP scale is indicated at around $1$--$10\TeV$ to solve the $R_{D^{(*)}}$ anomaly.
Among the NP models, the LQ particles have attracted a lot of interests.
Such particles have been searched for by studying direct pair-production channels at the ATLAS and CMS, and the LQ mass has been constrained to be $>1-1.5\TeV$. 
Although the next run will start at the LHC, if the mass is larger than $\gtrsim 2\TeV$, it is unlikely to discover the LQ directly in the near future. 
Nonetheless, thanks to the crossing symmetry of scattering amplitudes,
the NP contributions to $b\to c\tau\nubar$ processes lead to $b\overline{c}\to \tau\nubar$ scattering at the LHC.
Such a process was studied to probe the NP contributions indirectly
even if the LQ is heavier than the LHC collision energy. 

To amplify experimental sensitivities of such a non-resonant search,
we examined 
the impact of requiring an additional $b$-jet in the final state, \eg, $g\overline{c}\to \overline{b}\tau\nubar$.
We evaluated the current and future LHC sensitivities
based on both the EFT framework and the viable models of scalar- and vector-LQs;
S$_1$, R$_2$, and U$_1$ with/without the $U(2)$ flavor symmetry.
It was observed that the additional $b$-jet requirement and $\tau^-$ selection
can improve the LHC sensitivity on the NP searches by $\approx 40\%$ and $\approx 10\%$ versus those in the $\tau^\pm\nu$ search, respectively.
Furthermore,
the LQ mass dependence of the sensitivities is explicitly shown
in the LQ mass range of $M_{\rm LQ}=\mathcal{O}(1)\TeV$ for the $\tau^\pm\nu+b$ search as well as the $\tau^\pm\nu$ case. 
In particular, it was found that the sensitivity from the $\tau^\pm\nu+b$ search is better than that from $\tau^\pm\nu$ in the whole mass region for $R_{s/b} \lesssim 1$.

Based on those findings,
the LHC sensitivities are compared with the parameter regions 
that can accommodate the $R_{D^{(*)}}$ anomaly in several single LQ scenarios.
There are three types of viable leptoquark models responsible for the anomaly;
the $\text{R}_2$, $\text{S}_1$, and $\text{U}_1$ LQ scenarios.
We observed the following results:
\begin{itemize}
\item 
For the single-$\text{R}_2 (C_{S_2})$ LQ scenario, 
it is found that the current LHC data of $\int\mathcal{L}\,dt=139\ifb$ are enough to probe the $\text{R}_2$ LQ,
although the LQ mass dependence is crucial to claim whether 
the LQ is fully detectable.
For instance,
the $\tau^\pm\nu+b$ search with $\int\mathcal{L}\,dt=139\ifb$ fully  covers the single-$\text{R}_2 (C_{S_2})$ scenario 
with $M_\text{LQ}>2.5\TeV$  responsible for the $R_{D^{(*)}}$ anomaly, while the region of $1.5\TeV<M_\text{LQ}<2.5\TeV$ can be probed partially.
\item
For the $\text{S}_1$ LQ scenario,
the parameter region is already severely constrained from $B \to K^\ast \nu \nubar$ and $\Delta M_s$ measurements, and 
the current LHC data can not test the allowed region.
Larger luminosity such as the HL-LHC with requiring an additional $b$-jet is needed to probe these parameter regions.
\item
For the $\text{U}_1$ LQ scenarios, 
there are several parameter regions that can accommodate the $R_{D^{(\ast)}}$ anomaly 
depending on flavor structures of the LQ couplings and the LQ mass. 
It is found that the HL-LHC can probe the parameter regions
in both the single-U$_1$ ($C_{V_1}$) and $U(2)$-U$_1$ scenarios
by requiring an additional $b$-jet.
\item 
As mentioned in Sec.~\ref{Sec:selection_cut}, the ATLAS collaboration observed smaller number of events than the expected one in the $\tau^\pm\nu$ category
at the data of $\int\mathcal{L}\,dt=139\ifb$,
and provided stronger constraint than the expectation~\cite{ATLAS:2021bjk}.
Therefore, an experimental analysis with requiring an additional $b$-jet is of great importance.
Particularly, the single-$R_2$ LQ scenario could be probed immediately by using the data of $\int\mathcal{L}\,dt=139\ifb$.
\end{itemize}

The angular correlations among $\tau$-, $b$-jets
and missing transverse momentum 
were also studied.
It was shown that
the correlation between $\tau$- and $b$-jets 
in the pseudorapidity plane could be useful to discriminate the LQ scenarios.
Besides, it was found that the azimuthal angle distributions would also be helpful.
However, further studies especially with experimental information are needed for improving the analysis.
In this paper, models with light right-handed (sterile) neutrinos are not discussed.
In those scenarios, WCs are likely to be large to explain the $R_{D^{(*)}}$ anomaly, because there is no interference with the SM amplitude.
For instance,
the effective Hamiltonian analogous to that of $C_{V_1}$ is given as 
\begin{align}
 {\mathcal H}_{\rm{eff}}=   
 2 \sqrt 2 G_FV_{cb} 
  C^\prime_{V_1} (\overline{c} \gamma^\mu P_Lb)(\overline{\tau} \gamma_\mu P_R \nu_{\tau}) + \,\text{h.c.}\,,
  \label{Eq:effHR}
\end{align} 
and $C^\prime_{V_1}\approx 0.4\pm0.05$ can explain the $R_{D^{(*)}}$  anomaly.
Since the LHC searches are expected to be insensitive to the neutrino chirality, 
we can apply the bound/sensitivity obtained for $C_{V_1}$
to the right-handed neutrino scenario.
It was shown that the current data of $\int\mathcal{L}\,dt=139\ifb$ are enough to probe $C^\prime_{V_1} \gtrsim 0.2$ ($0.3$) in the EFT limit (for $M_{\rm LQ}=1.5\TeV$) in the $\tau^\pm \nu_R +b$ search, see Table~\ref{table:cutbsummary}.
Thus, the parameter region of $C^\prime_{V_1}$ favored by the $R_{D^{(*)}}$ anomaly can be tested immediately.  
Further improvements could be possible if larger amount of data is accumulated.
In this work,
we studied events in the region of $m\tr > 1\TeV$ to derive the NP sensitivities.
With larger data, one could push the $m\tr$ condition to a larger side, \eg, $m\tr > 2\TeV$. Then, the sensitivity would be improved by further suppression of the SM backgrounds.
Moreover, 
further suppression of $\epsilon_{j\to b}$ is expected to improve
the sensitivity, 
since a large amount of the SM background events coming from fake $b$-jets can be reduced.

In the aspect of the flavor physics, $q^2$ distribution, $D^*$ polarization and $\tau$ polarization in $\overline {B}\to D^{(*)}\tau{\nubar}$ as well as the other $b \to c \tau \overline{\nu}$ processes, \eg,
$B_c\to J/\psi \tau\nubar$, 
$\Lambda_b\to \Lambda_c\tau\nubar$, 
and $B_s\to D_s\tau\nubar$ will be important to cross check 
the NP scenarios in the next decade.
It would be exciting to see how the data evolves once we are moving to the higher precision.
Since the LHC and Belle II experiments enjoy the high statistic era in this and next decades, the interplay between the flavor physics and collider physics would become more significant.


\section*{\boldmath Acknowledgement}
\label{Sec:acknowledgement}
We thank Tomomi Kawaguchi and Yuta Takahashi 
for valuable comments and discussion on the flavor tagging at the LHC. 
We also thank Sho Iwamoto for useful comments on {\sc\small MadGraph}5\_a{\sc\small MC}@{\sc\small NLO}.
We appreciate Kazuhiro Tobe and Yuki Otsu for fruitful discussion on the relation between $\Delta F=2$ observables and LQs.
This work is supported by the Grant-in-Aid 
for
Scientific Research on Innovative Areas (No.\,21H00086 [ME] and No.\,19H04613~[MT]),
Scientific Research\,B (No.\,21H01086~[ME]), 
Scientific Research\,C (No.\,18K03611~[MT]),
Early-Career Scientists (No.\,16K17681 [ME] and No.\,19K14706~[TK]),
and
JSPS Fellows (No.\,19J10980~[SI])
from the Ministry of Education, Culture, Sports, Science, and Technology (MEXT), Japan.
The work of S.\,I. is supported by the World Premier International Research Center Initiative (WPI), MEXT, Japan (Kavli IPMU).
The work of S.\,I., T.\,K., and M.\,T. is also supported by the JSPS Core-to-Core Program (Grant No.\,JPJSCCA20200002).
R.\,W. is partially supported by the INFN grant ‘FLAVOR’ and the PRIN 2017L5W2PT. 
S.\,I is supported by the Deutsche Forschungsgemeinschaft (DFG, German Research Foundation) under grant 396021762–TRR\,257.
S.\,I thanks the Yukawa Institute for Theoretical Physics at Kyoto University, where this work was initiated during the YITP-W-19-05 on ``Progress in Particle Physics 2019''.
S.\,I appreciates Yuji Omura and C.\,-P.\,Yuan for the stimulating discussion at the initial stage.

\appendix
\section{Simulation details}
\label{Sec:App_cutflow}
Here, we present some details of our MC setup and the signal and background cut flows, whose final results are summarized in  Sec.~\ref{Sec:selection_cut}.

Event generations and hadronizations are done as described in Sec.~\ref{Sec:event_generation} with the following details; 
as for a jet matching scale, {\tt qCut = 45\,GeV} is used to obtain the merged cross section; 
regarding the SM background, a model of ``\texttt{sm-no\_b\_mass}'' in {\sc\small MadGraph}5 is used, which sets the bottom quark mass to be $0$ while keeping the Yukawa coupling non-vanishing. For the NP signal, the bottom mass is set to be $0$.

In the MC simulation, the following pre-cuts are imposed at the {\tt run\_card} level to reduce the computation cost:
\begin{align}
p^\tau\tr\ge200\GeV\,,
\quad E\tr^{{\rm{miss}}} \ge 200\GeV\,,
\quad |\eta_\tau| \le 5\,,
\quad {\tt maxjetflavor}=5\,,
\label{Eq:pre_cut}
\end{align}
and  {\tt JetMatching:nJetMax=-1} (default number) is set.
The number of generated background events are 
5M for $Wjj$, 
40M for $Zjj$, 
5M for $t\bar{t}$ with both $W$ bosons 
decaying to $\tau$,
5M for $t\bar{t}$ with one of the $W$ bosons decaying to $\tau$,
5M for $t+j$,
6M for $tW(1)$,
1M for $tW(2)$,
0.5M for $tZ(1)$,
5M for $Z,\gamma$ DY,
and 3M for each $WW$, $ZZ(\gamma)$, and $WZ(\gamma)$ categories. 
For the signal simulations,
100K events are generated 
in each model point of the NP signals.
We have checked that the $m\tr$ distributions after the {\bf cut a} and {\bf cut b} are well smooth for each SM background category and the LQ signal.
For the analysis of angular distributions discussed in Sec.~\ref{Sec:angular}, we increased the generated event numbers by factors to suppress the MC-statistical uncertainty appropriately.

Tables~\ref{table:AP_BG_cutaflow} and \ref{table:AP_BG_cutbflow} are detailed cut flows of the SM background for the {\bf cut a} and {\bf cut b}, respectively. 
As a comparison with literature, 
we show the results of Refs.~\cite{Sirunyan:2018lbg} and \cite{Marzocca:2020ueu} for {\bf cut a}
and Ref.~\cite{Marzocca:2020ueu} for {\bf cut b}.
It should be noted that some details in the analysis procedures are different from ours; particularly,  
the $b$-tagging efficiencies (different from Ref.~\cite{Marzocca:2020ueu} for {\bf cut b}), the jet cone size (different from Ref.~\cite{Sirunyan:2018lbg} for {\bf cut a} and {\bf cut b}), hadronic $\tau$ tagging method (not explained in Ref.~\cite{Marzocca:2020ueu} explicitly, and different from Ref.~\cite{Sirunyan:2018lbg}, for {\bf cut a} and {\bf cut b}), and so on.
As for {\bf cut a} the differences are expected to hardly affect the results, and we found that our result is consistent with those in Refs.~\cite{Sirunyan:2018lbg} and \cite{Marzocca:2020ueu}. 

On the other hand, out result for {\bf cut b} is well suppressed versus those in Ref.~\cite{Marzocca:2020ueu}.
This is mainly because we used a working point with smaller $b$ mis-tagging rates. 

Tables~\ref{table:AP_CV1_cutaflow}, \ref{table:AP_U2_cutaflow}, \ref{table:AP_CV1_cutbflow}, and \ref{table:AP_U2_cutbflow} 
are detailed cut flows of the LQ signal for {\bf cut a} and {\bf cut b}.
See the caption of Table~\ref{table:CV1_cuta} for the details.

\begin{table}[tb!]
\centering
\newcommand{\bhline}[1]{\noalign{\hrule height #1}}
\renewcommand{\arraystretch}{1.5}
   \scalebox{1.0}{
  \begin{tabular}{c| ccc ccc} 
  \bhline{1 pt}
  \rowcolor{white}
  BG ($\bf{cut~a}$) &$Wjj$ & $Zjj~(Z\to\nu\overline\nu)$ & $t\overline{t}$ & $Z,\gamma$ DY & $VV$ & single $t$   \\  \hline  
  $\tau$ cut (a-1) & 4613.3 & 562.0 & 241.8 & 1236.4 & 72.2 & 52.4\\ 
  lepton cut (a-2) & 4609.1 & 561.9 & 230.3 & 744.1 & 65.5 & 50.1\\ 
  MET cut (a-3)& 2933.0 & 471.9 & 190.8 & 83.9 & 42.8 & 42.6\\ 
  back-to-back (a-4)& 777.0 & 184.6 &  9.85 & 52.5 & 12.1 & 1.09\\ \hline
 $0.7$ $< m\tr < 1\TeV$  & 70.5 & 20.1 & 0.34 & 3.03 & 1.30 & 0.02\\ 
 $1\TeV$ $ < m\tr$ & 16.9 & 5.1 & 0.06 & 0.56 & 0.32 & 0.02\\ \hline \hline
$1\TeV$ $ < m\tr$~\cite{Sirunyan:2018lbg} & $22\pm 6.2$ & $0.9\pm0.5$ & $< 0.1$ & $<  0.1$ & $0.7\pm0.1$ & $< 0.1$\\ 
$1\TeV < m\tr$~\cite{Marzocca:2020ueu} & $18$ & $5.2$ & $0.44$ & $0.0025$ & $1.7$ & $0.1$\\ 
\bhline{1 pt}
   \end{tabular}
   }
    \caption{\label{table:AP_BG_cutaflow} 
    Cut flows of the SM background events in the $\bf{cut~a}$ category (the $\tau^\pm\nu$ search).
    The expected number of events corresponding to $\int\mathcal{L}\,dt=35.9\ifb$ at $\sqrt{s}=13\TeV$ are shown.
    The last two rows show the results by
    Refs.~\cite{Sirunyan:2018lbg} and \cite{Marzocca:2020ueu}.
    See, the main text for the detail.\vspace{0.7cm}
}
\end{table}
\begin{table}[tb!]
\centering
\newcommand{\bhline}[1]{\noalign{\hrule height #1}}
\renewcommand{\arraystretch}{1.5}
   \scalebox{.98}{
  \begin{tabular}{c| ccc ccc} 
  \bhline{1 pt}
  \rowcolor{white}
  BG ($\bf{cut~b}$) &$Wjj$ & $Zjj~(Z\to\nu\overline\nu)$ & $t\overline{t}$ & $Z,\gamma$ DY & $VV$ & single $t$   \\  \hline  
 number of jets & 6693.4 & 235099 & 346.7 & 1813.2 & 125.8 & 151.8\\ 
  number of $\tau$ & 3173.5 & 5617.1 & 73.9 & 894.9 & 59.7 & 34.0\\ 
 number of $b$ & 90.6 & 305.5 & 35.9 & 163.9 & 5.28 & 18.8\\ 
 isolated lepton & 90.5 & 305.5 & 29.7 & 10.4 & 1.38 & 17.0\\ 
 $\tau$ kinematics & 78.8 & 20.8 & 23.6 & 9.19 & 1.13 & 14.0\\ 
 MET cut & 71.2 & 4.62 & 20.9& 2.52 & 0.98 & 12.7\\
 back-to-back & 7.84& 3.61 & 1.67 & 0.57 & 0.18 & 0.54\\ \hline
   $0.7$ $< m\tr < 1\TeV$  & 0.58 & 0.37 & 0.056 & 0.28 & 0.018 & 0.029\\ 
  $1\TeV$ $ < m\tr$ & 0.16 & 0.06 & 0.01 & 0.007 & 0.005 & 0.005\\ \hline \hline
$1\TeV$ $ < m\tr$~\cite{Marzocca:2020ueu} & 0.18(5) & 0.21(12) & 0.29(3) & 4.2(4)$\times10^{-5}$ & 0.35(5) & 0.067(7)\\ 
\bhline{1 pt}
   \end{tabular}
   }
    \caption{\label{table:AP_BG_cutbflow} 
    Same as Table~\ref{table:AP_BG_cutaflow} but for $\bf{cut~b}$ (the $\tau^\pm\nu+b$ search).
    The last row shows the results by Ref.~\cite{Marzocca:2020ueu}.
    Note that their $b$-tagging efficiencies are different from ours (see, the footnote \ref{footnote:b-tagging}).
}
\end{table}
\begin{table}[tb!]
\centering
\newcommand{\bhline}[1]{\noalign{\hrule height #1}}
\renewcommand{\arraystretch}{1.5}
   \scalebox{1.04}{
  \begin{tabular}{c| cccc|c } 
  \bhline{1 pt}
  \rowcolor{white}
   &$C_{V_1, 1.5\TeV}$ & $C_{V_1, \text{EFT}}$ &$C_{V_1, 1.5\TeV}^{R_{s/b}=1}$ & $C_{V_1, \text{EFT}}^{R_{s/b}=1}$& BG \\  \hline 
   $\tau$ cut (a-1) & 889 & 1198 & 2182 & 2876 & 6778 \\
  lepton cut (a-2) & 888 & 1198 & 2180 & 2874 & 6261\\ 
  MET cut (a-3) & 539 & 783 & 1319 & 1861 & 3765\\ 
  back-to-back (a-4) & 452 & 577 & 1015 & 1483 & 1030\\ \hline 
 $0.7$ $< m\tr < 1\TeV$   & 90.0 & 139.4 & 225.9 & 351.4 & 95.3\\ 
 $1\TeV$ $ < m\tr$& 54.4 & 123.6 & 146.9 &345.8 & 23.0\\ 
\bhline{1 pt}
   \end{tabular}
   }
    \caption{\label{table:AP_CV1_cutaflow} 
    Cut flows of the signal event numbers in the $\bf{cut~a}$ category for several setups of the $C_{V_1}$ scenario
    with $C_{V_1}=1$ and $\int\mathcal{L}\,dt=35.9\ifb$.
   See the caption of Table~\ref{table:CV1_cuta} for the details.
}
\end{table}
\begin{table}[tb!]
\centering
\newcommand{\bhline}[1]{\noalign{\hrule height #1}}
\renewcommand{\arraystretch}{1.5}
   \scalebox{1.05}{
  \begin{tabular}{c| cccc|c } 
  \bhline{1 pt}
  \rowcolor{white}
   &$\text{U}_{1, 1.5\TeV}^{R_{s/b} =0}$ & $\text{U}_{1, \text{EFT}}^{R_{s/b} =0}$ & $\text{U}_{1, 1.5\TeV}^{R_{s/b} =1}$ & $\text{U}_{1, \text{EFT}}^{R_{s/b} =1}$& BG   \\  \hline  
   $\tau$ cut (a-1) &2875 & 4189 & 4106 & 6003 & 6778 \\
  lepton cut (a-2) & 2871 & 4184 & 4103 & 5999 & 6261\\ 
  MET cut (a-3) & 1863 & 2934 & 2672 &4123 & 3765 \\
  back-to-back (a-4) & 1530 & 2423 & 2108& 3409 & 1030\\ \hline 
 $0.7$ $< m\tr < 1\TeV$  & 361 & 582 & 502 & 809 & 95.3\\ 
 $1\TeV$ $ < m\tr$ & 204 & 571 & 279 & 799 & 23.0 \\
\bhline{1 pt}
   \end{tabular}
   }
    \caption{\label{table:AP_U2_cutaflow} 
    Same as Table~\ref{table:AP_CV1_cutaflow} but for the $\text{U}_1$ LQ scenario with the $U(2)$ flavor symmetry, where $C_{V_1}=1$ and $\int\mathcal{L}\,dt=35.9\ifb$.
}
\end{table}
\begin{table}[t!]
\centering
\newcommand{\bhline}[1]{\noalign{\hrule height #1}}
\renewcommand{\arraystretch}{1.5}
   \scalebox{1.08}{
  \begin{tabular}{c| cccc|c } 
  \bhline{1 pt}
  \rowcolor{white}
   &$C_{V_1, 1.5\TeV}$ & $C_{V_1, \text{EFT}}$ &$C_{V_1, 1.5\TeV}^{R_{s/b}=1}$ & $C_{V_1, \text{EFT}}^{R_{s/b}=1}$& BG  \\  \hline   
  number of jets & 1529 & 1873 & 3290 & 4283 & 244230 \\
 number of $\tau$ & 693 & 907 & 1576 & 2114 & 9853\\ 
 number of $b$ & 144 & 182 & 178 & 237 & 620.0\\ 
 isolated lepton & 142 & 180 & 177 & 234 & 454.5\\ 
 $\tau$ kinematics & 128 & 165 & 156 & 210 & 147.5\\ 
 MET cut & 99.5 & 131 & 125 & 169 & 112.9\\ 
 back-to-back & 48.5 & 84.3 & 76.0 & 111 & 14.4\\ \hline
   $0.7$ $< m\tr < 1\TeV$  &11.6 & 16.6 & 13.9 & 21.7 & 1.33\\
 $1\TeV$ $ < m\tr$ & 6.51 &14.6 & 9.39 & 21.6 & 0.25\\ 
\bhline{1 pt}
   \end{tabular}
   }
    \caption{\label{table:AP_CV1_cutbflow} 
    Same as Table~\ref{table:AP_CV1_cutaflow} but in the $\bf{cut~b}$ category.
    See the caption of Table~\ref{table:CV1_cutb} for the details.
}
\end{table}
\begin{table}[tb!]
\centering
\newcommand{\bhline}[1]{\noalign{\hrule height #1}}
\renewcommand{\arraystretch}{1.5}
   \scalebox{1.07}{
  \begin{tabular}{c| cccc|c } 
  \bhline{1 pt}
  \rowcolor{white}
   &$\text{U}_{1, 1.5\TeV}^{R_{s/b} =0}$ & $\text{U}_{1, \text{EFT}}^{R_{s/b} =0}$ & $\text{U}_{1, 1.5\TeV}^{R_{s/b} =1}$ & $\text{U}_{1, \text{EFT}}^{R_{s/b} =1}$& BG   \\  \hline  
 number of jets & 4245 & 6085 & 5966 & 8376 & 244230\\ 
 number of $\tau$ & 2024 & 2941 & 2898 & 4168 & 9853\\ 
 number of $b$ & 460 & 692 & 535 & 754 & 620.0\\ 
 isolated lepton & 454 & 685 & 485 & 747 & 454.5\\ 
 $\tau$ kinematics & 424 & 637 &451 & 692 & 147.5\\ 
 MET cut &350 & 540 & 371 & 590 & 112.9\\ 
 back-to-back & 258 & 402 & 263 & 443 & 14.4\\ \hline
   $0.7$ $< m\tr < 1\TeV$  & 53.9 & 86.4 & 55.8 & 92.0 & 1.33\\
 $1\TeV$ $ < m\tr$ & 26.0 & 71.6 & 30.7 & 101 & 0.25\\ 
\bhline{1 pt}
   \end{tabular}
   }
    \caption{\label{table:AP_U2_cutbflow} 
    Same as Table~\ref{table:AP_CV1_cutbflow} but for the $\text{U}_1$ LQ scenario with the $U(2)$ flavor symmetry, where $C_{V_1}=1$ and $\int\mathcal{L}\,dt=35.9\ifb$.
}
\end{table}

\section{Flavor observables}
\label{Sec:App_flavor}

In this appendix, the $\text{S}_1$ LQ contributions to $B \to K^{(\ast)}\nu\nubar$
and $B_s$--$\Bb_s$ mixing are discussed.

A ratios between the measured branching fractions of  $B\to K^{(\ast)} \nu\nubar$ 
and the  SM predictions is represented by $\mathcal{R}_{K^{(\ast)}}^{\nu \nubar}$~\cite{Buras:2014fpa}.
For a case of the minimal coupling of the $\text{S}_1$ LQ scenario,
we obtain~\cite{Carvunis:2021dss}
\beq
\mathcal{R}_{K^{(\ast)}}^{\nu \nubar} = \frac{2}{3} + \frac{1}{3}\frac{\left|C_{L,sb}^{{\rm SM},33} + C^{{\rm NP},33}_{L,sb}\right|^2}{\left|C_{L,sb}^{{\rm SM},33}\right|^2}\,,
\eeq 
with
\beq
C^{{\rm NP},33}_{L,sb} \simeq + 2 \frac{\pi}{\alpha} C_{V_1}\,,\qquad 
C_{L,sb}^{{\rm SM},33} \simeq -\frac{1.47}{\sin^2 \theta_{\rm W}}\,,
\eeq 
and
\beq
\mathcal{H}_{\rm eff}^{\nu \nu} = - \frac{ G_F \alpha}{\sqrt{2}\pi} V_{tb} V^\ast_{ts} C_{L,sb}^{fi} \left( \overline{s}\gamma^\mu P_L b\right) \left( \nubar_f \gamma_\mu (1-\gamma_5) \nu_i \right)
+\text{h.c.}\,,
\eeq
where there are no QCD corrections from the RG evolution.
Note that the WC, $C_{V_1}$, is defined by the effective Hamiltonian in Eq.~\eqref{Eq:effH}.
The Belle collaboration has provided a severe upper bound on $B \to K^\ast \nu \nubar$ as 
 $\mathcal{R}_{K^\ast}^{\nu \nubar} < 2.7$ at the $90\%$ C.L.~\cite{Belle:2017oht}.
 From these numbers,  we obtain
\beq
-0.011 < C_{V_1} < 0.027\,,
\label{Eq:BKnunu}
\eeq
for the $\text{S}_1$ LQ scenario. 
It is clearly seen that the $\text{S}_1$ LQ scenario is severely constrained (see Fig.~\ref{Fig:WC_bound_S1}).
It is known, however, that 
adding $SU(2)_L$ triplet scalar LQ $\text{S}_3$ can alleviate the constraints from the $b \to s \nu \nubar$ processes due to a destructive interference~\cite{Crivellin:2017zlb,Crivellin:2019dwb,Gherardi:2020qhc}.\kf\footnote{
Such a singlet-triplet LQ model can also explain the $b \to s \ell^+ \ell^- $ anomaly~\cite{LHCb:2017avl,LHCb:2021trn} and the muon $g-2$ anomaly~\cite{Muong-2:2021ojo}, simultaneously~\cite{Crivellin:2019dwb}.}

Next, the $\text{S}_1$ LQ contribution  to $\Delta M_s$ (via LQ--$\nu_{\tau}$ box)
is given as~\cite{Crivellin:2019dwb,Crivellin:2021lix}
\beq
\frac{\Delta M_s}{\Delta M_s^{\rm SM}}
= \left| 1 + \frac{C_1^{\rm NP}}{C_1^{\rm SM}}\right|\,,
\eeq
with 
\begin{align}
C_1^{\rm NP} \simeq \left(\frac{\alpha_s (M_{\rm LQ})}{\alpha_s (M_W)}\right)^{\frac{2}{7}} \frac{\left(V_{cb} G_F  M_{\rm LQ}\right)^2 }{4\pi^2} C_{V_1}^2\,,
\qquad C_1^{\rm SM} = 2.35 \frac{\left( V_{tb}V_{ts}^\ast G_F M_W\right)^2}{4 \pi^2}\,, 
\end{align}
and
\beq
\mathcal{H_{\rm eff}}= C_1 \left(\overline{s}\gamma^{\mu} P_L b\right)
\left(\overline{s}\gamma_{\mu} P_L b\right)\,.
\eeq
Here, the WC, $C_1$, is given at the electroweak scale, and the prefactor $\left[{\alpha_s (M_{\rm LQ})}/{\alpha_s (M_W)}\right]^{\frac{2}{7}}$ is the leading QCD correction from the RG evolution~\cite{Bagger:1997gg}.
Using the experimental data $\Delta M_s^{\rm exp} = (17.741 \pm 0.020)\,\text{ps}^{-1}$~\cite{Zyla:2020zbs} and 
the SM prediction is $ \Delta M_s^{\rm SM}=(18.4^{+0.7}_{-1.2})\,\text{ps}^{-1}$~\cite{DiLuzio:2019jyq}, 
one obtains the upper bound, ${\Delta M_s}/{\Delta M_s^{\rm SM}} < 1.11$, at $2\,\sigma$ level.


\bibliographystyle{utphys28mod}

\bibliography{ref}

\end{document}